\documentclass[twocolumn]{aastex63}

\accepted{\today}
\submitjournal{ApJ}

\usepackage{comment}
\usepackage{makecell}
\usepackage{amsmath}

\begin{document}

\author[0000-0002-8770-6764]{R\'eka K\"onyves-T\'oth}
\affiliation{Konkoly Observatory, CSFK, Konkoly-Thege M. ut 15-17, Budapest, 1121 Hungary}

\author[0000-0001-8764-7832]{J\'ozsef Vink\'o}
\affiliation{Konkoly Observatory, CSFK, Konkoly-Thege M. ut 15-17, Budapest, 1121 Hungary}
\affiliation{Department of Optics \& Quantum Electronics, University of Szeged, D\'om t\'er 9, Szeged, 6720 Hungary}
\affiliation{ELTE E\"otv\"os Lor\'and University, Institute of Physics, P\'azmany P\'eter s\'et\'any 1/A 1117, Budapest, Hungary}

\shorttitle{Ejecta mass of SLSNe I}
\shortauthors{K\"onyves-T\'oth et al.}

\correspondingauthor{R\'eka K\"onyves-T\'oth}
\email{konyvestoth.reka@csfk.mta.hu}

\graphicspath{{./}{}}

\title{Photospheric Velocity Gradients and Ejecta Masses of Hydrogen-poor Superluminous Supernovae --  Proxies for Distinguishing between Fast and Slow Events }

\begin{abstract}

We present a study of 28 Type I superluminous supernovae (SLSNe) in the context of the ejecta mass and photospheric velocity. We combine photometry and spectroscopy to infer ejecta masses via the formalism of radiation diffusion equations. We show an improved method to determine the photospheric velocity by combining spectrum modeling and cross correlation techniques. 
We find that Type I SLSNe can be divided into two groups by their pre-maximum spectra. Members of the first group have the W-shaped absorption trough in their pre-maximum spectrum, usually identified as due to O II. This feature is absent in the spectra of supernovae in the second group, whose spectra are similar to SN~2015bn. We confirm that the pre- or near-maximum photospheric velocities correlate with the velocity gradients: faster evolving SLSNe have larger photosheric velocities around maximum. We classify the studied SLSNe into the Fast or the Slow evolving group by their estimated photosheric velocities, and find that all those objects that resemble to SN~2015bn belong to the Slow evolving class, while SLSNe showing the W-like absorption are represented in both  Fast and Slow evolving groups. We estimate the ejecta masses of all objects in our sample, and obtain values in the range of 2.9 ($\pm$0.8) - 208 ($\pm$61) $M_\odot$, with a mean of $43 (\pm 12)~ M_\odot$. We conclude that Slow evolving SLSNe tend to have higher ejecta masses compared to the Fast ones. Our ejecta mass calculations suggests that SLSNe are caused by energetic explosions of very massive stars, irrespectively of the powering mechanism of the light curve.

\end{abstract}


\section{Introduction} \label{sec:intro}

A new class of transients, the so-called superluminous supernovae (SLSNe), was discovered and extensively studied in the past two decades. These extremely luminous events have at least ${\sim}10^{51}\, {\rm erg}$ total radiated energy, leading to  an absolute brightness of $M < -21$ in all bands of the optical wavelengths  \citep{galyam12,galyam19}. It has also been reported that these supernovae (SNe) prefer to explode in dwarf galaxies having low metallicity and  high specific star-formation rate \citep{Lunnan13, Lunnan14, Leloudas15, Angus16, japelj16, i, Schulze18,hatsu18}, although  some counterexamples are also known. For example, PTF10tpz \citep{arab19}, PTF10uhf \citep{i} and SN~2017egm \citep{, chen17,bose18,izzo18,yan18,hatsu20} occured in  relatively bright and metal-rich, or, at least not metal-poor, host galaxies.  The recent publications of \citet{s1, p1}, and \citet{Angus19} revealed that this population is quite multitudinous: some lower luminosity transients (e.q. DES14C1rhg with $M_{r} = -19.4$; \citealt{Angus19}) have also been classified as superluminous supernovae, because of the similar photometric or spectroscopic evolution to known, well-observed SLSNe \citep[e.g.][]{quimby18}.

Similarly to the traditional/normal supernovae, SLSNe can also be separated into two main sub-classes: the H-poor Type I, and the H-rich Type II SLSN group \citep{BW17}. 
SLSNe-II are divided into two distinct populations: the luminosity of Type IIn SLSNe is powered by the strong interaction with the surrounding, massive circumstellar medium (CSM, e.q. SN~2006gy; \citealt{smith07} or CSS121015; \citealt{benetti14}), and have similar spectroscopic properties and evolution to normal Type IIn SNe \citep{BW17}. The representatives of the second group, called normal Type II SLSNe show no visible signs of the CSM-interaction (e.g. SN~2013hx and PS15br; \citealt{inserra18}). 

This study focuses on several events belonging to the H-poor SLSN-class. The members of SLSNe-I are usually revealed to be spectroscopically similar to normal Ic or BL-Ic SNe \citep[e.g.][]{Pastorello10,e1, Yan17}, with the difference that events in the former class have larger luminosities. SLSNe-I can be also separated into two groups \citep{inserra18}: the Fast (e.g. SN~2005ap; \citealt{Lunnan13}) and the Slow evolving events (e.g. SN~2010kd; \citealt{kumar20,ktr20-2}), with an average light curve (LC) rise-time of $\sim 28$ days, and $\sim 52$ days, respectively. \citet{inserra18} examined a sample of SLSNe statistically, and showed that Slow evolving SLSNe exhibit lower, and slowly evolving, or nearly constant photospheric velocities ($v \lesssim 12000$ km~s$^{-1}$) from the maximum to +30 days phase, compared to the Fast evolving events having $v \gtrsim 12000$ km~s$^{-1}$, and larger velocity gradients. However, some studies suggest that the transition between Fast and Slow events is continuous: e.g. Gaia16apd (SN~2016eay) was found to be a SLSN with LC time-scale in between those of the two groups \citep{kangas17}.

In many cases the pre-maximum, photospheric phase spectra of Type I SLSNe can be distinguished from lower luminosity Type Ic and BL-Ic events by a peculiar W-like absorption blend between 3900 and 4500 \AA, which is identified to be due to O II \citep[e.g.][]{liu17}. Alternatively, this feature can be modeled using the mixture of different ions, e.g. O III and C III \citep{quimby07, dessart19, galyam19b, ktr20-2}.

In this paper, we present ejecta mass calculations for a sample of 28 Type I SLSNe, using publicly available photometric and spectroscopic data. Our sample selection process is described in  Section \ref{sec:sample}.

Recently, a similar study of SLSNe was carried out by \citet{nicholl15} who inferred the ejecta mass ($M_{\rm ej}$) of 24 SLSNe-I from bolometric LC modeling using the magnetar powering mechanism of the LC \citep{maeda07}, resulting in an average $M_{\rm ej}$ of 10 $M_\odot$, with a range of 3 and 30 $M_\odot$ for their sample. \citet{yu17} also inferred the ejecta mass of 31 SLSNe by fitting their bolometric LCs utilizing the  magnetar engine model. On the other hand, from pair instability supernova (PISN; e.g. \citealt{galyam09,kasen11}) models, \citet{p1} showed that the ejecta mass of some SLSNe may exceed far the values inferred by \citet{nicholl15} from the magnetar model: for example, the initial mass of iPTF16eh was estimated to be 115 $M_\odot$.

In our study the ejecta masses were inferred directly from the formulae derived by \citet{arnett80} (shown in detail in Section \ref{sec:theory}), instead of full bolometric LC modeling. Our approach has the advantage of being independent from the assumed powering mechanism as long as the heating source is centrally located and the ejecta is optically thick, which are probably valid assumptions during the pre-maximum phases.  

In our calculations the photospheric velocities ($v_{\rm phot}$) of the examined SLSNe measured before or near maximum light play crucial role. In Section \ref{sec:vphot} we show photospheric velocity estimates for each object using a method that can provide reasonable $v_{\rm phot}$ values in a computationally less expensive way  
than modeling all available spectra individually. We use a combination of spectrum modeling and the cross-correlation technique, similar to \citet{liu17} (see also e.g. in \citealt{takats12}). We also find that the W-shaped feature, typically observed in the pre-maximum spectra of SLSNe-I, is not always present, and the spectra without it seem reminiscent of SN~2015bn. We infer post-maximum photospheric velocities as well (see Section \ref{subsec:maxutan}) in order to classify the studied SLSNe into the Fast or the Slow evolving SLSN-I sub-classes via their velocity gradients \citep{inserra18}.

The ejecta mass calculations are presented in Section \ref{sec:mej} as well as the comparison of our results with those of \citet{nicholl15}. We discuss our findings in Section \ref{sec:discussion}, and summarize them in Section \ref{sec:summary}.

\section{Estimating the mass of an optically thick SN ejecta}\label{sec:theory}

The analytical description of the light variation of supernovae (SNe) was first described by \citet{arnett80}, then extended by \citet{arnett82} and \citet{arnettfu}. This simple semi-analytical treatment has been applied for many SN subtypes including SNe II-P \citep{popov93, arnettfu, nagy14}, Ia \citep{pinto1, pinto2},  Ib/c \citep{valenti08} and SLSNe \citep{manos12, manos13}. \citet{BW17} presents a concise, yet in-depth summary of these analytical models (referred to as ``Arnett-models'' hereafter), which we follow here for our purposes. 

The model assumes a homologously expanding ($v(r) \sim r$) ejecta having constant density profile ($\rho(r,t) = \rho_0 t^{-3}$). Shortly after explosion the ejecta is very hot, implying that radiation pressure dominates the gas pressure and the internal energy is governed by the radiation energy density ($u \sim T^4$). Within this context the energy conservation law can be written as 
\begin{equation}
{{du} \over {dt}} + P {{dV} \over {dt}} ~=~ \varepsilon - {{\partial L} \over {\partial m}}, 
\label{eq-energy}
\end{equation}
where $V = 1/\rho$ is the specific volume (i.e. volume of unit mass), $u$ is the specific internal energy, $\varepsilon$ is the specific energy injection rate, $L$ is the luminosity and 
$m$ is the Lagrangian mass coordinate ($dm = 4 \pi r^2 \rho dr$).  
 
Another very important, simplifying assumption is that the opacity of the ejecta is constant in space and also in time as long as there is no recombination. Since the density profile of the ejecta has been already set up as a constant in space, in first approximation this is a physically self-consistent assumption, if the opacity is dominated by Thomson scattering on free electrons as it frequently happens in hot SN envelopes. This assumption, however, ignores the chemical stratification within the SN ejecta that may cause significant spatial variation in the number density of free electrons even if the mass density profile is flat. See e.g. \citet{nagy18} for further details on the opacity variations in different SN types. The effect of recombination is taken into account by \citet{arnettfu} \citep[see also][]{nagy16}. 

A consequence of the simplifying assumptions is that in Eq. \ref{eq-energy} the spatial and temporal parts are separable, and the solution leads to an eigenvalue problem \citep{arnett80}. The 
temperature profile inside the ejecta has a fixed spatial profile of $\psi(x) = \sin(\sqrt{\alpha} x) / (\sqrt{\alpha} x)$, where $x = r / R_{\rm SN}$ is the normalized radial coordinate and $\alpha$ is the eigenvalue of the problem. \citet{arnett80} showed that $\alpha = \pi^2$ corresponds to the so-called ``radiative zero'' solution that goes to zero at the surface of the ejecta ($\psi(1) = 0$). 
It is important to note that the Arnett-model {\it assumes} that such a temperature profile is valid as early as $t=0$, which is also true for the onset of the homologous expansion. Thus, this model ignores the initial ``dark phase'' between the explosion and the moment of first light \citep[e.g.][]{piro14}. This and other limitations of the Arnett-models are thoroughly discussed by \citet{kk19}.    

Shortly after explosion, when the whole ejecta is hot and dense, it is optically thick, thus, the photosphere is located near the outer boundary (denoted as $R_{\rm SN}$ above). Photons that are generated inside the ejecta, {\it regardless of the physical nature of the powering mechanism}, must diffuse out to the photosphere in order to escape. Following \citet{arnett80}, the timescale of the photon diffusion can be expressed as 
\begin{equation}
t_{\rm d} ~=~ {{3 R_{\rm SN}^2 \rho \kappa} \over {\alpha c}},
\label{eq-tdif}
\end{equation}
where $\alpha = \pi^2$ is the eigenvalue of the radiative zero solution. In the diffusion approximation the luminosity inside the ejecta is 
\begin{equation}
L(r) ~=~ - 4 \pi r^2 {{\lambda c} \over 3} {{du} \over {dr}} ~=~ - 4 \pi r^2 {{c} \over {3 \kappa \rho}} {{du} \over {dr}} ,  
\label{eq-lr}
\end{equation}
where $\lambda = (\kappa \rho)^{-1}$ is the photon mean free path. Eq. \ref{eq-lr} is similar to the expression for radiative energy transport 
within stellar interiors. 

The other characteristic timescale of the problem is the expansion timescale (also called as ``hydrodynamic timescale'') that is simply
\begin{equation}
t_h ~=~ {R_{SN} \over v_{SN}},
\label{eq-th}
\end{equation}   
where $v_{\rm SN}$ is the expansion velocity at $R_{\rm SN}$.  Since real SN ejecta have no constant density profiles, $v_{SN}$ cannot be related unambiguously to measured SN velocities. Therefore, it is often referred to as the ``scaling velocity'' that characterizes only the approximate analytic solution. 

Since $R_{\rm SN} \sim t$ while $\rho \sim t^{-3}$, $t_d \sim t^{-1}$ is decreasing in time, while $t_h \sim t$ is increasing.  At the start of the expansion $t_d >> t_h$, thus, later there is a moment when $t_d$ and $t_h$ become equal. At this moment the diffusing photons have the same effective speed as the expanding ejecta, thus, the thermalized photons from the instantaneous energy input (the heating source) are no longer trapped inside the ejecta. In another words, the escaping luminosity is equal to the instantaneous energy input, which occurs when the luminosity reaches its maximum, $L_{\rm max}$ (``Arnett's rule'', see also \citet{kk19}). If $t_{\rm max}$ is the moment of maximum light in the observer's frame, and $t_0$ denotes the moment of explosion (actually, the moment of the start of homologous expansion, see above), then the rise time to maximum light in the SN rest frame is 
\begin{equation}
t_{\rm rise} ~=~ {{t_{\rm max} - t_0} \over {1 + z}}, 
\label{eq-trise}
\end{equation}
where $z$ is the redshift of the SN. 

Close to $t_{\rm max}$, when $t_d \approx t_h$, the optical depth of the whole constant density ejecta can be written as \citep{BW17}
\begin{equation}
\tau ~=~ \kappa \rho R_{SN} ~=~ {{\pi^2 c} \over {3 v_{SN}} } \approx {{3 c} \over v_{SN}}. 
\label{eq-tau}
\end{equation} 
Because $c >> v_{\rm SN}$, $\tau >> 1$, i.e. at $t \sim t_{\rm max}$  most of the ejecta is still optically thick, as expected. As a consequence, the photosphere, where the ejecta becomes transparent, must be located close to $R_{\rm SN}$. i.e. $R_{\rm phot} \approx R_{\rm SN}$.  

Eq. \ref{eq-tau} allows a possibility for estimating the ejecta mass, in particular the mass of the optically thick part inside the photosphere \citep[e.g.][]{ktr20-2}. Due to the constant density profile $\rho = 3 M_{\rm ej} (4 \pi)^{-1} R_{\rm phot}^{-3}$. Inserting this into Eq. \ref{eq-tau} one may get
\begin{equation}
M_{\rm ej} ~=~ 4 \pi {c \over \kappa} v_{ \rm ph} t_{\rm rise}^2, 
\label{eq-mej1}
\end{equation}
where we used the photospheric velocity at maximum light, $v_{\rm ph}$, to approximate the scaling velocity, $v_{\rm SN}$, of the optically thick ejecta, and $R_{\rm ph} = v_{\rm ph} t_{\rm rise}$ in the SN rest frame.

Eq. \ref{eq-mej1} is very similar to the original expression introduced by \citet{arnett80}, which gives the total ejecta mass from the ``mean light curve timescale'' $t_m = \sqrt{2 t_h t_d}$ in the following form:
\begin{equation}
M_{\rm ej} ~=~ {{\beta c} \over {2 \kappa}} v_{\rm SN} t_m^2, 
\label{eq-mej2}
\end{equation}
where $\beta \approx 13.8$ is an integration constant, slightly depending on the ejecta density profile. Even though $t_m$ cannot be measured directly, its value is similar to the rise time of the light curve, thus Eq. \ref{eq-mej1} and \ref{eq-mej2} provide approximately the same ejecta mass for a given SN, with  the systematic difference of a constant multiplier: the quotient of the two formulae is
\begin{equation}
    4 \pi \cdot {{2} \over {\beta}}~=~ 1.82 .
\end{equation}

In the rest of this paper we apply Eq. \ref{eq-mej1} and \ref{eq-mej2} to observational data of SLSNe-I to derive constraints for their ejecta mass. We note that these estimates do not use any assumption on the physics of the powering mechanism (magnetar, radioactivity, etc.) as long as the heating source is centrally located, thus, the thermalized photons must diffuse through the whole ejecta.

\section{Sample selection} \label{sec:sample}

We constructed a sample of SLSNe from the events listed in the Open Supernova Catalog\footnote{https://sne.space/} \citep{guill17} before 2020, having at least 10 epochs of observed photometric data. From the identified 98 objects,  18 were immediately excluded from the sample because of being  Type II SLSNe.  As the main goal of this study is to determine the ejecta masses of Type I SLSNe using Eq. \ref{eq-mej1} and Eq. \ref{eq-mej2},  spectra taken before or shortly after the moment of the maximum light are crucial to identify the typical SLSN-I features and estimate the value of the photospheric velocity ($v_{\rm phot}$). Without knowing $v_{\rm phot}$ at maximum, the ejecta mass calculations based on the formulae presented in Section \ref{sec:theory} would not lead to reasonable results. Out of pre-selected 80 SLSNe-I, 39 did not pass the criterion of possessing pre-maximum spectra. From the remaining 41 objects, 13 additional SLSNe-I had to be removed from the sample because of several reasons listed in the Appendix. All SLSNe excluded from our analysis are collected in Table \ref{tab:removed} in the Appendix, for completeness.

Table \ref{tab:basic} contains the basic observational data of our final sample (28 SLSNe) obtained from the Open Supernova Catalog.

Before the analysis,  all downloaded spectra were normalized to the flux at 6000 \AA, and corrected for redshift and Milky Way extinction.

\begin{table*}
\caption{Basic data of the studied SLSNe.}
\label{tab:basic}
\begin{center}
\scriptsize
\begin{tabular}{lcccccccc}
\hline
\hline
SLSN & $t_0$ & $t_{\rm max}$ & $M_{\rm max}$  & R.A. & Dec. & $z$ & $E(B-V)$  & References\\ 
&  MJD & MJD & mag &  & & & mag & \\
\hline
SN2005ap & 53415   & 56122 & 18.16 & 13:01:14.84 & +27:43:31.4 & 0.2832 & 0.0072 &a, b, c, d\\
SN2006oz & 53415 & 54068.2 & 19.8  & 22:08:53.56 & +00:53:50.4 & 0.376 & 0.0403 & b, e, f, g, h \\
SN2010gx & 53415 & 55277.2 & 17.62 & 11:25:46.71 & -08:49:41.4 & 0.2299 & 0.0333 &a, b, i, j, k, l\\
SN2010kd  & 55499.5 &  55552.2 & 16.16  & 12:08:01.11 & +49:13:31.1 & 0.101 &    0.0197 & a,  b, j \\
SN2011kg & 55907  & 55938.2 & 18.39  & 01:39:45.51 & +29:55:27.0 & 0.1924 & 0.0371 & b, i, j, m, n\\
SN2015bn  & 57000 & 57101.2 & 15.69  & 11:33:41.57 & +00:43:32.2 & 0.1136 & 0.0221 & b, o, p \\
SN2016ard & 57424 & 57454.2 & 18.39  & 14:10:44.55 & -10:09:35.4 & 0.2025 & 0.0433 & b, q \\
SN2016eay & 57509 & 57530.2  & 15.2  & 12:02:51.71 & +44:15:27.4 & 0.1013 & 0.0132 & b, r, s\\
SN2016els & 57578 & 57605.2 & 18.31 & 20:30:13.920 & -10:57:01.81 & 0.217 & 0.0467 & b, j \\
SN2017faf & 57908 & 57941.2 & 16.78 & 17:34:39.98 & +26:18:22.0 & 0.029 & 0.0482  & b, t\\
SN2018bsz & 58197 & 58275.2 & 13.99  & 16:09:39.1 & -32:03:45.73 & 0.02667 & 0.2071 & b, u, v\\
SN2018ibb & 58336 & 58466.2  & 17.66  & 04:38:56.96 & -20:39:44.01 & 0.16 & -- & w \\
SN2018hti & 58197 & 58486.2 & 16.46 & 03:40:53.75 & +11:46:37.29 & 0.063 & -- & x, y \\
SN2019neq & 58701 & 58766.2 & 17.79  & 17:54:26.736 & +47:15:40.56 & 0.1075 & -- & z , a1\\
DES14X3taz & 57021 & 57093.2 & 20.54  & 02:28:04.46 & -04:05:12.7 & 0.608 & 0.022  & b, b1, c1\\
iPTF13ajg & 56348 & 56430.2 & 19.26 & 16:39:03.95 & +37:01:38.4 & 0.74 & 0.0121 & b, k, d1 \\
iPTF13ehe & 56565 & 56676.2 & 19.6 & 06:53:21.50 & +67:07:56.0 & 0.3434 & 0.0434 & b ,k, e1 \\
LSQ12dlf & 56098 & 56150.2 & 18.46  & 01:50:29.80 & -21:48:45.4 & 0.255 & 0.011 & b, f1, g1, h1, i1 \\
LSQ14an & 56639 & 56660.2 & 18.6 & 12:53:47.83 & -29:31:27.2 & 0.163 & 0.0711  & b, j1, k1\\
LSQ14mo & 56659  & 56693.2 & 18.42  & 10:22:41.53 & -16:55:14.4 & 0.253 & 0.0646 & b, j, l1 \\
LSQ14bdq & 56735 & 56660.2 & 19.16  & 10:01:41.60 & -12:22:13.4 & 0.345 & 0.0559 & b, m1, n1\\
PS1-14bj & 56597 & 56808.2 & 21.19  & 10:02:08.433 & +03:39:19.0 & 0.5215 & 0.0205 & b, o1, p1, q1\\
PTF09atu & 54999 & 55062.2 & 19.91  & 16:30:24.55 & +23:38:25.0 & 0.5015 & 0.0409 &b, i, k, r1, s1, t1\\
PTF09cnd & 55017 & 55085.2 & 17.08  & 16:12:08.94 & +51:29:16.1 & 0.2584 & 0.0207 & i, j, k, f1, t1 \\
PTF10nmn & 55267 & 55385.2 & 18.52  & 15:50:02.81 & -07:24:42.38 & 0.1237 & 0.1337 &b, i, r1 \\
PTF12dam & 56021 & 56091.2 & 15.66  & 14:24:46.20 & +46:13:48.3 & 0.1074 & 0.0107 &b, i, j, k, u1, v1 \\
PTF12gty & 56082  &56139.2 & 19.45  & 16:01:15.23 & +21:23:17.4 & 0.1768 & 0.06 & b, k, r1\\
SSS120810 & 56122 & 56159.2 & 17.38  & 23:18:01.80 & -56:09:25.6 & 0.156 & 0.0158 &b, g1, h1, w1, x1 \\
\hline
\end{tabular}
\tablecomments{a:\citet{a}; 
b: \citet{b};
c: \citet{c};
d: \citet{d};
e: \citet{e};
f: \citet{f};
g: \citet{g};
h: \citet{h};
i: \citet{i};
j: \citet{j};
k: \citet{k};
l: \citet{l};
m: \citet{m};
n: \citet{n};
o: \citet{o};
p: \citet{p};
q: \citet{q};
r: \citet{r};
s: \citet{s};
t: \citet{t};
u: \citet{u};
v: \citet{v};
w: \citet{w};
x: \citet{x};
y: \citet{y};
z: \citet{z};
a1:\citet{a1}; 
b1: \citet{b1};
c1: \citet{c1};
d1: \citet{d1};
e1: \citet{e1};
f1: \citet{f1};
g1: \citet{g1};
h1: \citet{h1};
i1: \citet{i1};
j1: \citet{j1};
k1: \citet{k1};
l1: \citet{l1};
m1: \citet{m1};
n1: \citet{n1};
o1: \citet{o1};
p1: \citet{p1};
q1: \citet{q1};
r1: \citet{r1};
s1: \citet{s1};
t1: \citet{t1};
u1: \citet{u1};
v1: \citet{v1};
w1: \citet{w1};
x1: \citet{x1}
}
\end{center}
\end{table*}

\section{Photospheric velocity measurement}\label{sec:vphot}

In this section, we describe a method for estimating the photospheric velocity of SLSNe-I in our sample. The $v_{\rm phot}$ value before or near the moment of maximum light plays a major role in the ejecta mass calculations (see Section~\ref{sec:theory}). Post-maximum photospheric velocities are needed also in order to infer velocity gradients, and classify these events into the Fast or the Slow evolving SLSN-I subgroups.

\begin{figure}
\centering
\includegraphics[width=8cm]{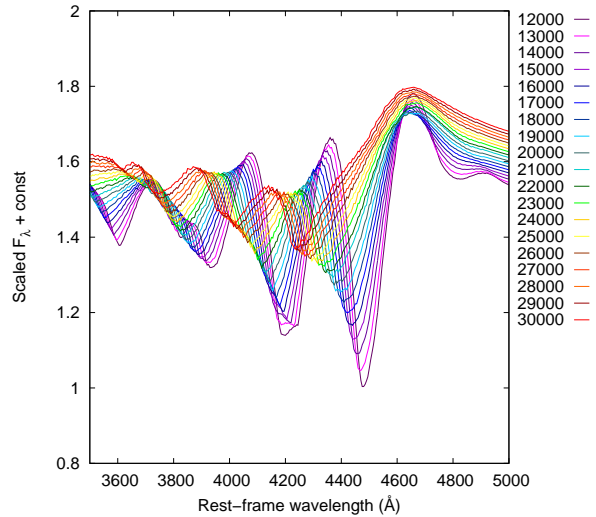}
\caption{SYN++ models built with $T_{\rm phot} = 17000$ K for the W shaped O II blend appearing typically between 3900 and 4500 \AA\ in the pre-maximum spectra Type I SLSNe. Different colors code the models having  different $v_{\rm phot}$  values ranging from 10000 to 30000 km s$^{-1}$.  }
\label{fig:OIImodels}
\end{figure}

\begin{figure*}
\centering
\includegraphics[width=12cm]{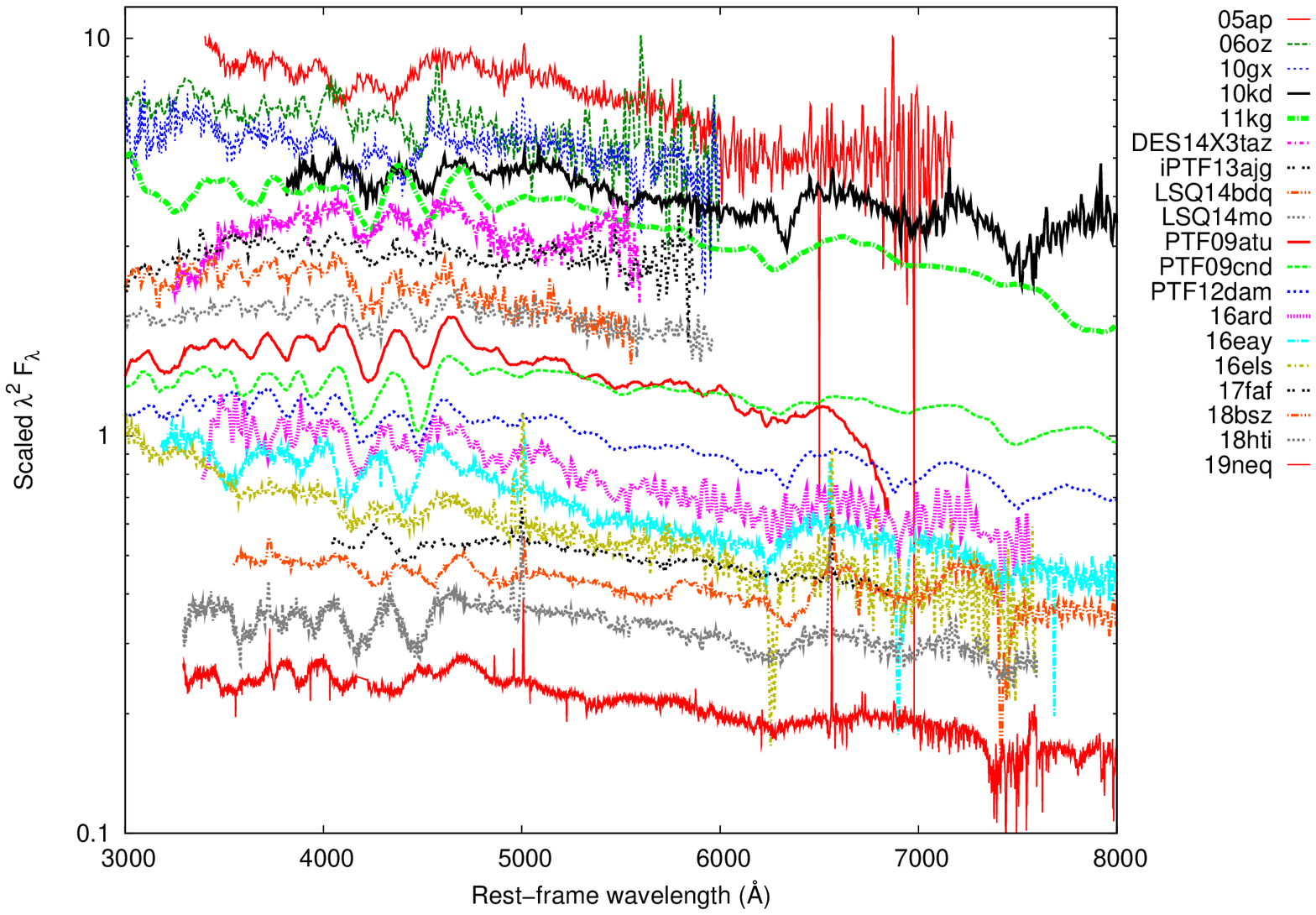}
\caption{Observed pre-maximum spectra of Type W SLSNe. The colors code the individual objects, and the spectra are shifted vertically for clarity. }
\label{fig:sp_slsne_w}
\end{figure*}

\begin{figure*}
\centering
\includegraphics[width=12cm]{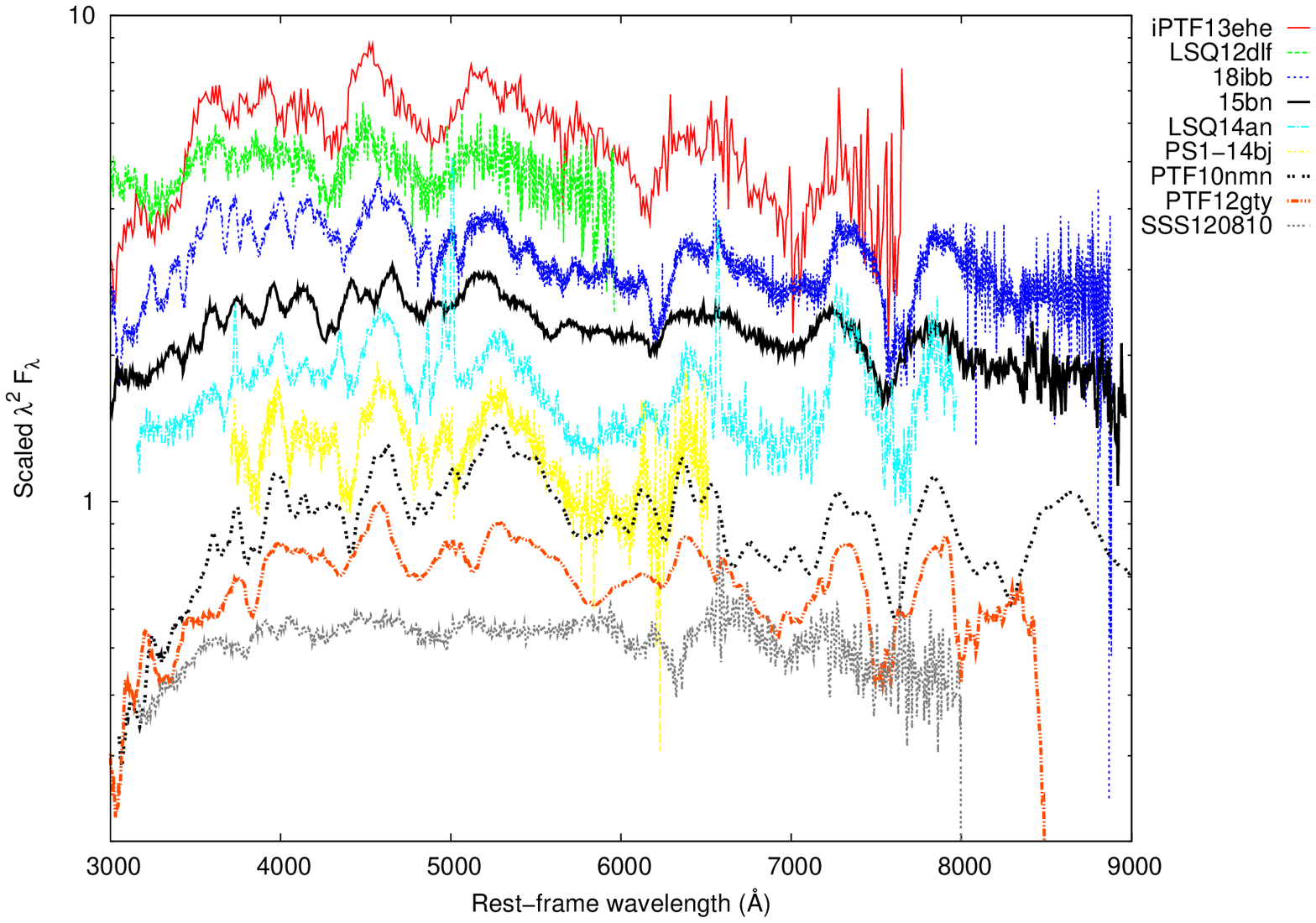}
\caption{Observed pre-maximum spectra of Type 15bn SLSNe. The colors code the individual objects, and the spectra are shifted vertically for clarity. }
\label{fig:sp_slsne_15bn}
\end{figure*}

However, getting realistic $v_{\rm phot}$ estimates is not a trivial problem, as a typical SLSN spectrum contains broad and heavily blended features instead of isolated and easily identifiable P Cygni profiles. In this case a spectrum synthesis code is required to reliably identify the spectroscopic features and the ejecta composition, but even this method suffers from ambiguity: occasionally, the absorption blends can be fitted equally well with features of different ions \citep[see e.g.][]{ktr20-2}. Furthermore, modeling each available spectrum in our sample would be very time consuming. Thus, in Section \ref{subsec:method} we present a faster and reasonably accurate method by combining spectrum synthesis and cross-correlation \citep[see also e.g.][]{takats12, liu17} to estimate the $v_{\rm phot}$ of the 28 SLSNe we studied. 

\subsection{Methodology}\label{subsec:method}

According to e.g. \citet{quimby18}, and \citet{perley19}, a W-shaped absorption feature appearing between $\sim$3900 and $\sim$4500 \AA\  is typically present in the pre-maximum spectra of Type I SLSNe. It is usually modeled as a blend of O II lines, and assumed to appear in all spectra of Type I SLSNe. \citet{liu17} examined a large set of normal and superluminous SNe, and noticed that this W-shaped O II feature can be found in all Type I SLSNe, but missing from the spectrum of normal Type Ic or broad-lined Ic SNe. They proposed the presence/absence of the W-feature as a tool for distinguishing between SLSNe and normal Ic SN events using only pre-maximum spectra.

Motivated by these previous findings, we assumed that the W-shaped feature plays a significant role in the spectrum formation of all SLSNe in our sample. We built a series of SYN++ models \citep{thomas11} containing only O II features (see Figure \ref{fig:OIImodels}).
These models share the same local parameters, e.g. the photospheric temperature ($T_{\rm phot}$) of 17000~K, but have different $v_{\rm phot}$ values ranging from 10000 to 30000 km s$^{-1}$, as shown in Figure \ref{fig:OIImodels} with different colors. The fixed value of all global ($a_0$, $v_{\rm phot}$, $T_{\rm phot}$) and local ($\log \tau$, $v_{\rm min}$, $v_{\rm max}$, $aux$, $T_{\rm exc}$)  SYN++ model parameters can be found in Table \ref{tab:syn_alltypes}. in the Appendix. Here, we utilize the name {\it global} to the parameters referring to the whole model spectrum, and {\it local} to the ones fitting the lines of individual elements in the spectrum.

Next, we cross-correlated the O II models to each other using the {\tt fxcor} task in the {\tt onedspec.rv} package of IRAF\footnote{IRAF is distributed by the National Optical Astronomy
Observatories, which are operated by the Association of Universities for Research in Astronomy, Inc., under cooperative agreement with the National Science Foundation. http://iraf.noao.edu} (Image Reduction and Analysis Facility). We chose the model corresponding to $v_{\rm phot}~=~10000$ km s$^{-1}$ as the template spectrum, and computed the cross-correlation velocity differences ($\Delta v_X$) between the template and all other model spectra. Then, by comparing $\Delta v_{\rm X}$ with the real velocity differences between the models ($\Delta v_{\rm phot}$), we obtained a formula to convert the velocity differences inferred by {\tt fxcor} to real, physical velocity differences between the SYN++ models. Having this correction formula we are able to use the cross-correlation method to determine reliable velocities for the observed spectra, despite the well-known issues with applying cross-correlation to spectra with P Cygni features \citep[e.g][]{takats12}. Our method is similar to the one developed by \citet{liu17}, but we focused on the more pronounced pre-maximum O~II features in the model instead of the Fe~II $\lambda5169$ feature in post-maximum spectra. 

Afterwards, we cross-correlated the 28 observed spectra in the sample with the template O~II model spectrum (i.e. the one having $v_{\rm phot}~=~10000$ km s$^{-1}$). 
We derived $v_{\rm phot}$ for the observed spectra by getting $\Delta v_X$ from {\tt fxcor}, then applying the correction formula between $\Delta v_{\rm phot}$ and $\Delta v_X$ (see above).  As a cross-check, we also plotted together the observed spectra with the SYN++ model having the nearest $v_{\rm phot}$ to the corrected velocity from {\tt fxcor} (see Section~\ref{subsec:W} and \ref{subsec:15bn}).

\subsection{New subtypes of SLSNe-I}

\begin{figure}
\centering
\includegraphics[width=8cm]{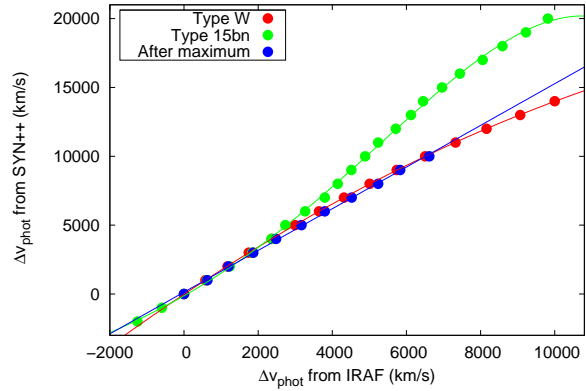}
\caption{Comparison of the relative velocities between the SYN++ models and the template model given by {\tt fxcor} ($\Delta v_{\rm X}$, horizontal axis) and the real velocity differences between the model $v_{\rm phot}$ values ($\Delta v_{\rm phot}$, vertical axis). Red and green colors represent the models for the ``Type W'' and ``Type 15bn'' subclasses, respectively, while the blue symbols correspond to the post-maximum spectra. The best-fit polynomials are also shown (see text).}
\label{fig:keresztkorr_all}
\end{figure}

Applying the method described above, we found that it did not work for about one-third of the sample, i.e. their derived photospheric velocities turned out to be physically impossible. Closer inspection of those spectra revealed the cause of this inconsistency: the W-shaped O II feature was not present in their spectra at all, therefore, the cross-correlation process did not work properly. 

After collecting the spectra without the W-shaped absorption feature, we found that they are similar to each other. The best-observed prototype of these SLSNe is SN~2015bn. 

Thus, we define two distinct groups of Type I SLSNe in our sample, characterized by the presence/absence of the W-shaped O II feature between 3900 and 4500 \AA. Hereafter we refer them as ``Type W'' and ``Type 15bn'' SLSNe (see Table \ref{tab:mej}). 

The observed spectra taken before maximum of all ``Type W'' SLSNe can be seen in Figure \ref{fig:sp_slsne_w}, while the same for ``Type 15bn'' events is shown in Figure \ref{fig:sp_slsne_15bn} with different colors representing each object in the given subclass.

For the latter subclass, we estimated their correct $v_{\rm{phot}}$ values by applying a different SYN++ model template in the cross-correlation process. The formula for correcting their $\Delta v_{\rm{X}}$ to $\Delta v_{\rm{phot}}$ was also re-calculated accordingly.

Further details on the cross-correlation analysis of
Type W and Type 15bn SLSNe are given in Section \ref{subsec:W} and \ref{subsec:15bn}, respectively.

In Section \ref{subsec:maxutan}, we present the $v_{\rm phot}$ estimates after the maximum for 9 objects in our sample, which had observational data in between +25 and +35 rest-frame days after maximum besides the pre-maximum data. Although e.g. \citet{galyam12} and \citet{inserra18} defined the   Fast and the Slow evolving subgroup of Type I SLSNe by their light curve evolution time-scales, the date of explosion is weakly defined in  several cases, thus the rise-time of these SLSNe remains uncertain. Therefore we utilized the photospheric velocity evolution by $\sim$30 days after the maximum for classification (see the details in Section \ref{subsec:F/s}).

\subsection{Type W SLSNe}\label{subsec:W}

\begin{figure*}
\centering
\includegraphics[width=4.8cm]{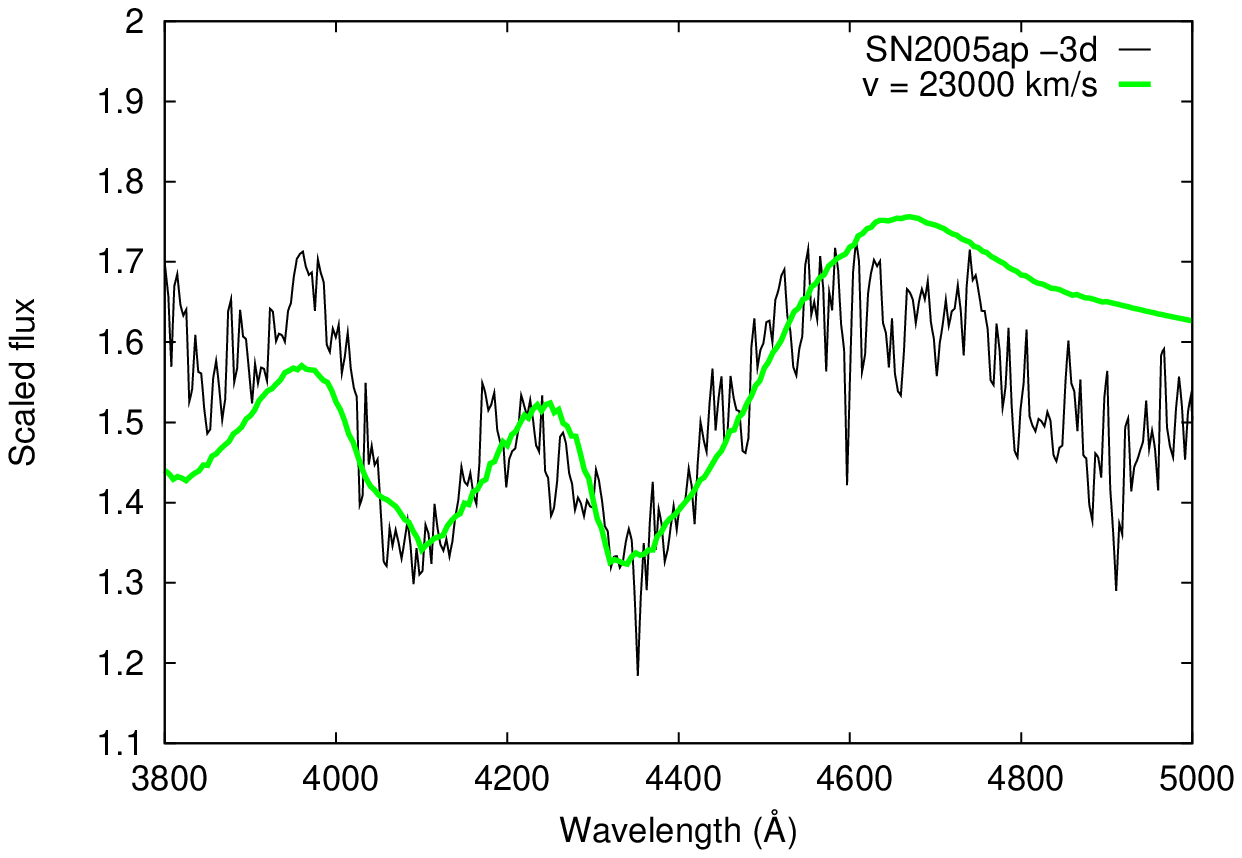}
\includegraphics[width=4.8cm]{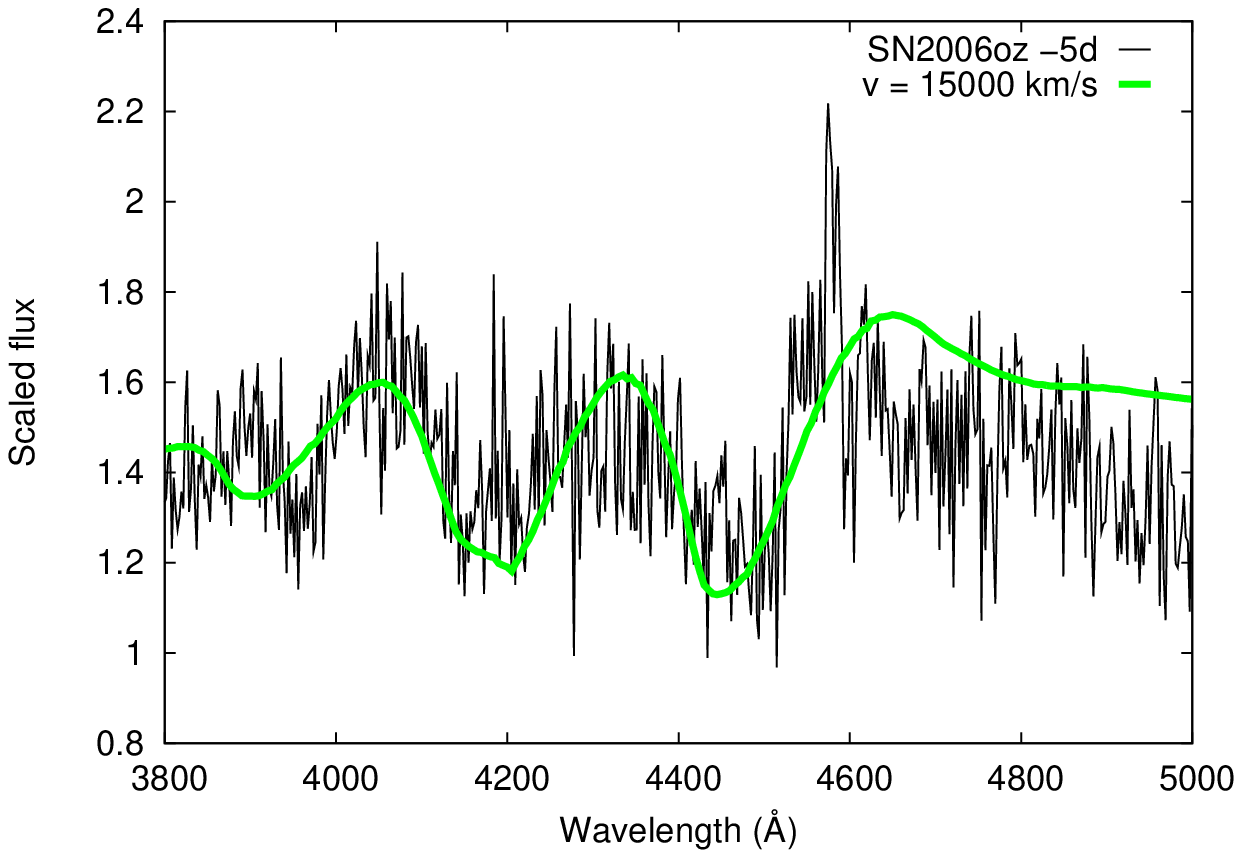}
\includegraphics[width=4.8cm]{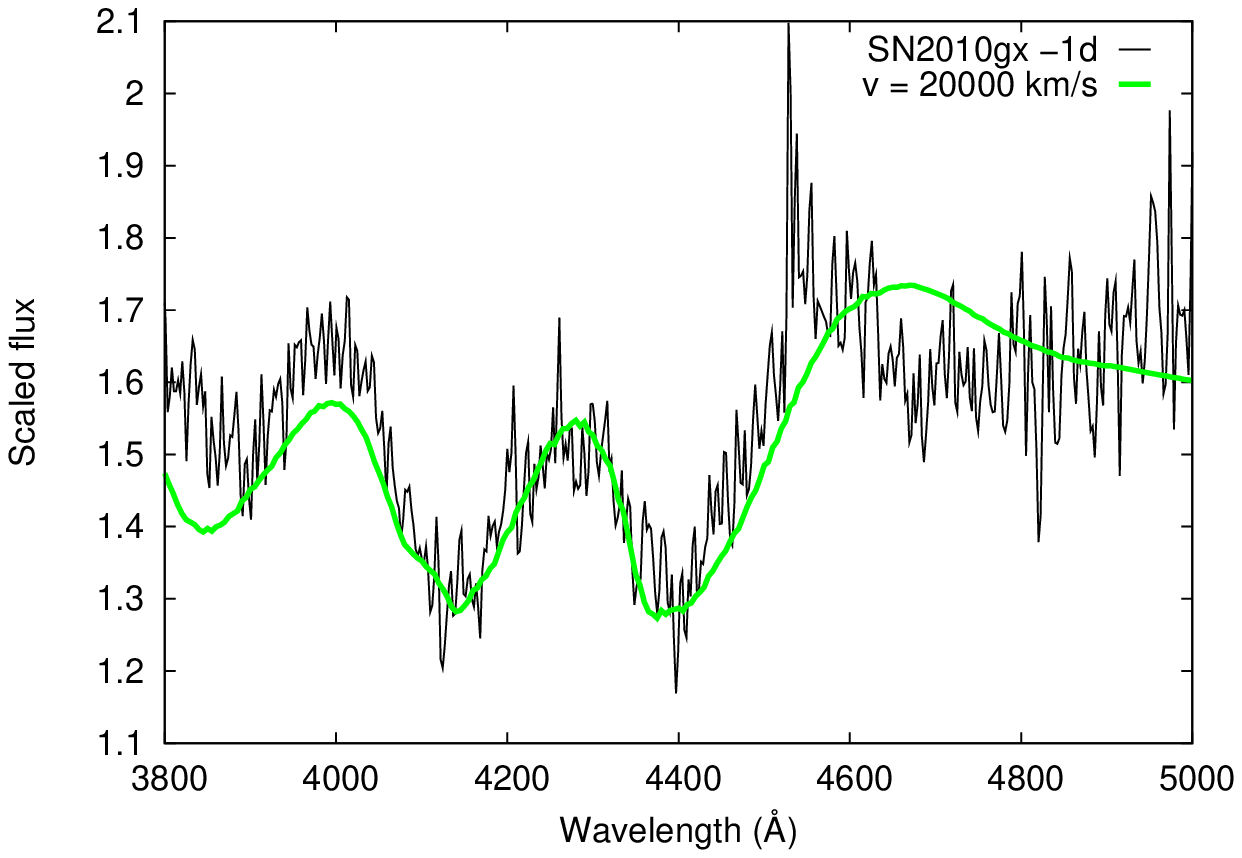}
\includegraphics[width=4.8cm]{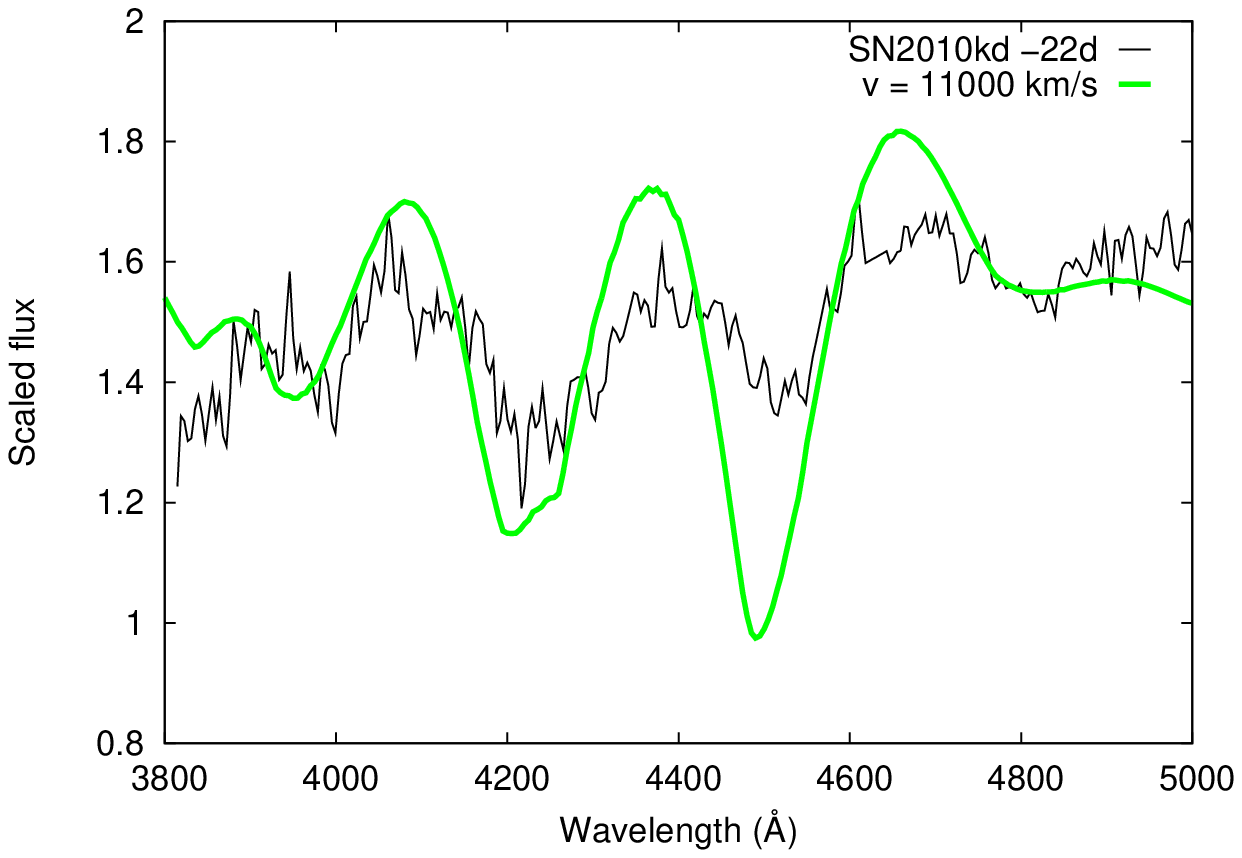}
\includegraphics[width=4.8cm]{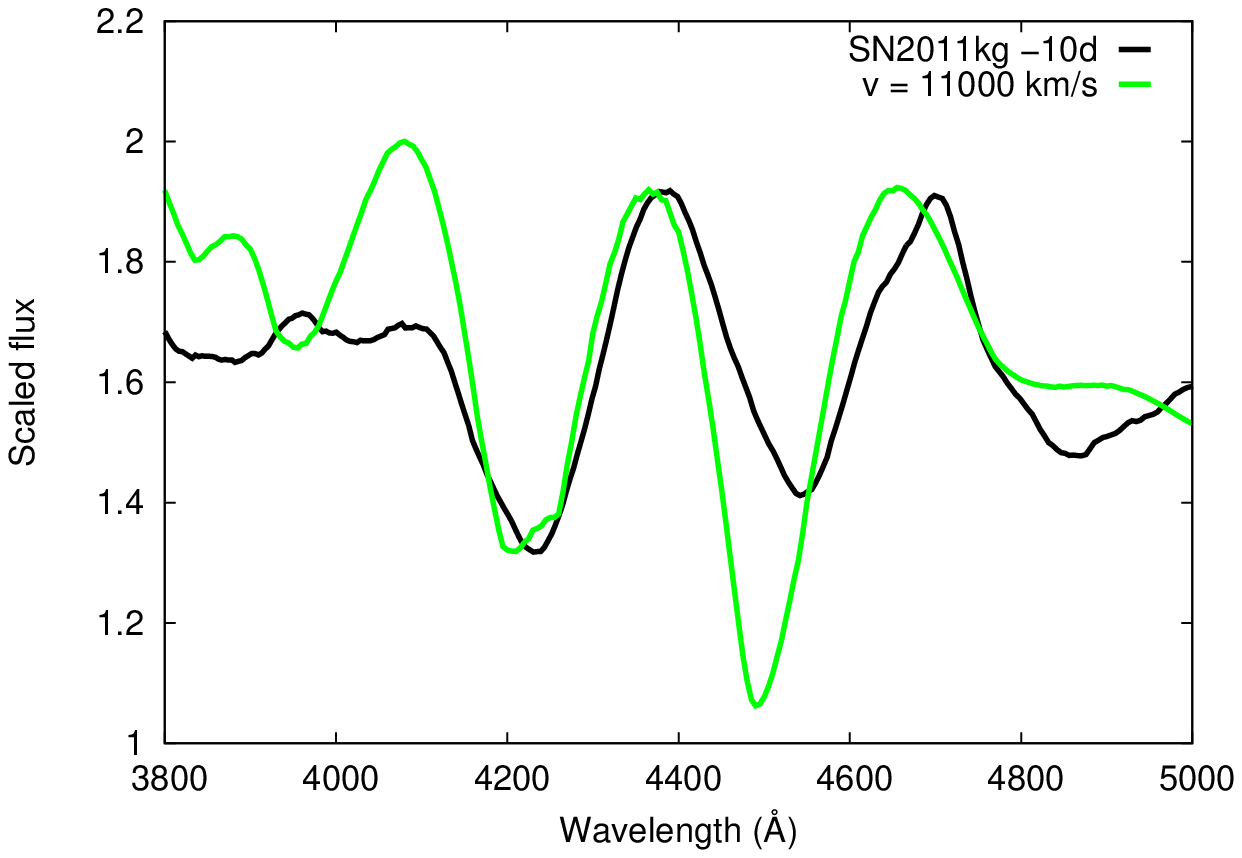}
\includegraphics[width=4.8cm]{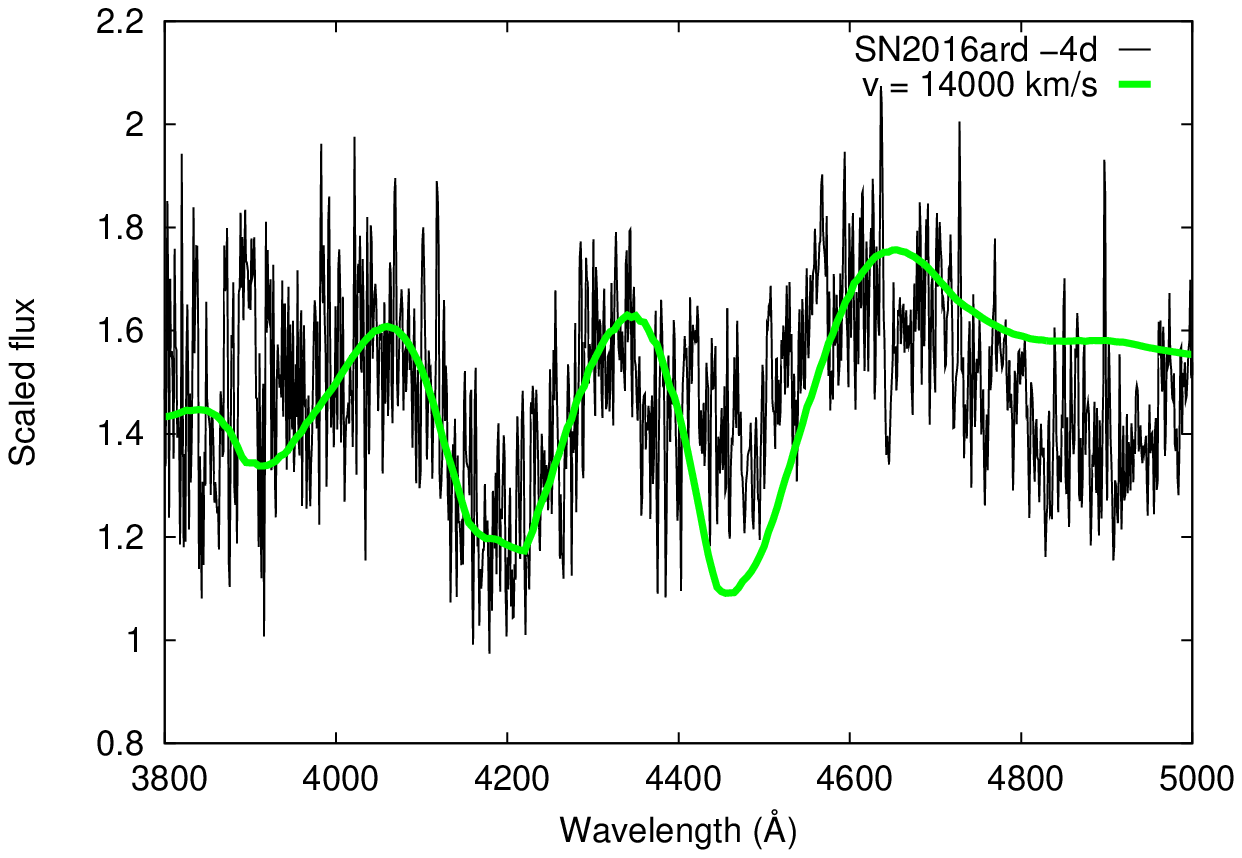}
\includegraphics[width=4.8cm]{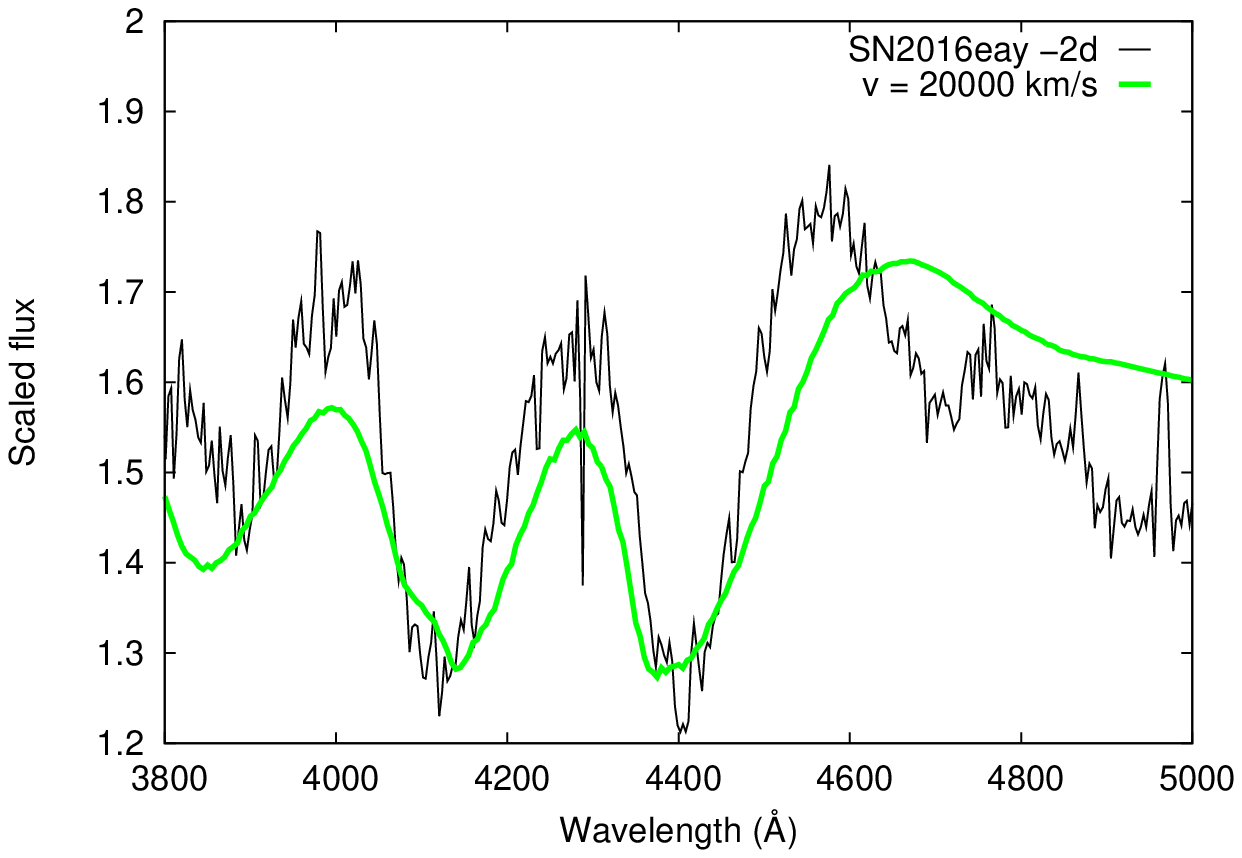}
\includegraphics[width=4.8cm]{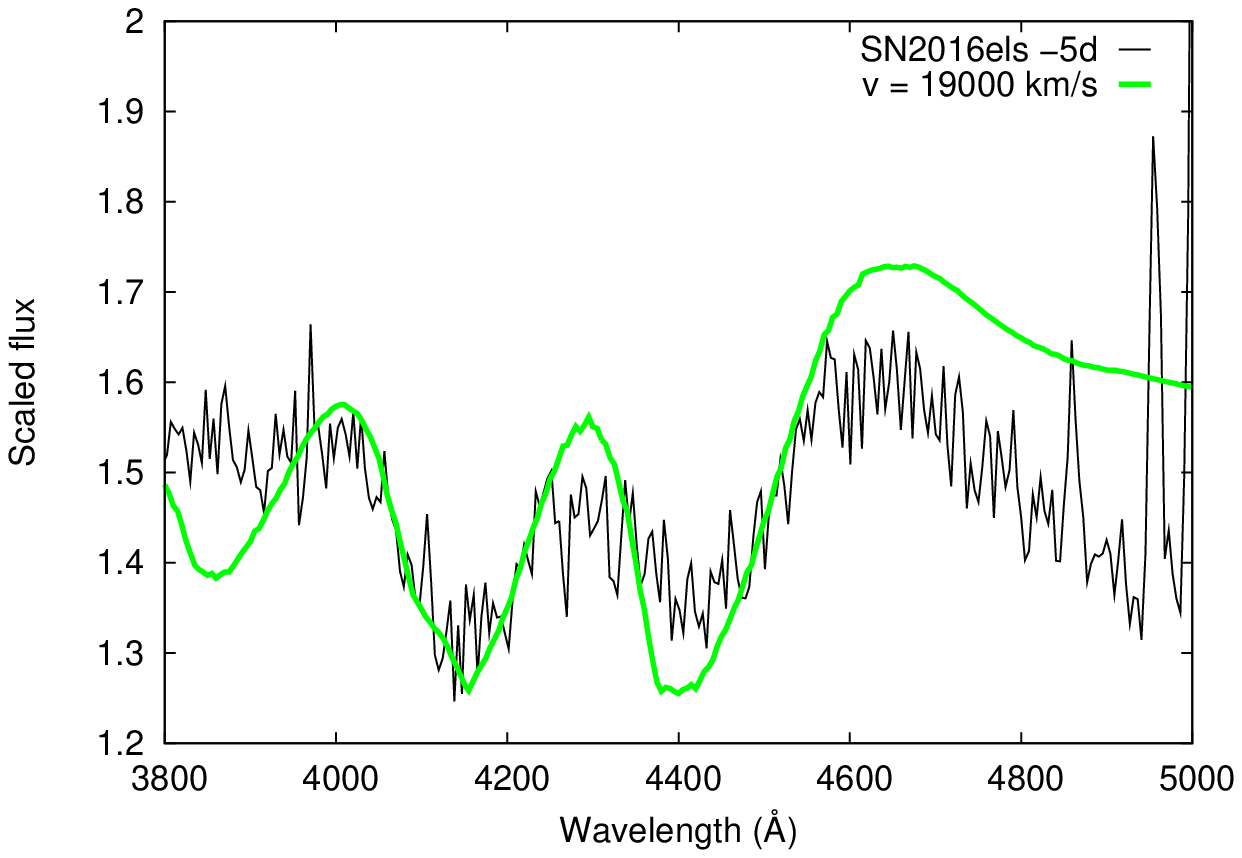}
\includegraphics[width=4.8cm]{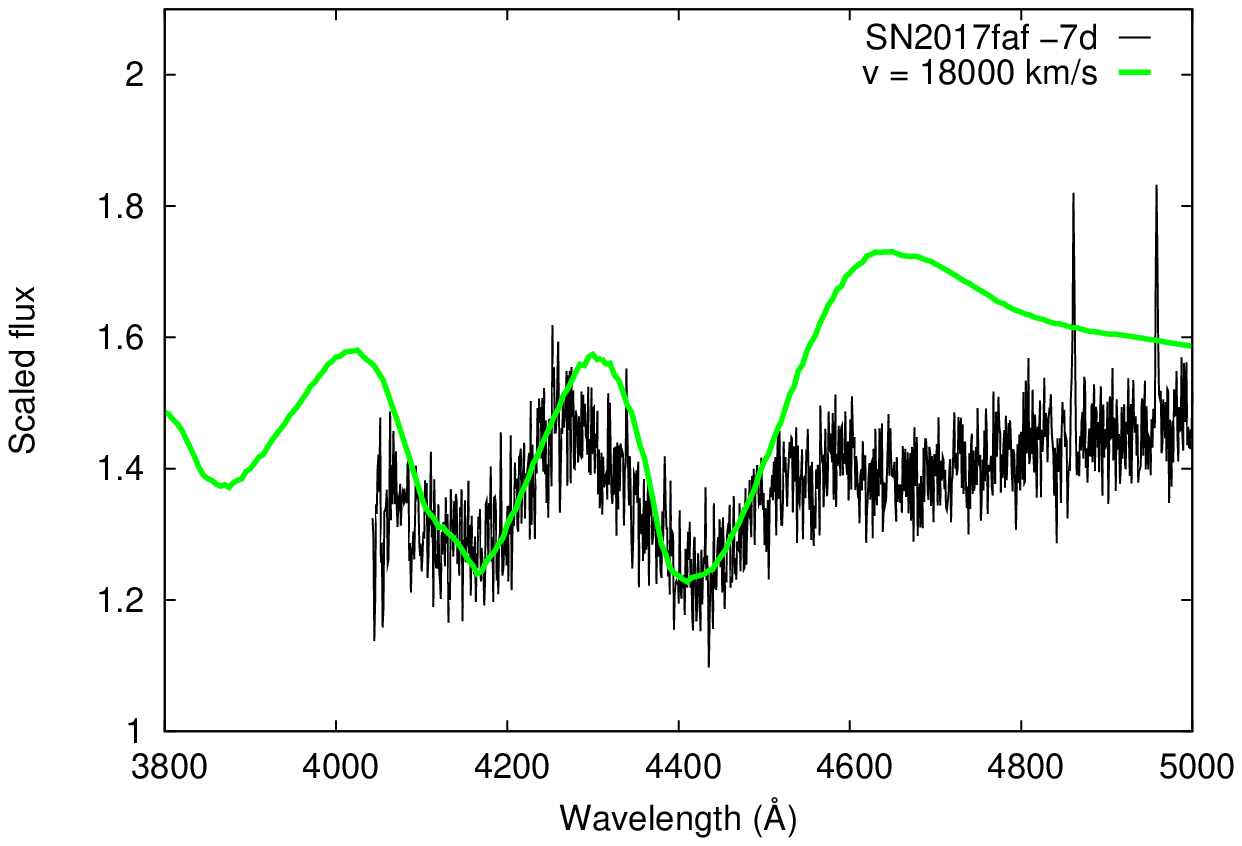}
\includegraphics[width=4.8cm]{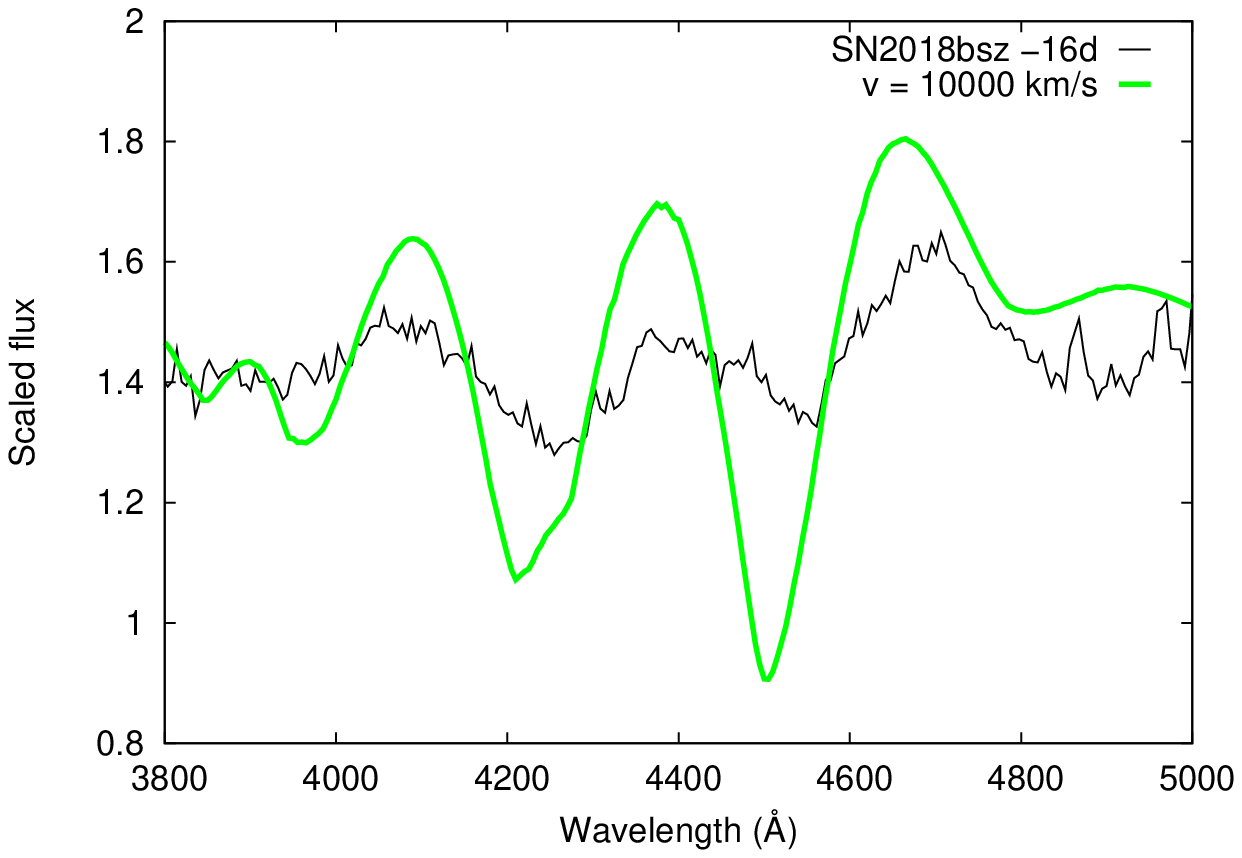}
\includegraphics[width=4.8cm]{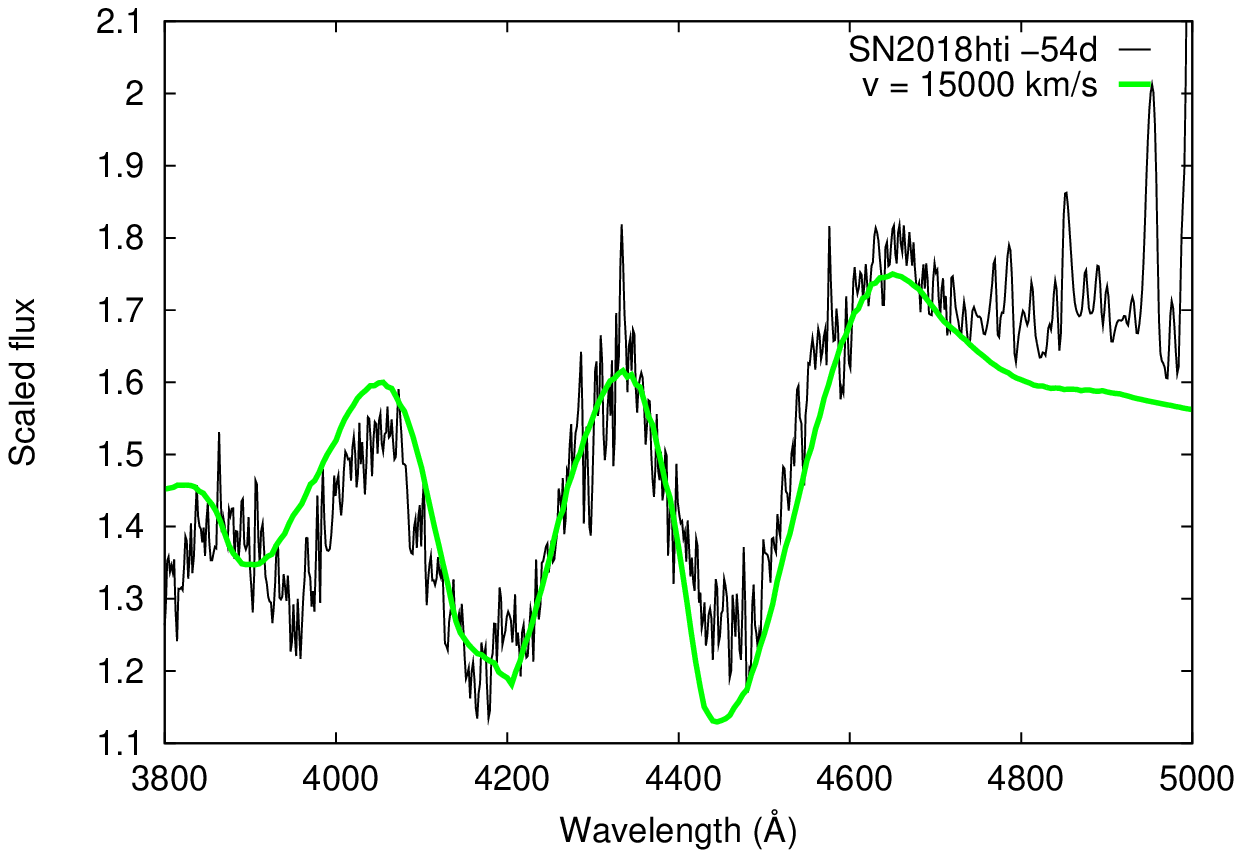}
\includegraphics[width=4.8cm]{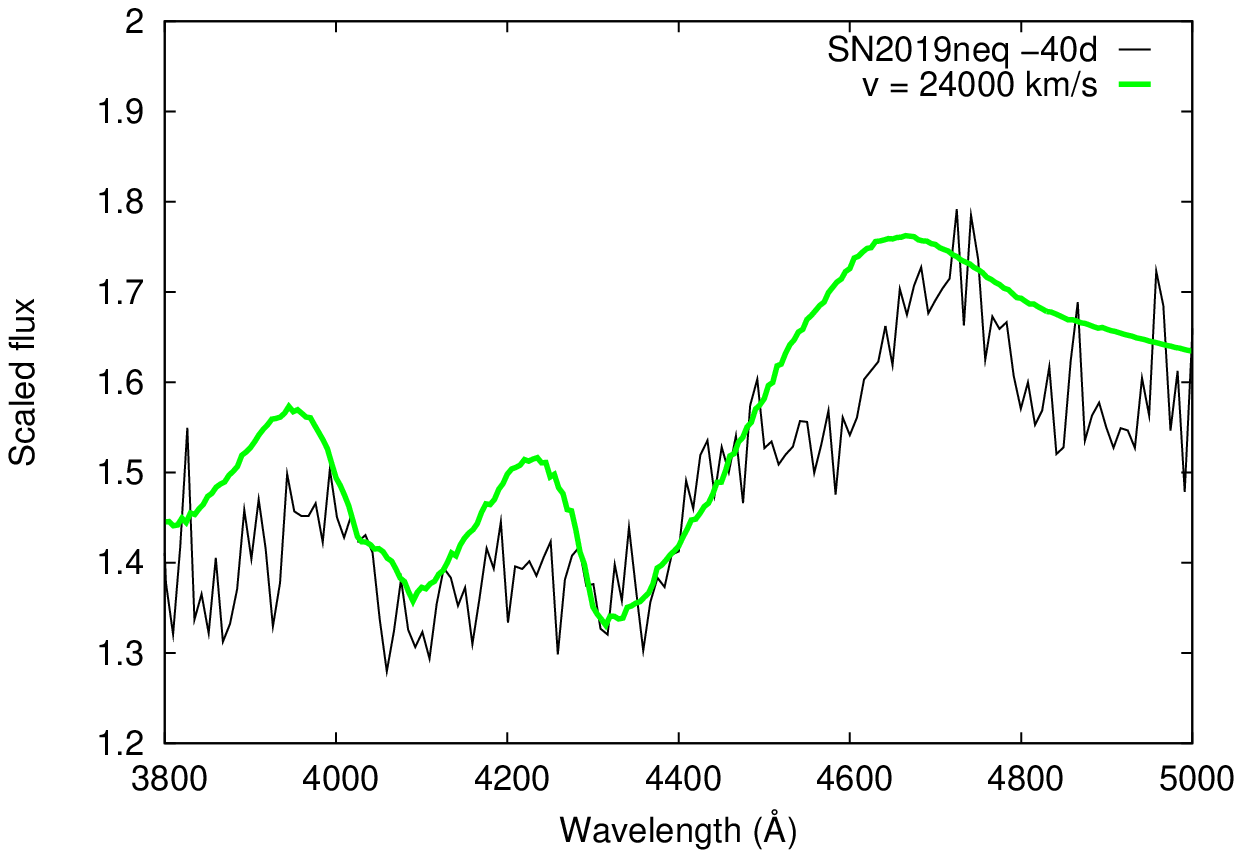}
\includegraphics[width=4.8cm]{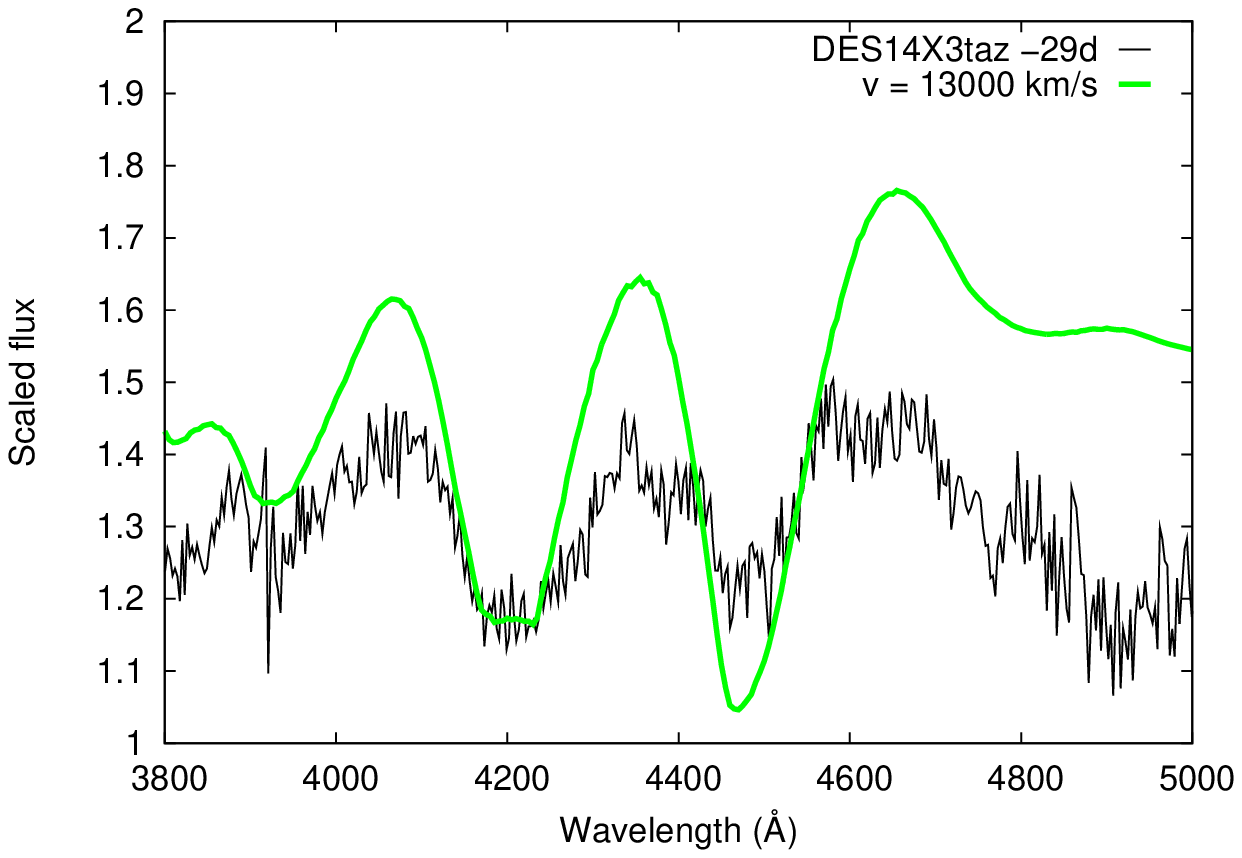}
\includegraphics[width=4.8cm]{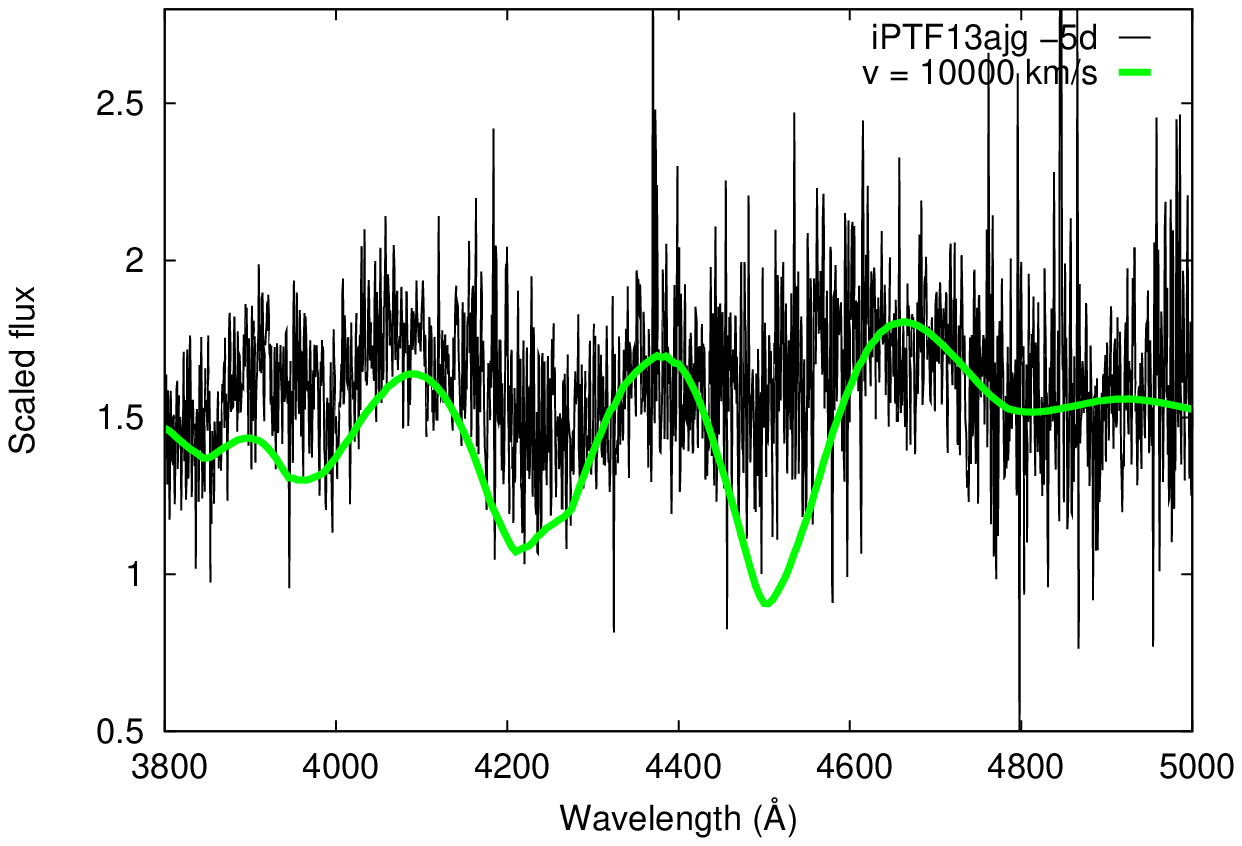}
\includegraphics[width=4.8cm]{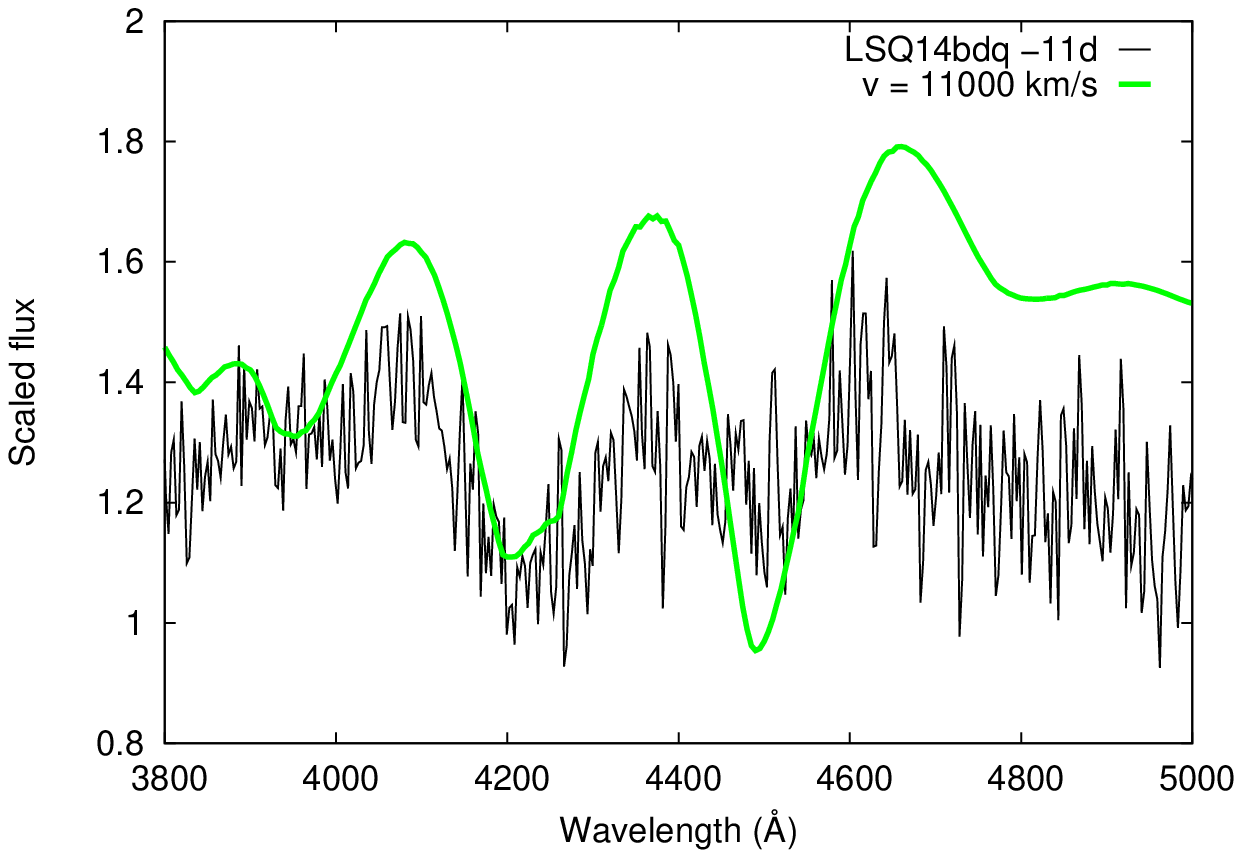}
\includegraphics[width=4.8cm]{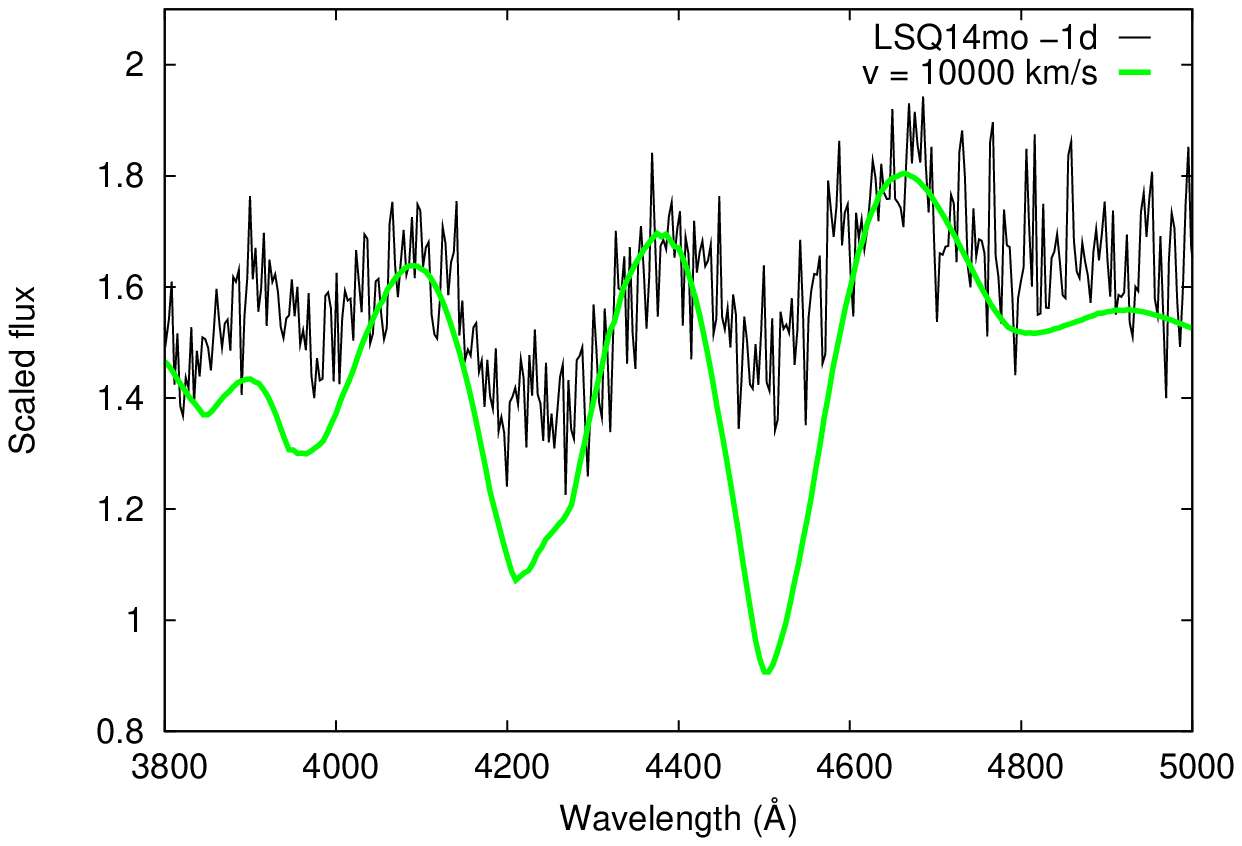}
\includegraphics[width=4.8cm]{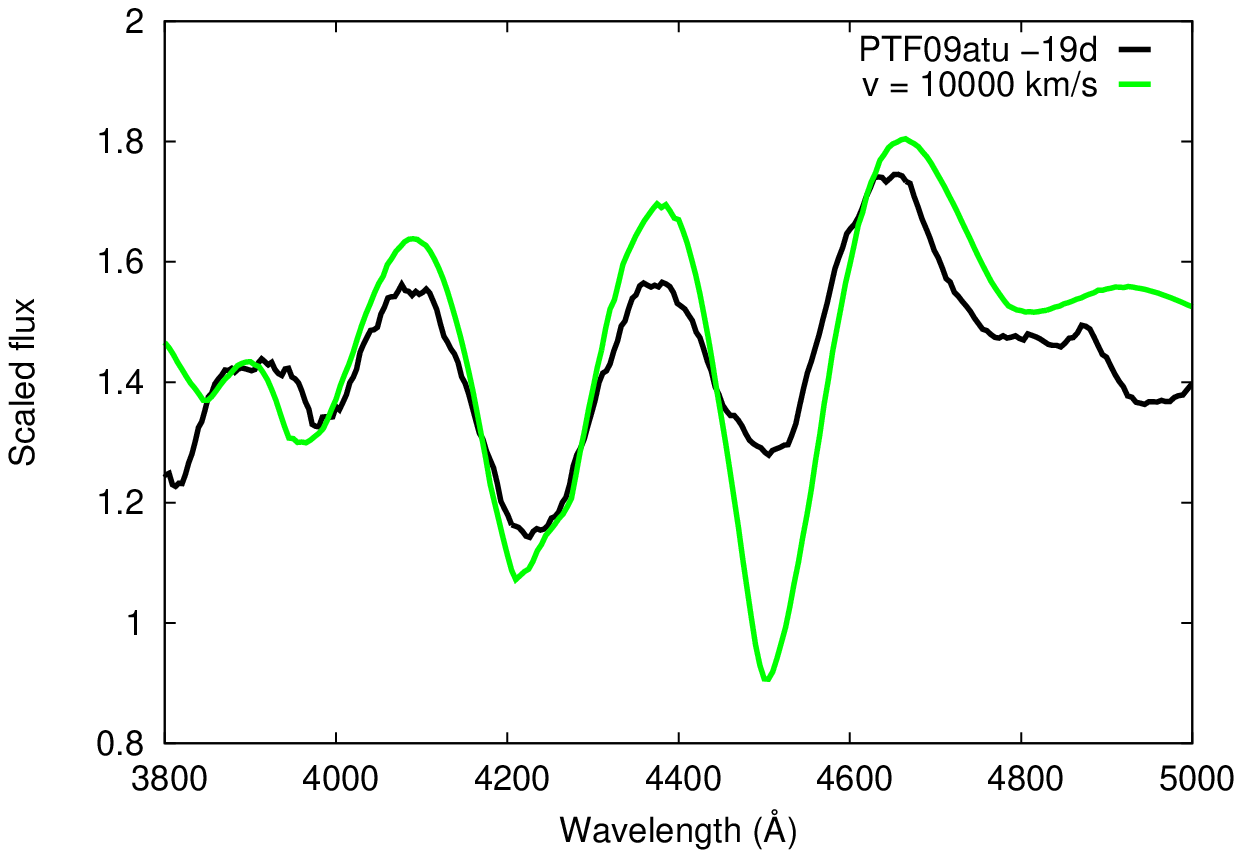}
\includegraphics[width=4.8cm]{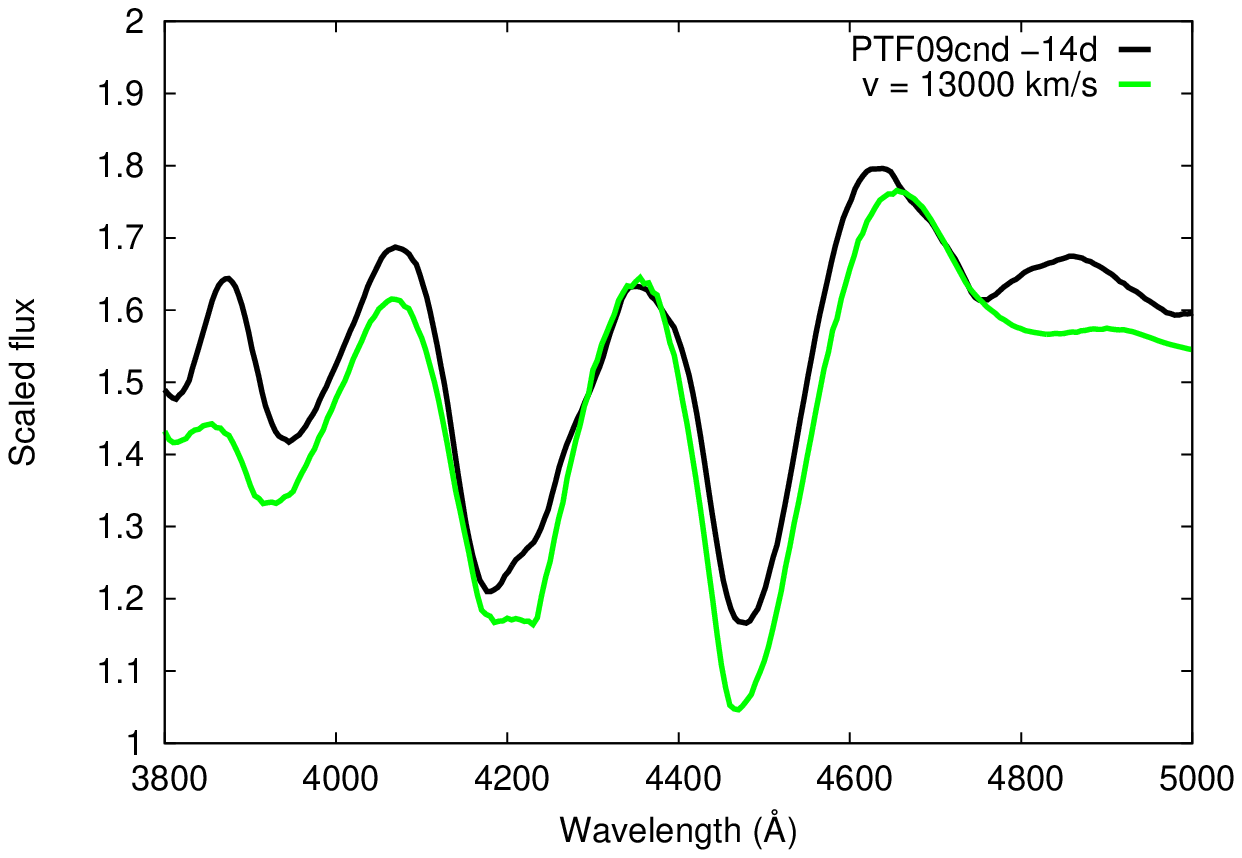}
\includegraphics[width=4.8cm]{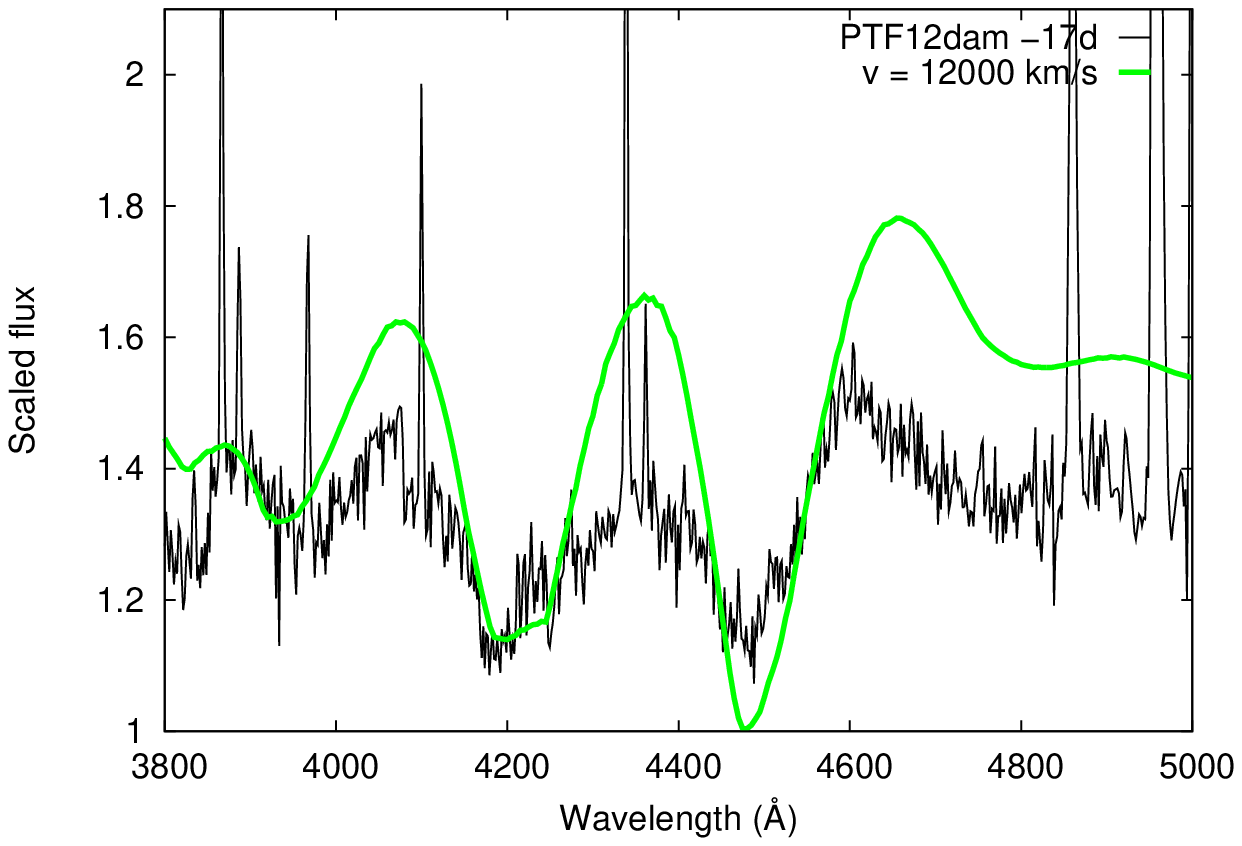}
\caption{The observed pre-maximum spectra of ``Type W'' SLSNe (black), together with their best-fit OII model spectra obtained in SYN++ (green). }
\label{fig:w_tip_spmodell}
\end{figure*}

The real, physical velocity differences
($\Delta v_{\rm phot}$) between the models having $v_{\rm phot}$ ranging from 10000 to 30000 km~s$^{-1}$ and the template model spectrum of 10000 km~s$^{-1}$ can be seen in Figure \ref{fig:keresztkorr_all} as a function of the velocity difference calculated by the {\tt fxcor} task in IRAF ($\delta v_{\rm X}$). The data for the
``Type W'' subclass (red circles) were fitted by a second-order polynomial as 
\begin{equation}
\Delta v_{\rm{phot}} ~=~ a_0 ~+~ a_1 \cdot \Delta v_{\rm{X}} ~+~ a_2 \cdot \Delta v_{\rm{X}}^2, 
\label{eq:keresztkorr_W}
\end{equation}
and obtained $a_0 = 155.01~(\pm 82.64)$, $a_1 = 1.68~(\pm 0.03)$ and $a_2 = -2.78 \cdot 10^{-5} ~(\pm 1.63 \cdot 10^{-6})$.

Finally, after cross-correlating the observed spectra with the model template, we applied Eq.~\ref{eq:keresztkorr_W} 
to infer the final $v_{\rm phot}$ values, which are shown in Table \ref{tab:mej}, together with epochs of the observations and their rest-frame phases. 

In Figure \ref{fig:w_tip_spmodell}, the observed pre-maximum spectra of the ``Type W'' sample are plotted with black lines, together with the best-fit SYN++ model spectrum (green line) that has the most similar photospheric velocity to the inferred $v_{\rm phot}$. Note that since this analysis aims at measuring only the expansion velocity, the spectra appearing in Fig.~\ref{fig:w_tip_spmodell} are flattened, and neither the continuum, nor the feature depths are fitted. Thus, only the wavelength positions of the features are expected to match.

\subsection{Type 15bn SLSNe}\label{subsec:15bn}

\begin{figure}
\centering
\includegraphics[width=8cm]{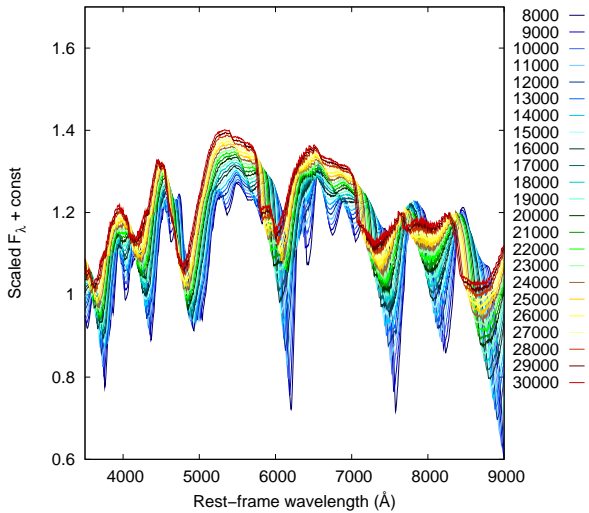}
\caption{ SYN++ models of ``Type 15bn'' SNe having $T_{\rm phot}$= 11000~K and $v_{\rm phot}$ ranging between 8000 and 30000 km s$^{-1}$.}
\label{fig:15bnmodels}
\end{figure}

\begin{figure*}
\centering
\includegraphics[width=5.5cm]{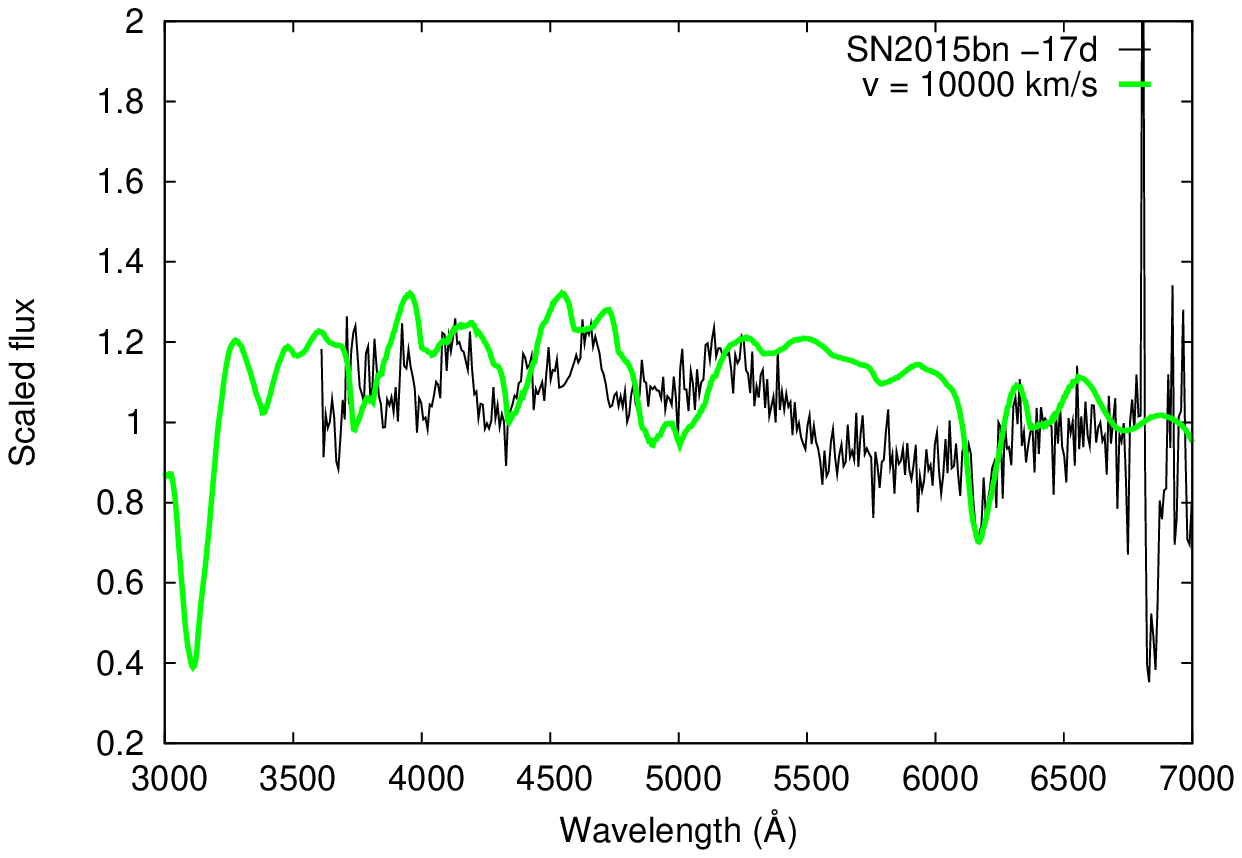}
\includegraphics[width=5.5cm]{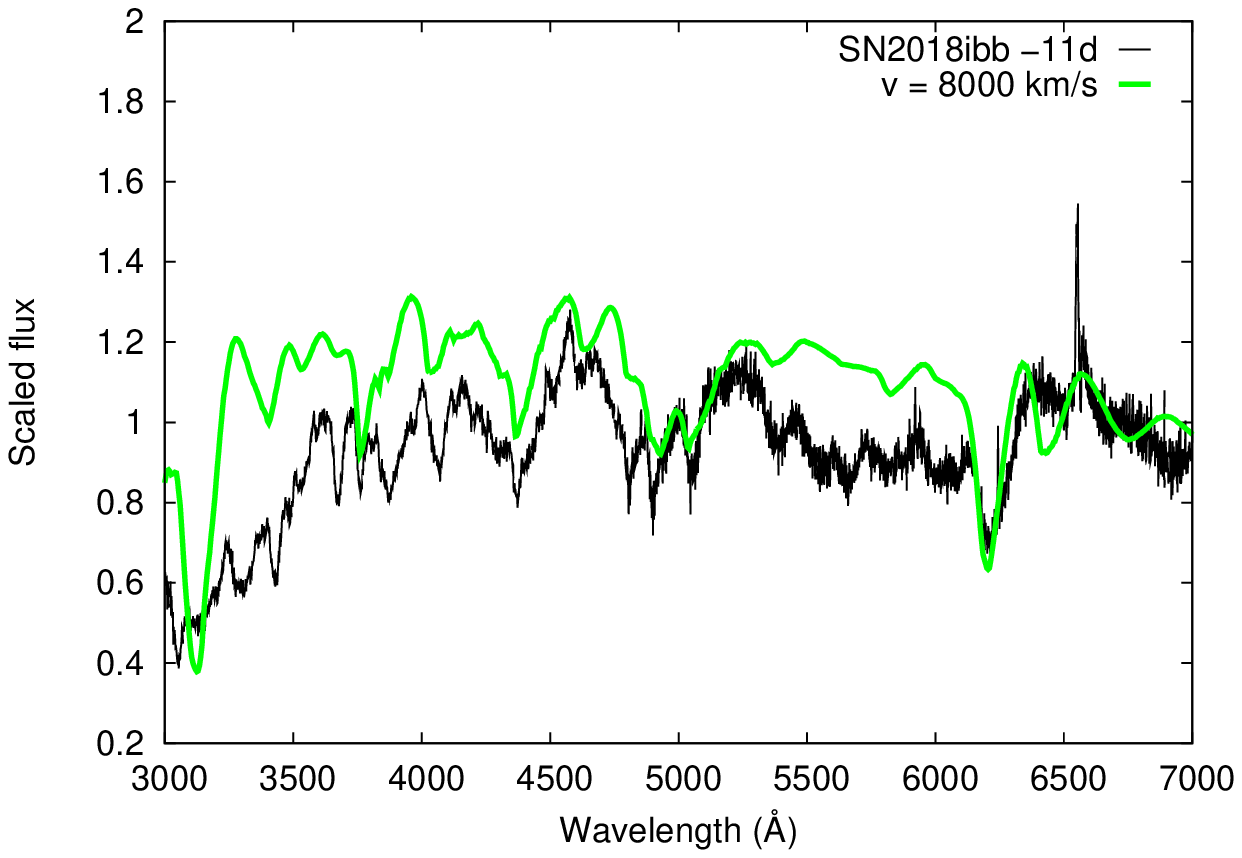}
\includegraphics[width=5.5cm]{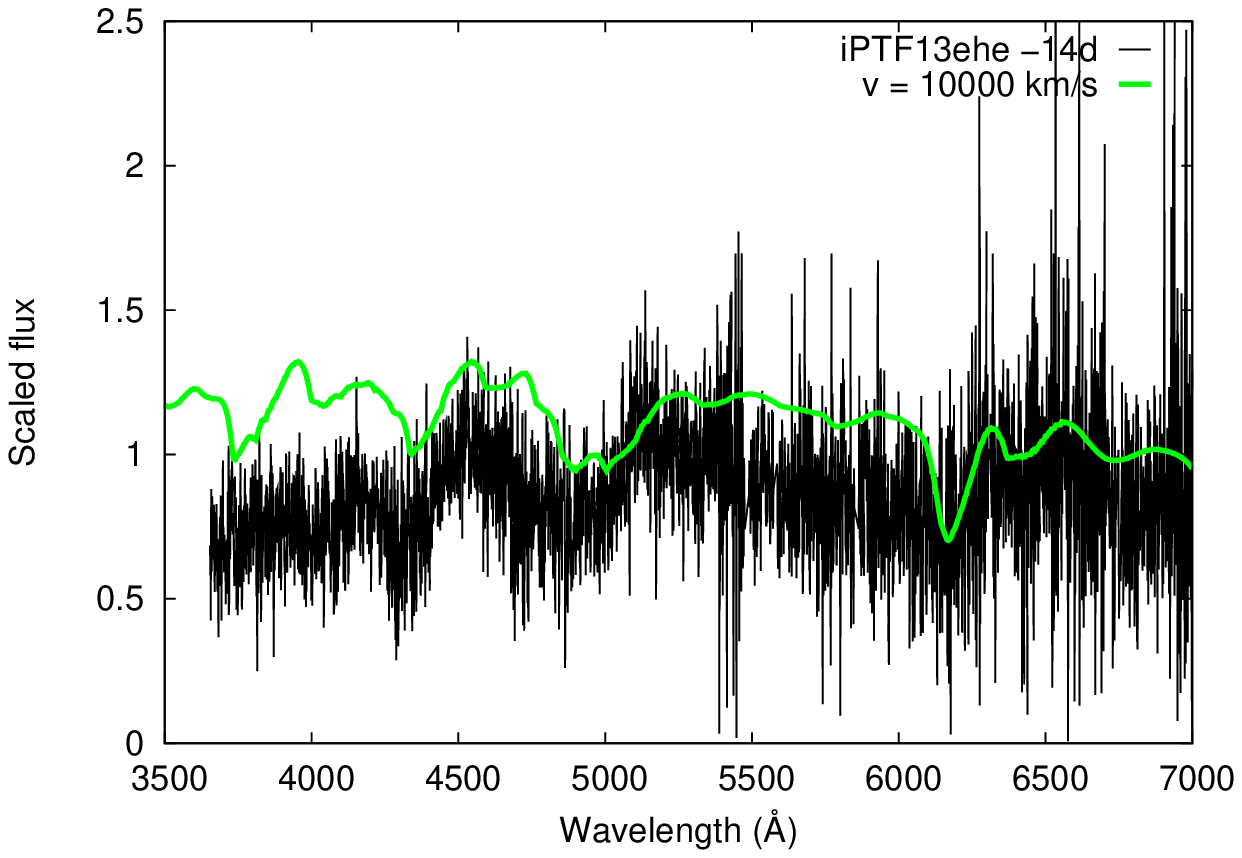}
\includegraphics[width=5.5cm]{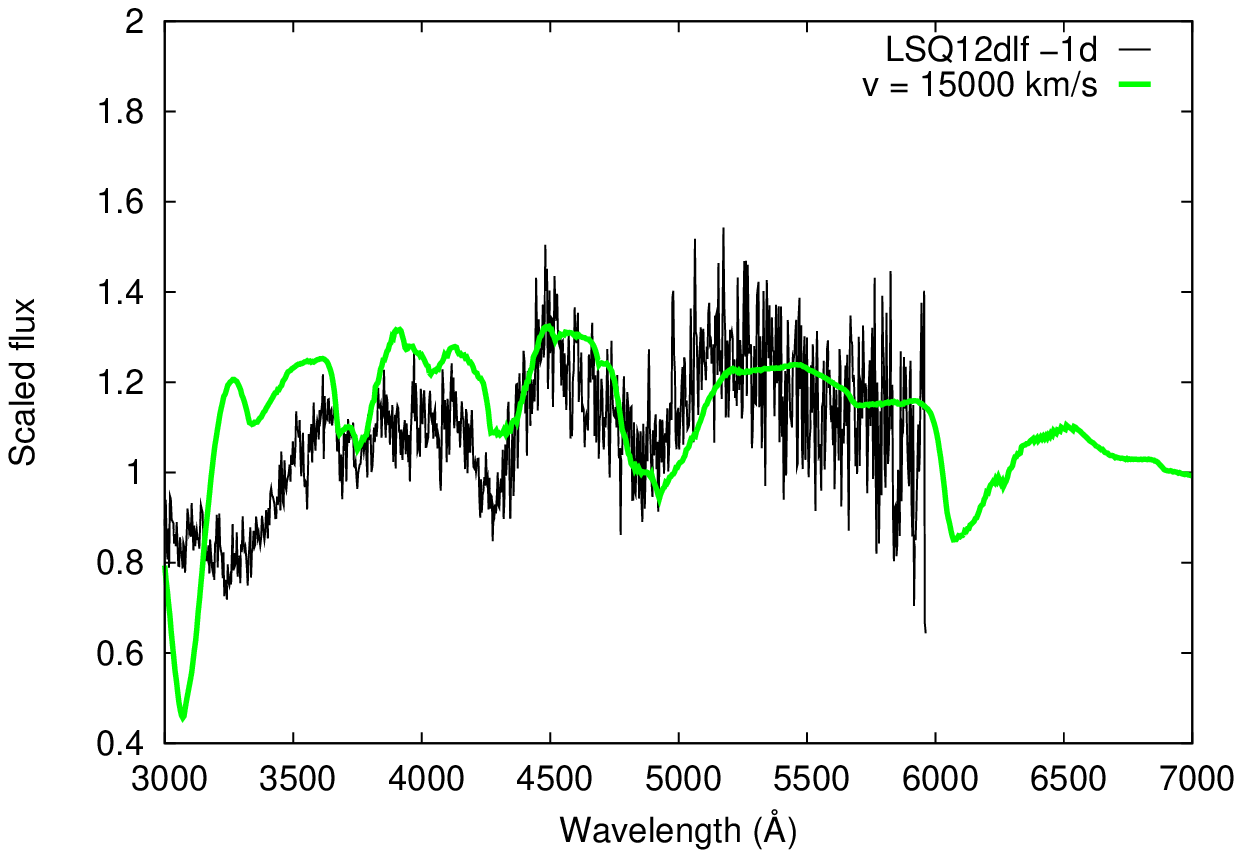}
\includegraphics[width=5.5cm]{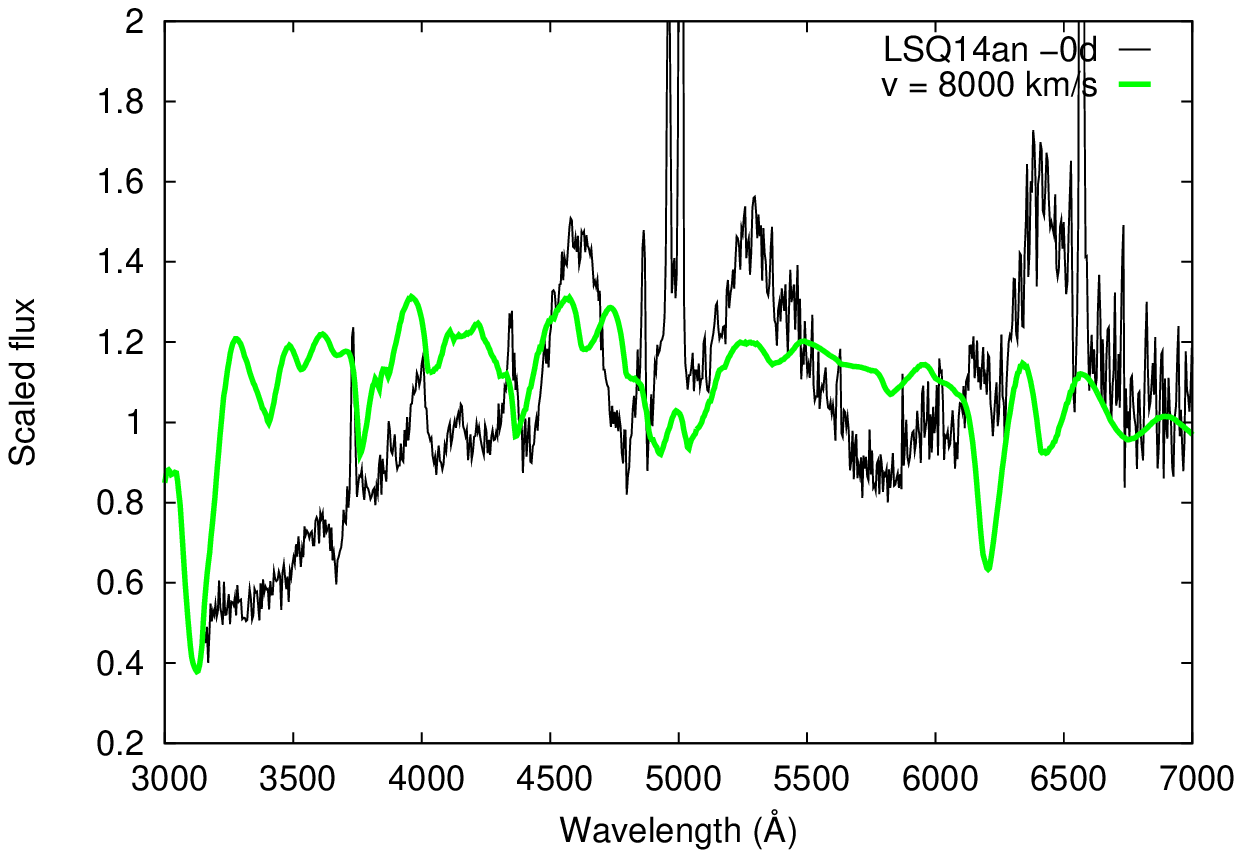}
\includegraphics[width=5.5cm]{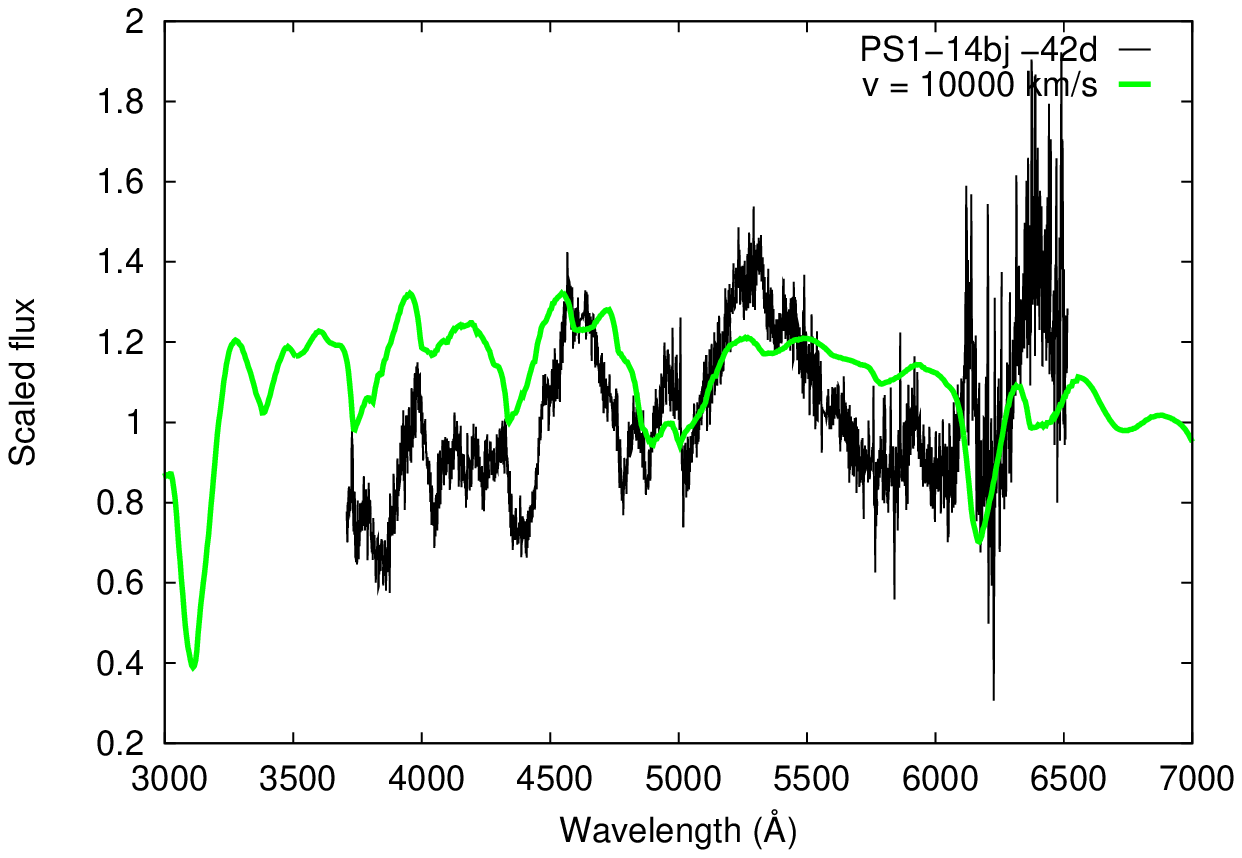}
\includegraphics[width=5.5cm]{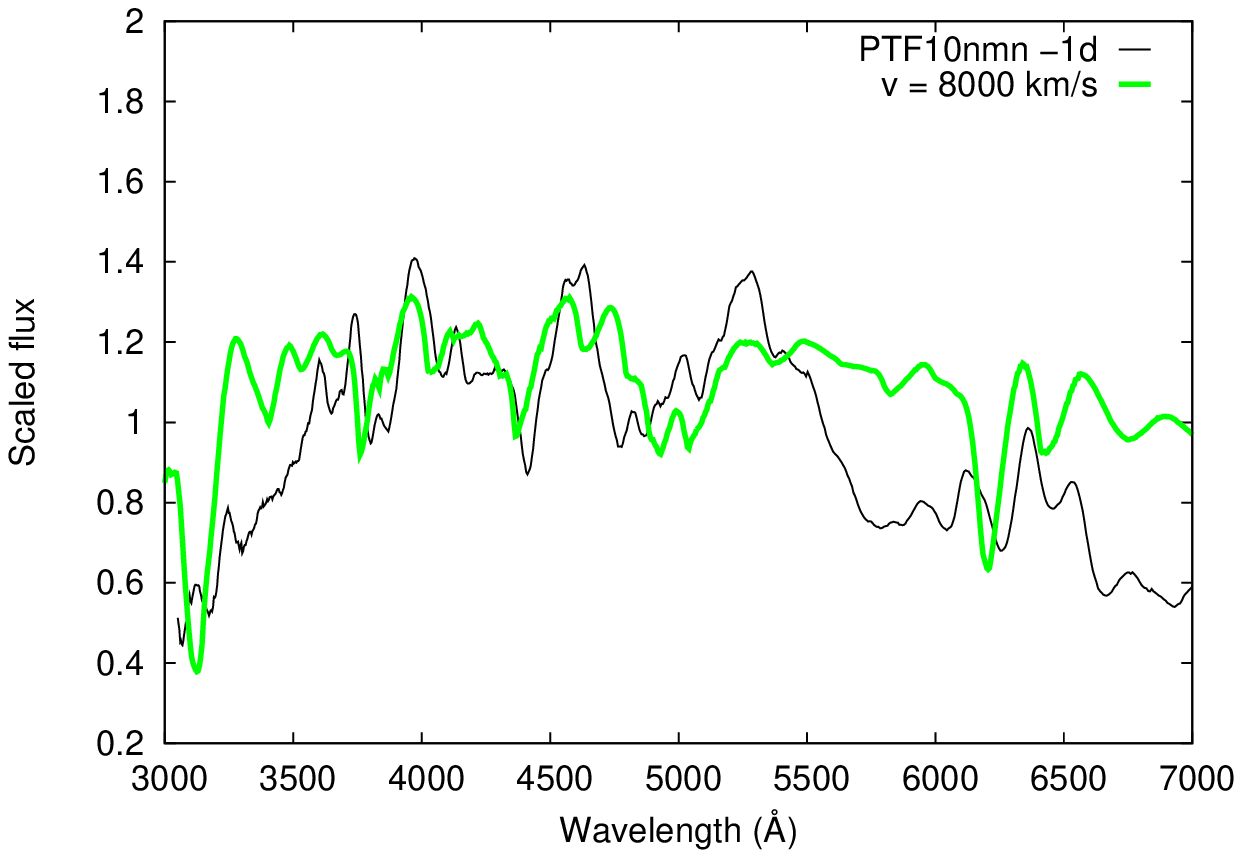}
\includegraphics[width=5.5cm]{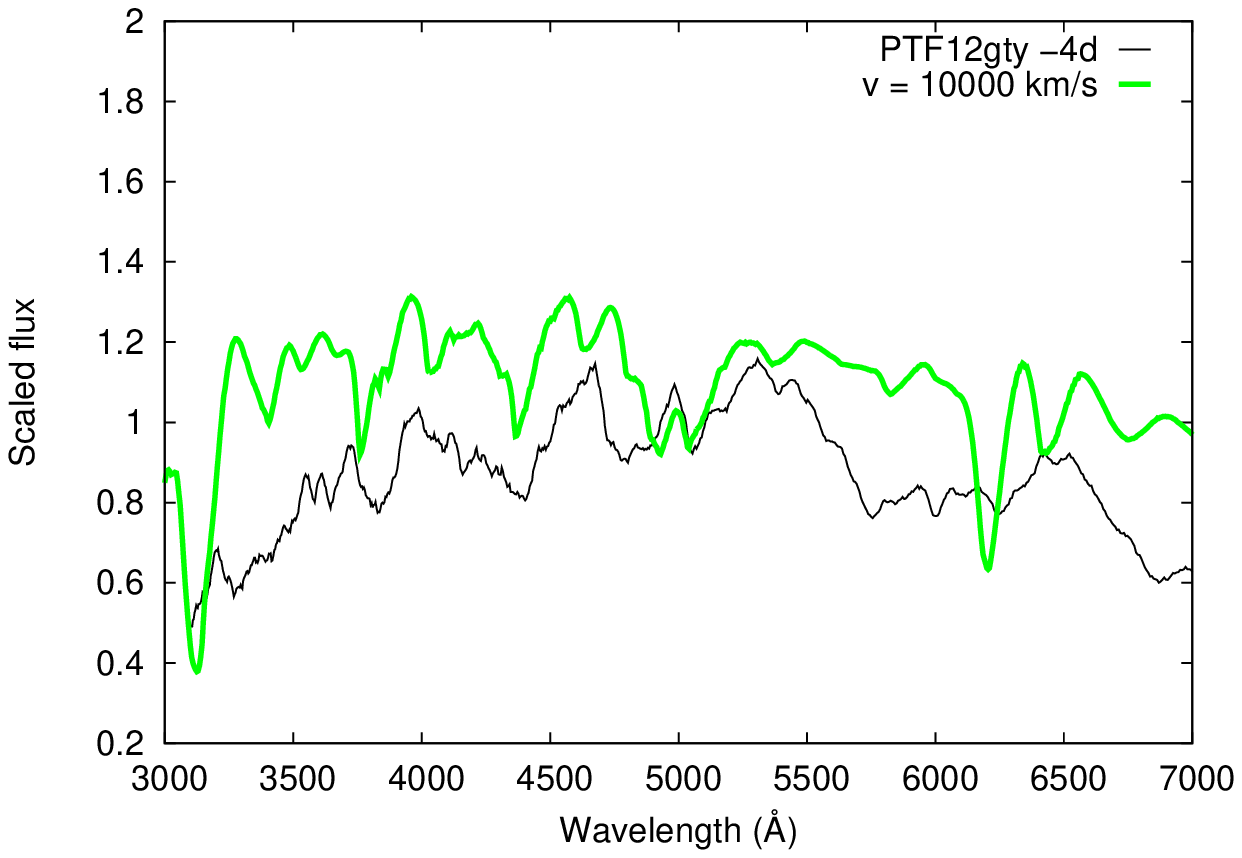}
\includegraphics[width=5.5cm]{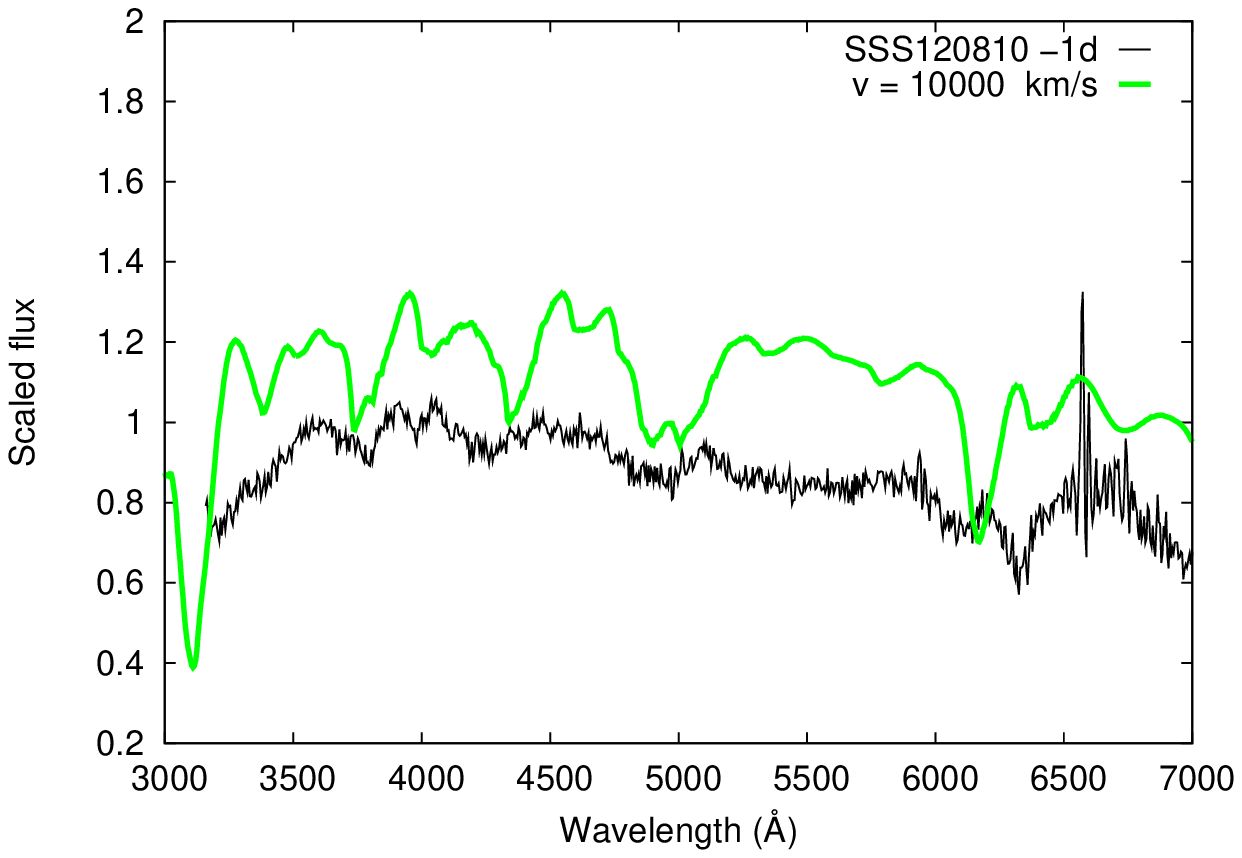}
\caption{ The  observed  pre-maximum  spectra  of  ``Type 15bn''  SLSNe  (black),  together  with the Doppler-shifted best-fit model built for SN~2018ibb (green) in accordance with the inferred $v_{\rm phot}$ for each object.}
\label{fig:15bn_tip_spmodell}
\end{figure*}

For each object belonging to the ``Type 15bn'' subclass, 
the photospheric velocity was determined using the same method as discussed in the previous Section~\ref{subsec:W}. However, in this case the modeling of the the whole optical spectrum was necessary to get reliable estimate for $v_{\rm phot}$, since no typical and easily identifiable feature can be found in those spectra in contrast with the ``Type W'' SLSNe. 

Therefore, we built a SYN++ model to describe the spectrum of a well-observed representative of the ``Type 15bn'' group, which was selected to be SN~2018ibb. The observed spectrum of SN~2018ibb taken at $-11$ rest-frame days relative to maximum light can be seen in the left panel of Figure \ref{fig:18ibb15bn_models} in the Appendix (black line), together with its best-fit SYN++ model (red line). The single-ion contributions to the overall model spectrum are also presented as orange curves, shifted vertically for better visibility. The photospheric temperature and velocity of the best-fit model is $T_{\rm phot} = 11000$~K and 
$v_{\rm phot} ~=~ 8000$ km s$^{-1}$, respectively. The spectrum contains C II, O I, Mg II, Si II, Ca II, Fe II and Fe III ions. 
The full set of the global and local parameter values for the SN~2018ibb model can be found in Table \ref{tab:syn_alltypes} in the Appendix.

Thereafter, we synthesized model spectra with the same local and global parameters as the best-fit SYN++ model of the pre-maximum spectrum of SN~2018ibb, but having different $v_{\rm phot}$ in between 8000 and 30000 km s$^{-1}$ (see Figure~\ref{fig:15bnmodels}). These models were cross-correlated with the one having $v_{\rm phot}~=~10000$ 
km~s$^{-1}$. 
Then, a similar correction formula between the velocity differences was computed as previously, resulting in 

\begin{equation}
\Delta v_{\rm{phot}} ~=~ a_0 ~+~ \sum_{n=1}^{4} a_n \cdot \Delta v_{\rm{X}}^n,
\label{eq:keresztkorr_15bn}    
\end{equation}
with $a_0= -128.61 ~(\pm 79.92)$, $a_1 = 1.53~(0.06)$, 
$a_2 = 1.09 \cdot 10^{-4}~(3.88 \cdot 10^{-5})$, 
$a_3 =5.45 \cdot 10^{-9}~(7.44 \cdot 10^{-9})$ and
$a_4 = -1.16 \cdot 10^{-12}~(4.27 \cdot 10^{-13})$. 

The resulting $\Delta v_{\rm phot}$ values  are plotted with green dots in Figure \ref{fig:keresztkorr_all}, and the best-fit polynomial (Eq. \ref{eq:keresztkorr_15bn}.) is shown also with a green line. 

Finally, after applying Eq. \ref{eq:keresztkorr_15bn} to the  observed pre-maximum spectra in the ``Type 15bn'' subclass, the $v_{\rm phot}$ velocities 
are collected in Table \ref{tab:mej}. 

The observed pre-maximum spectra of ``Type 15bn'' SLSNe are shown in Figure \ref{fig:15bn_tip_spmodell} with black lines, together with the best-fit SYN++ model for SN~2018ibb (green) Doppler-shifted to the inferred $v_{\rm phot}$ for each object.

It is seen in Table \ref{tab:mej} that SLSNe in the ``Type 15bn'' group have lower photospheric velocities compared to the ``Type W'' SLSNe in general. This suggests that ``Type 15bn'' SLSNe are similar to each other, not only in the appearance of their spectra, but also in their $T_{\rm phot}$ and $v_{\rm phot}$ parameters as well. It is suspected that they are forming a subgroup of SLSNe-I that is different from the ``Type W'' subclass, because the latter have faster ejecta expansion velocities and hotter photospheres during the pre- and near-maximum phases. In Sections \ref{sec:mej} and \ref{sec:discussion}, we discuss additional differences between these two subclasses in details. 

\subsection{Post-maximum spectra}\label{subsec:maxutan}

\begin{figure*}
\centering
\includegraphics[width=12cm]{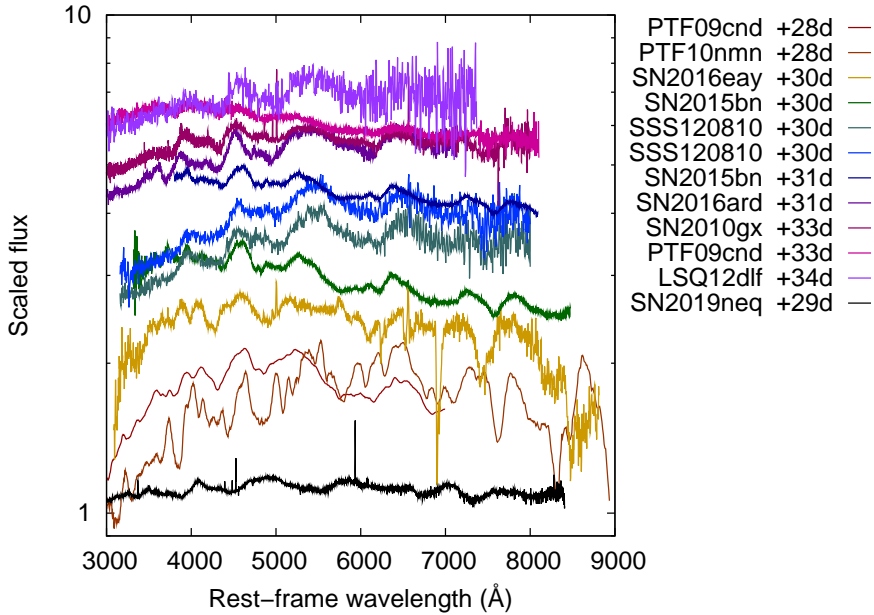}
\caption{Spectra taken between +25 and +35 days rest-frame phases after maximum light for 9 SLSNe in our sample.}
\label{fig:maxutan}
\end{figure*}

In order to classify the events in our sample into the Fast- or the Slow-evolving subgroup of Type I SLSNe \citep{inserra18}, photospheric velocities determined from the spectra taken at  $\sim$+30 rest-frame days after maximum are required. By comparing the post-maximum velocities to the $v_{\rm phot}$ estimated near the moment of maximum light, it can be decided unambiguously if a SLSN belongs to the Fast or the Slow SLSNe-I.  

From the 28 SLSNe in our sample, 9 possessed post-maximum spectra between +25 and +35 days phase, and  both ``Type W'' and ``Type 15bn'' objects were represented amongst them,. These spectra are collected and shown in Figure \ref{fig:maxutan}.

\begin{table*}
\caption{Photospheric velocities before and after the maximum for the 9 SLSNe having post-maximum spectra.}
\label{tab:vphotok}
\begin{center}
\begin{tabular}{lcccccc}
\hline
\hline
SLSN & Pre-max phase  & Pre-max $v_{\rm phot}$  & Post-max phase & Post-max $v_{\rm phot}$  & W/15bn & Fast/Slow \\ 
     &  (day)  & (km s$^{-1}$) & (day) & (km s$^{-1}$) & &  \\
\hline
SN2010gx & -1 & 20371 & +33 & 9926 & W & F \\
SN2015bn & -17 & 9870 & +30  & 8136 & 15bn & S\\
SN2016ard & -4 & 14398 &  +31 & 11585 & W & S\\
SN2016eay & -2 & 20362 & +30  & 9814 & W & F \\
SN2019neq & -4 & 23000 & +29 & 9972 & W & F \\
LSQ12dlf & -1 & 15000 &  +34 & 7916 & 15bn & S \\
PTF09cnd &   -14& 13200 & +28 & 7593 & W & S \\
PTF10nmn & -1 & 7871 & +28 & 4307 & 15bn & S\\
SSS120810 & -1  & 9870 & +30 & 8136 & 15bn & S \\
\hline
\end{tabular}
\end{center}
\end{table*}

From the available post-maximum spectra, the one taken at +30 rest-frame days phase of SN~2015bn was chosen for modeling. It can be seen together with its best-fit SYN++ model in the right panel of Figure \ref{fig:18ibb15bn_models} in the Appendix, with the same color coding as the model of SN~2018ibb. The best-fit model was found to heve $T_{\rm phot}~=~9000$~K, and $v_{\rm phot}~=~ 8000$ km~s$^{-1}$, and it contains O I, Na I, Mg II, Si II, Si II v, Ca II and Fe II ions. The "v" next to Si II refers to the high velocity of this feature. It has higher velocity than $v_{\rm phot}$, as it is formed above the photosphere in the outer regions of the SN ejecta. The parameters of the best-fit SYN++ model can be found in Table \ref{tab:syn_alltypes} in the Appendix.

Since the $v_{\rm phot}$ of  an expanding SN atmosphere decreases with time as the ejecta becomes more and more transparent, in case of the post-maximum spectra we utilized and cross-correlated models having lower velocities compared to the models built for the pre-maximum phases.

\begin{figure}
\centering
\includegraphics[width=8cm]{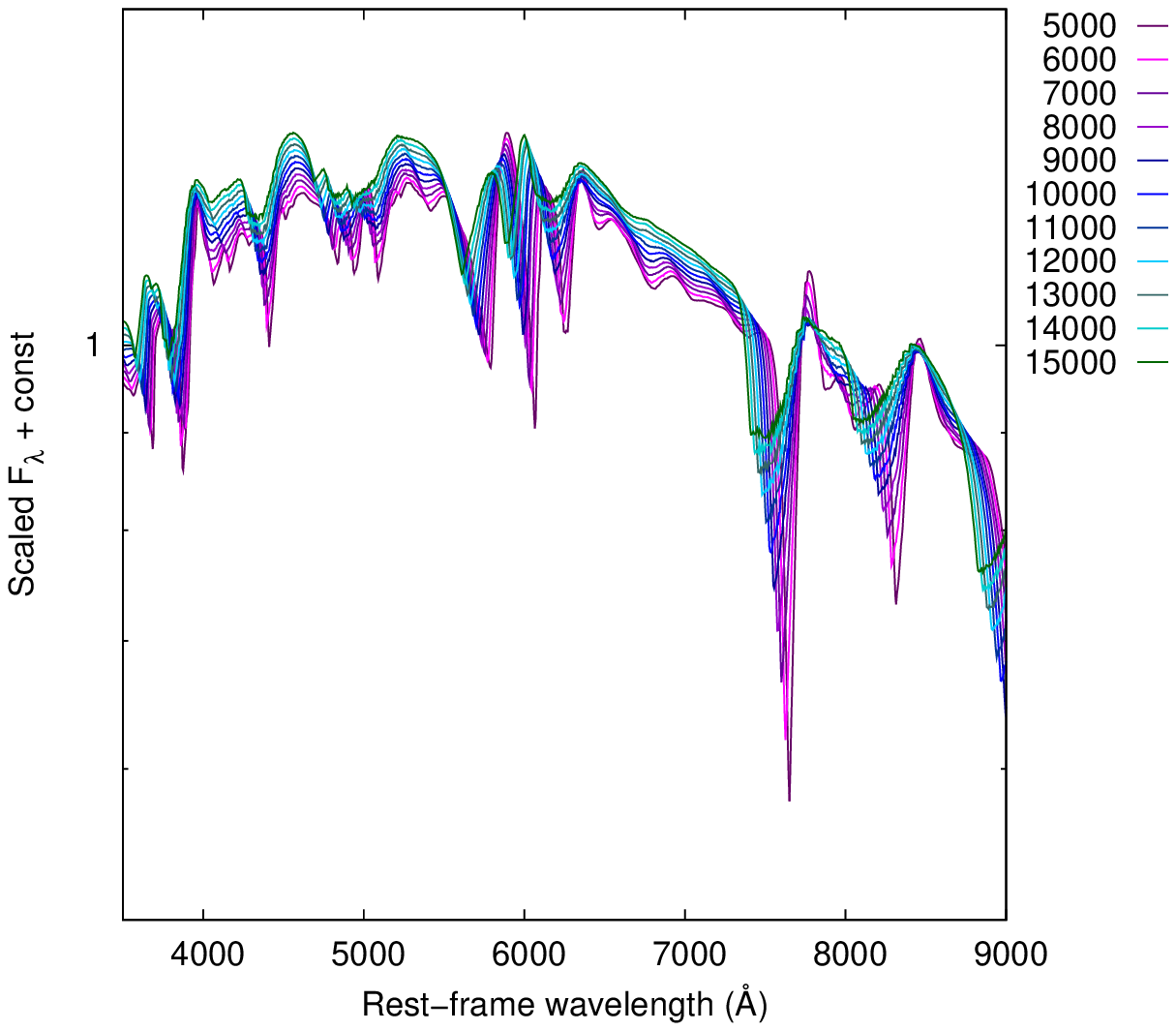}
\caption{ SYN++ models having $T_{\rm phot}~=~11000$~K, and $v_{\rm phot}$ in between 5000 and 15000 km~s$^{-1}$ for the spectra taken around $\sim$30 days after maximum.}
\label{fig:maxutan_models}
\end{figure}

We created 11 variants of the best-fit model of the +30 days phase spectrum of SN~2015bn, with $v_{\rm phot}$ in between 5000 and 15000 km s$^{-1}$ (see Fig. \ref{fig:maxutan_models}). 
After cross-correlating them, we reached to a similar velocity correction formula, as discussed in Sections \ref{subsec:W} and \ref{subsec:15bn}, namely
\begin{equation}
    \Delta v_{\rm phot} ~=~ a_0 + a_1 \Delta v_{\rm X}
    \label{eq:keresztkorr_maxutan}
\end{equation}
with $a_0 = 135.62 ~(\pm 54.81)$ and $a1 = 1.51 (0.01)$. 
The data as well as the best-fit straight line are
shown in Figure \ref{fig:keresztkorr_all} with blue color.

\begin{figure*}
\centering
\includegraphics[width=5.5cm]{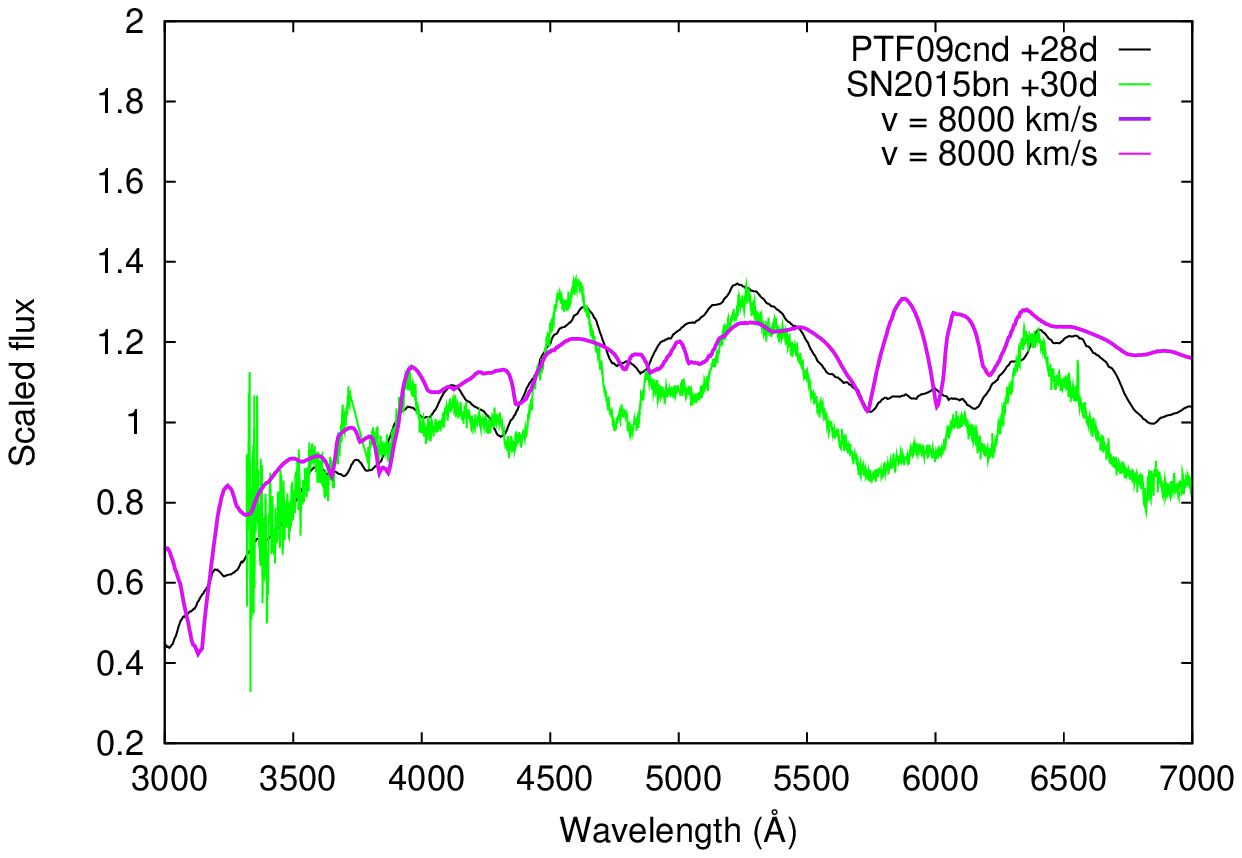}
\includegraphics[width=5.5cm]{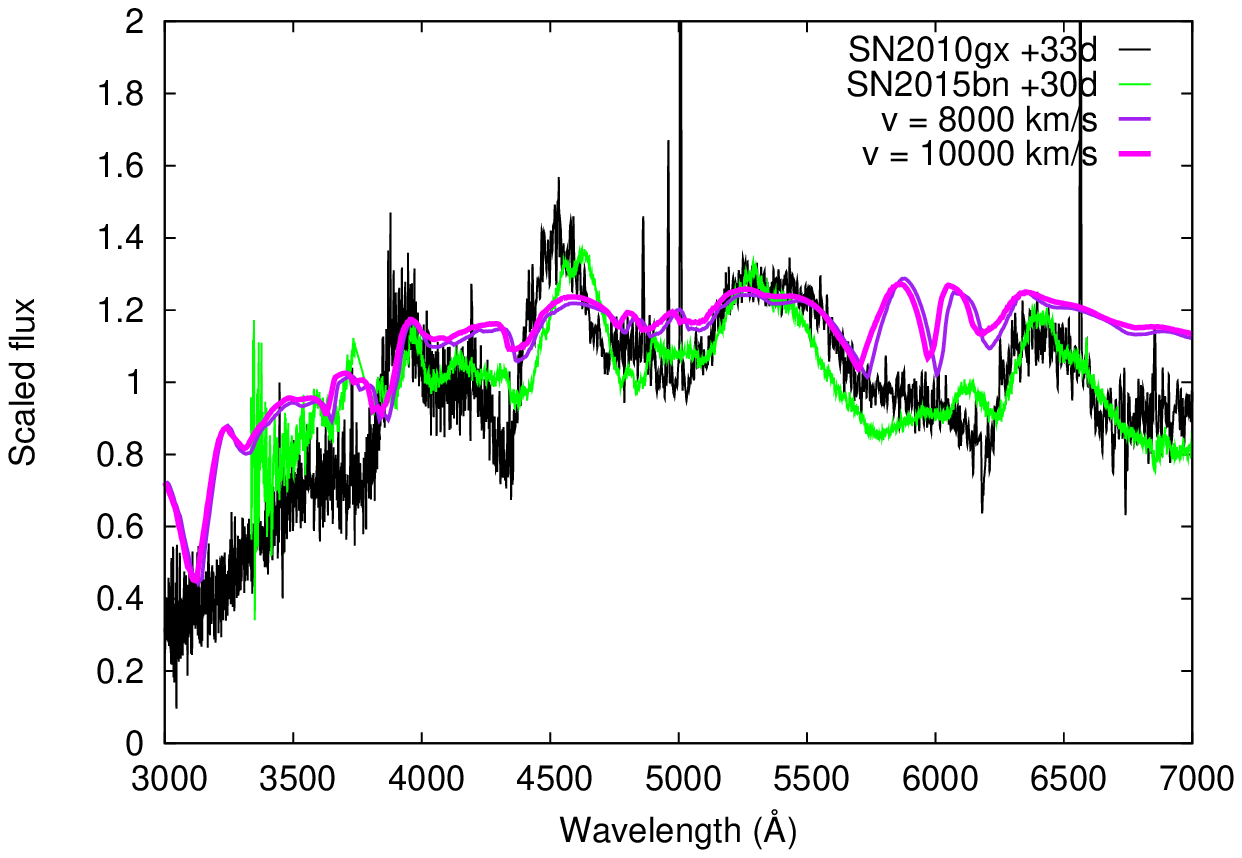}
\includegraphics[width=5.5cm]{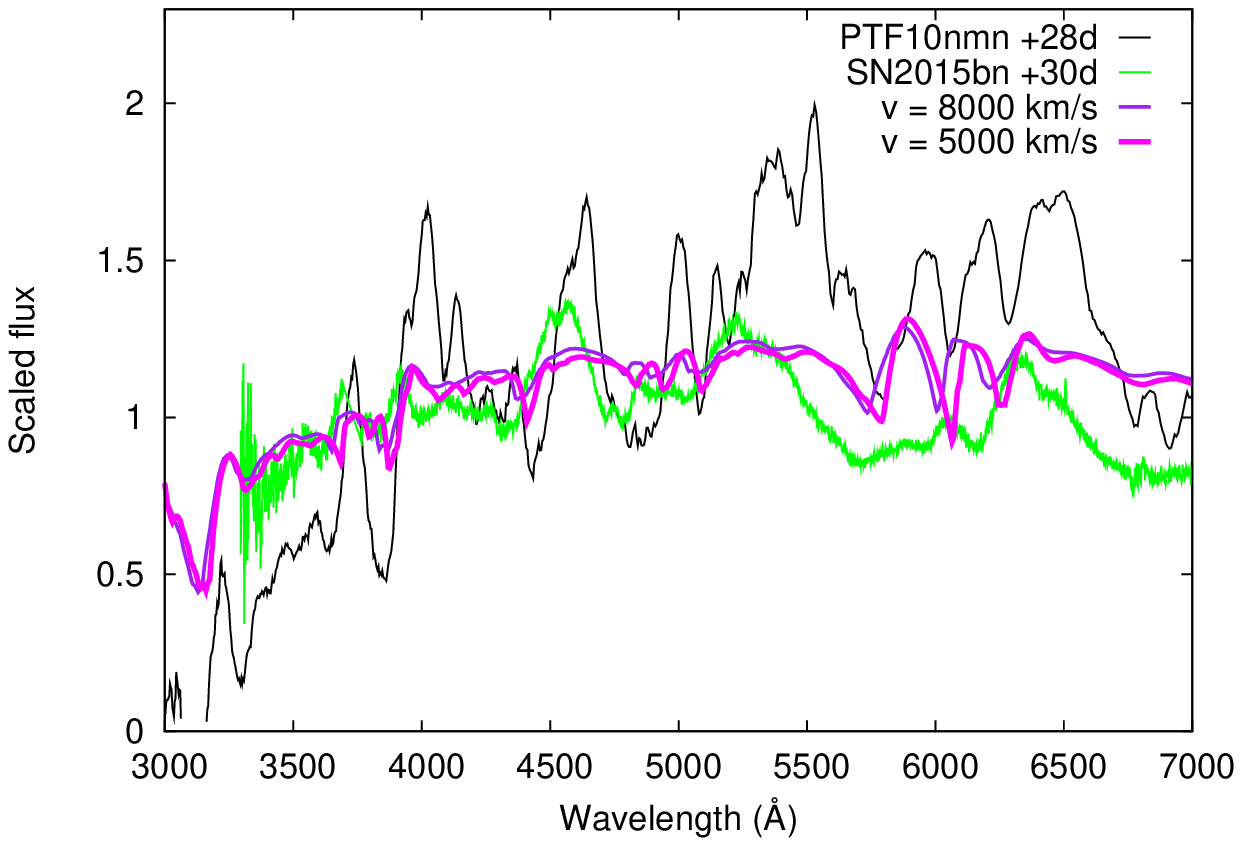}
\includegraphics[width=5.5cm]{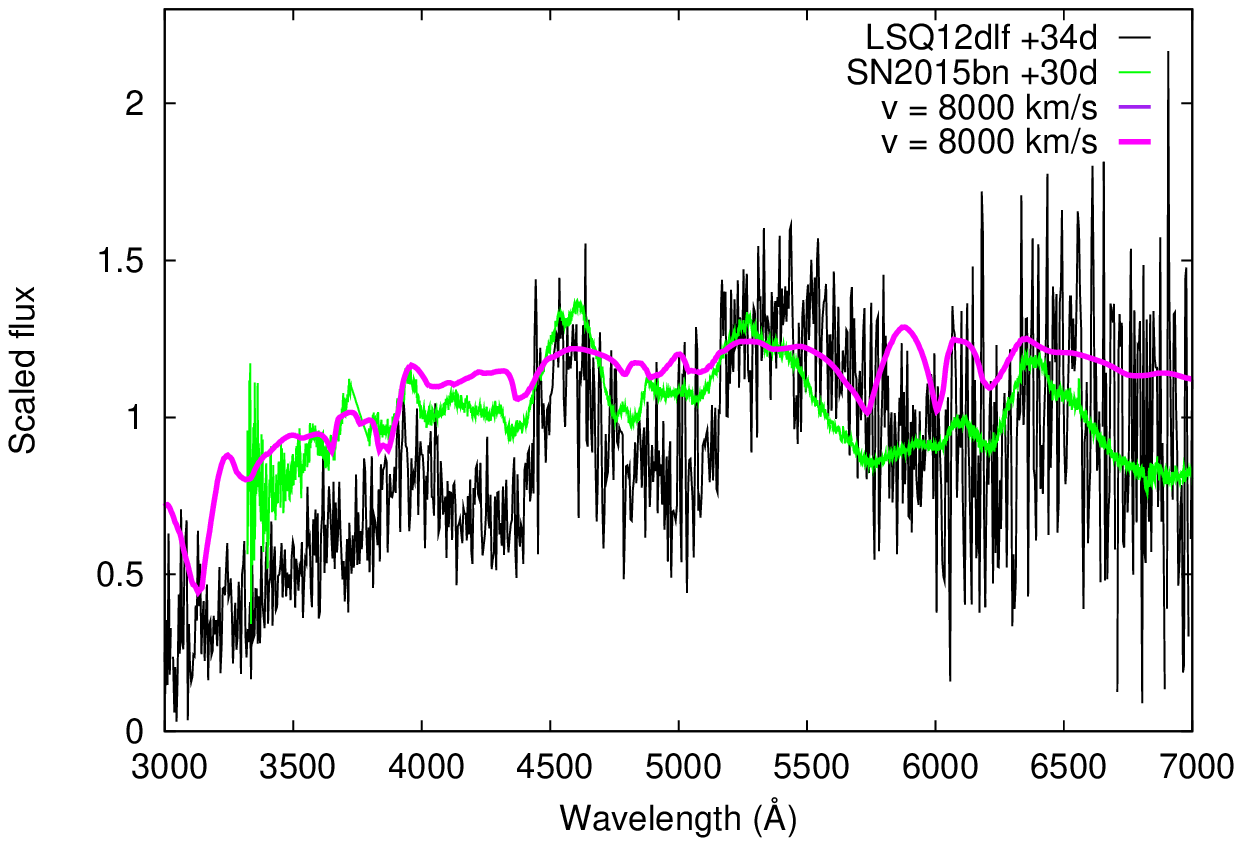}
\includegraphics[width=5.5cm]{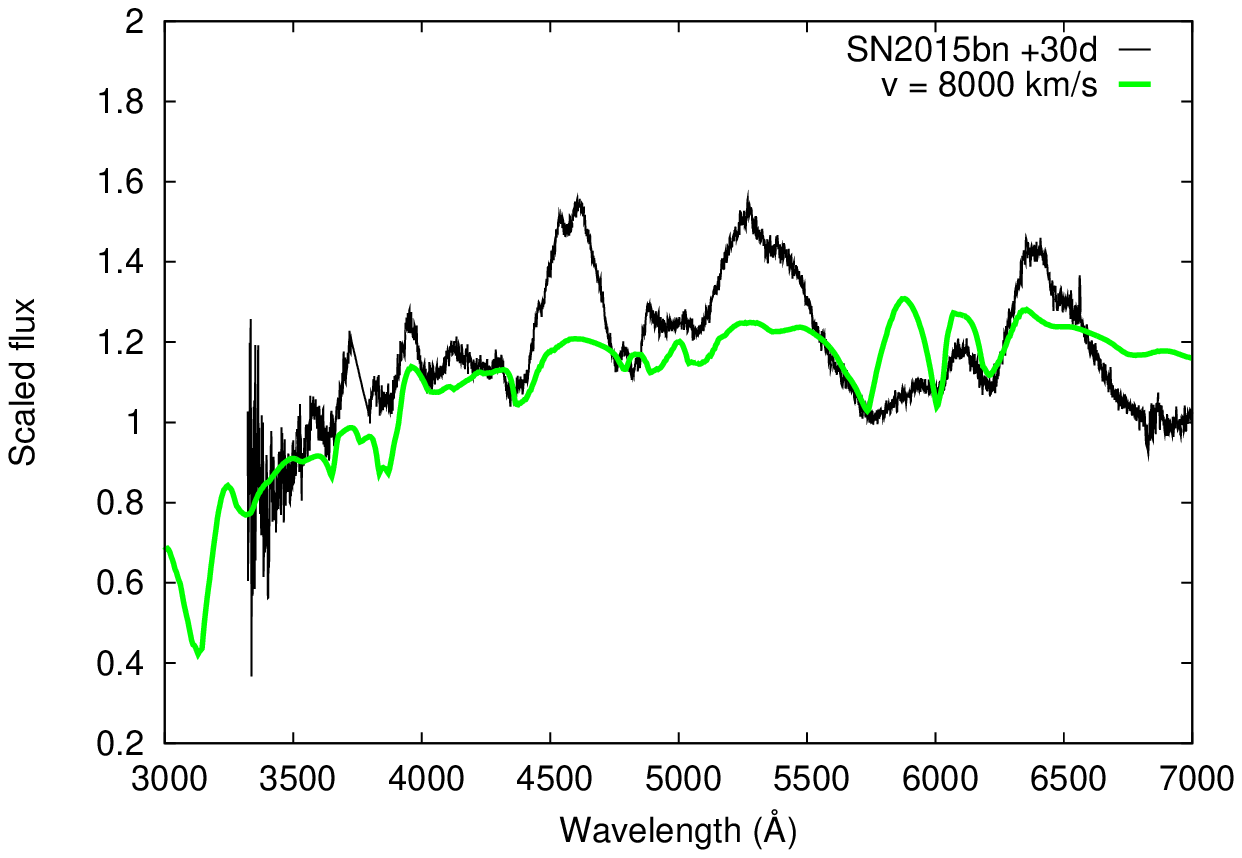}
\includegraphics[width=5.5cm]{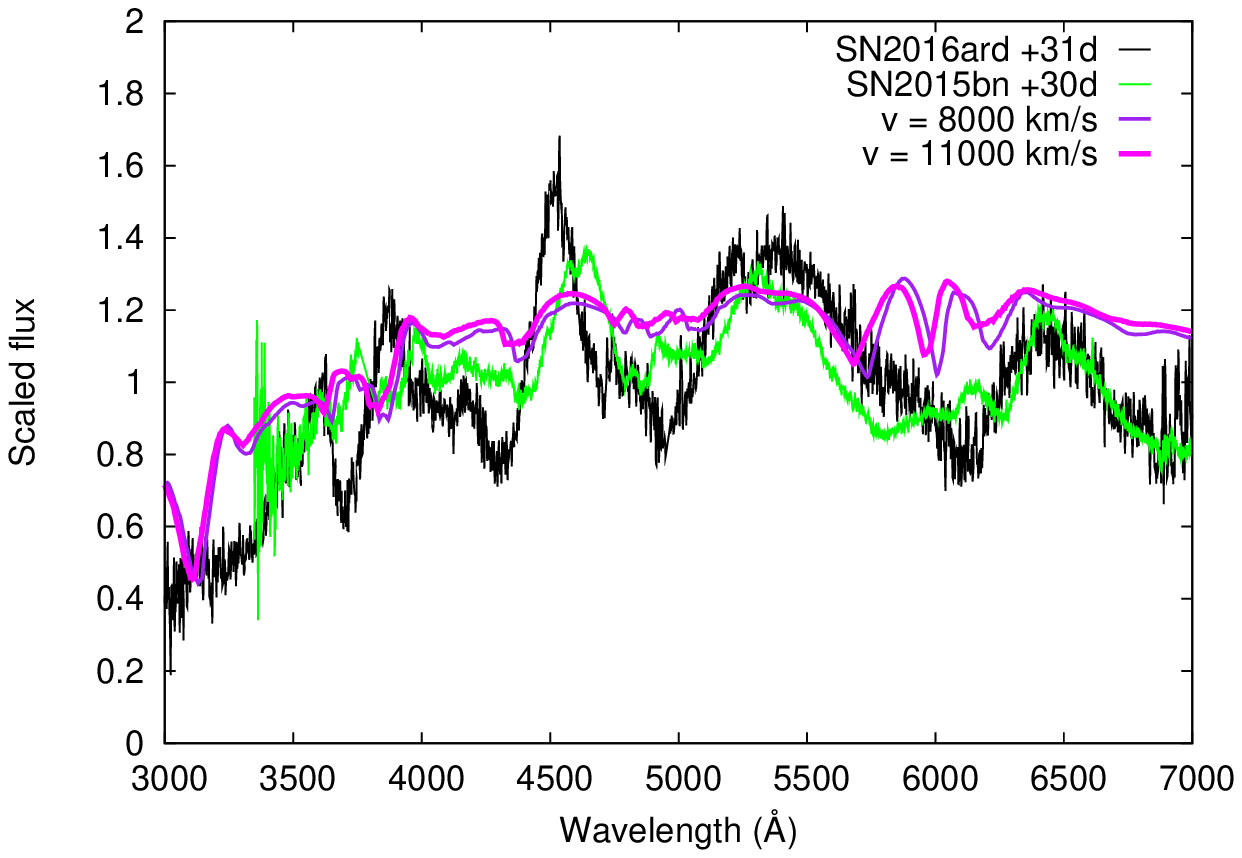}
\includegraphics[width=5.5cm]{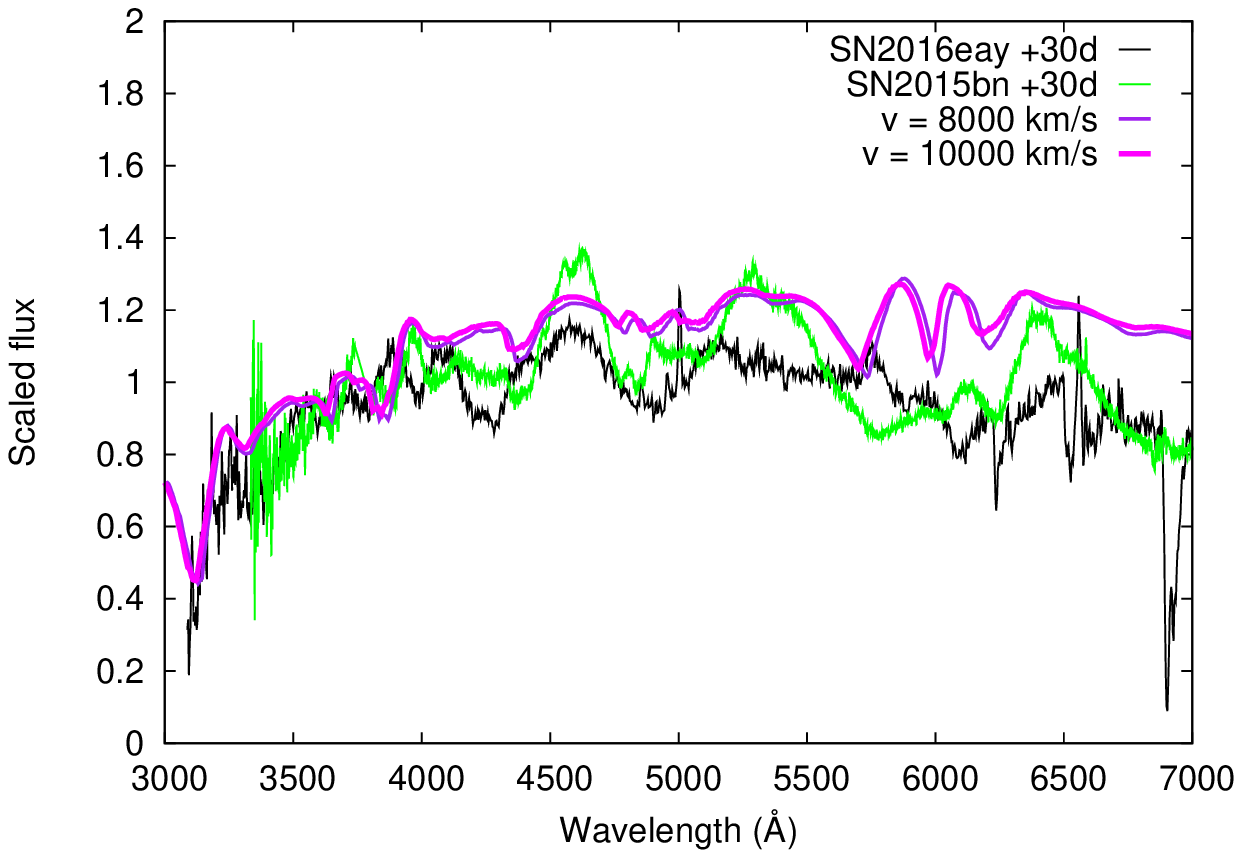}
\includegraphics[width=5.5cm]{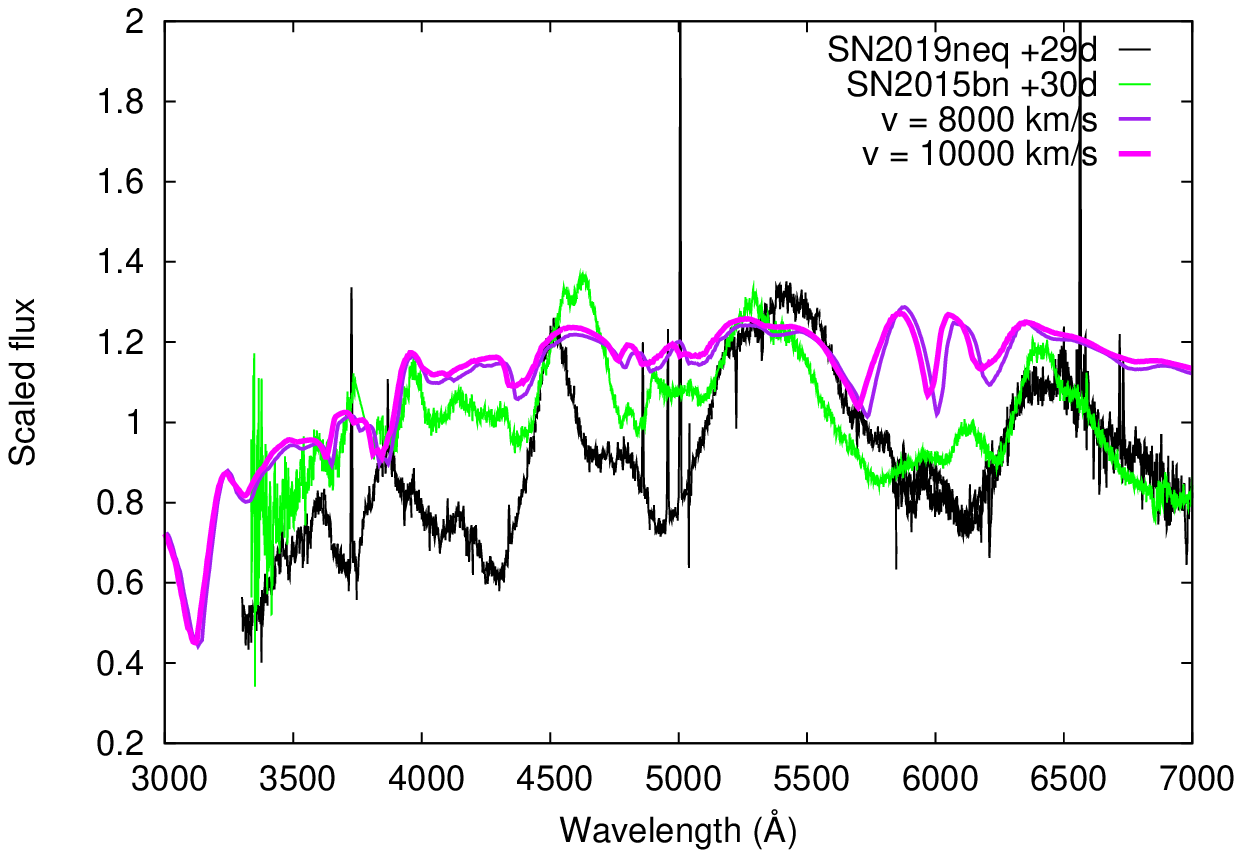}
\includegraphics[width=5.5cm]{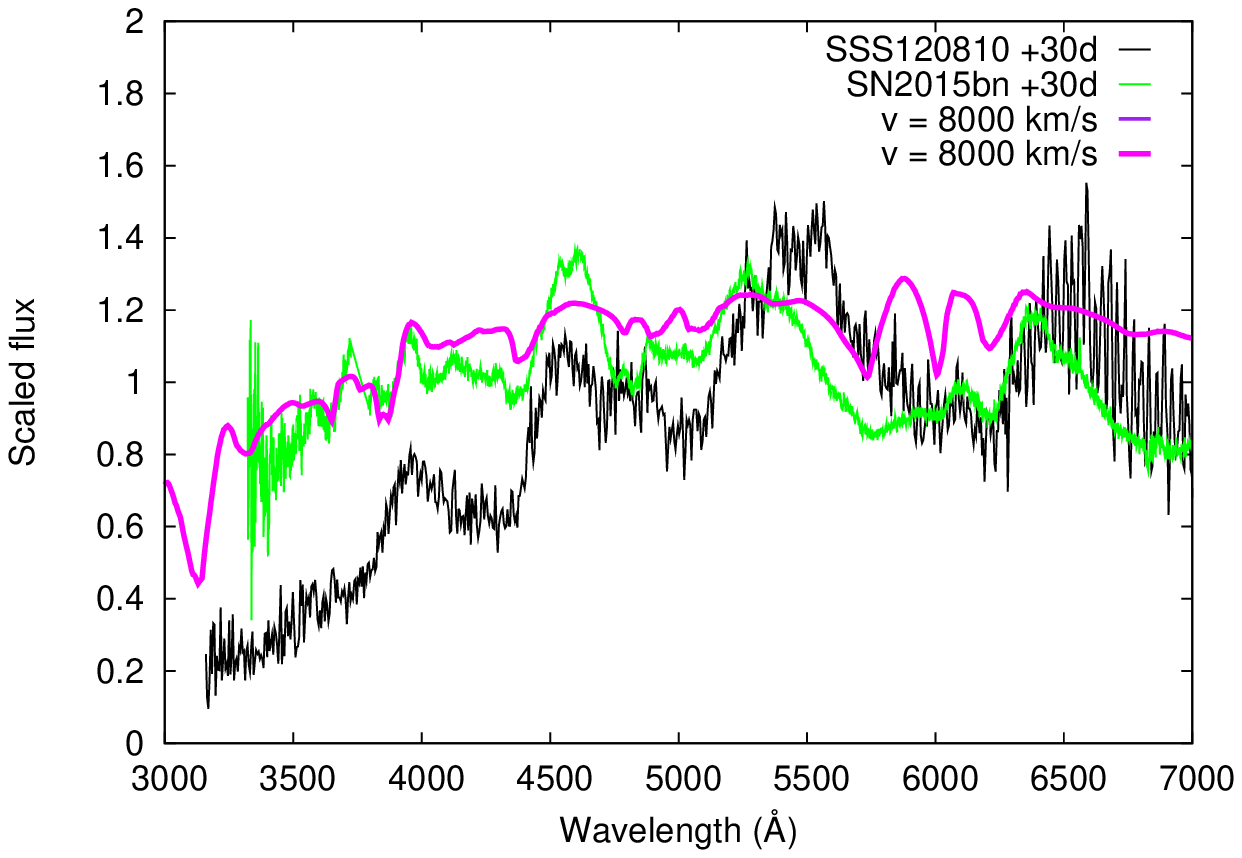}
\caption{The available post-maximum spectra of 9 SLSNe in our sample (black), with the +30 days phase spectrum of SN~2015bn Doppler-shifted with the velocity difference obtained with IRAF (green). The purple curves refer to the best-fit SYN++ model for SN~2015bn, but Doppler-shifted to $v_{\rm phot}~=~ 8000$ km s$^{-1}$. The magenta spectrum is the same SYN++ model shifted to the inferred velocity of each SLSN.}
\label{fig:maxutan_spmodel}
\end{figure*}

Afterwards, we cross-correlated the observed post-maximum spectra with the +30 days phase spectrum of SN~2015bn, instead of a SYN++ model, and then Doppler-shifted them with the $\Delta v_{\rm phot}$ differences from SN~2015bn calculated via {\tt fxcor} and Eq.\ref{eq:keresztkorr_maxutan}. Figure \ref{fig:maxutan_spmodel} displays the available post-maximum spectra of 9 SLSNe in our sample (black), together with the +30 days phase spectrum of SN~2015bn Doppler-shifted with the velocity difference obtained with IRAF for each objects (green). The best-fit SYN++ model for SN~2015bn having  $v_{\rm phot}~=~ 8000$ km s$^{-1}$, and the best-fit model referring to the particular SLSN are also plotted with purple and magenta curves, respectively. 

The photospheric velocity estimates for the post-maximum phase spectra of the 9 available objects can be found in Table \ref{tab:vphotok}.

\subsection{Fast/Slow classification}\label{subsec:F/s}

In the case of the 9 events, for which both pre- and post-maximum spectra were available, the classification into the Fast or Slow category was unambiguous. The estimated $v_{\rm phot}$ values before and after maximum can be found in Table \ref{tab:vphotok}. 

\begin{figure}
\centering
\includegraphics[width=8cm]{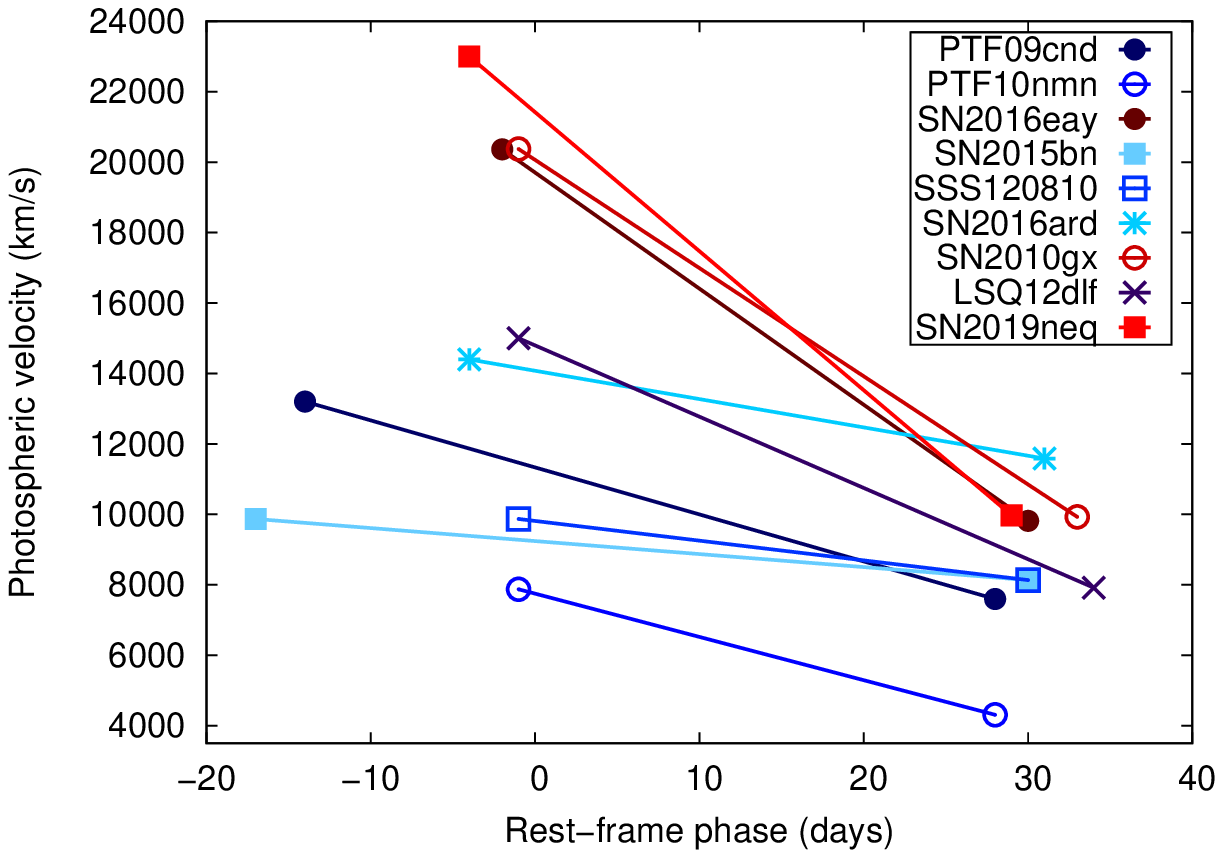}
\caption{Photospheric velocity evolution of the 9 SLSNe that possessed both pre-maximum and $\sim$30 days post-maximum spectra as well. Red symbols denote objects having  $v_{\rm phot} \geq 20000 $ km~s$^{-1}$ near maximum (Fast SLSNe-I), while blue colors code the Slow SLSNe-I exhibiting $v_{\rm phot} \leq 16000 $ km~s$^{-1}$ and almost constant velocity evolution. }
\label{fig:fazis_vs_vphot}
\end{figure}

Figure \ref{fig:fazis_vs_vphot} displays the photospheric velocity evolution of the 9 SLSNe as a function of rest-frame phase relative to the moment of the maximum light. It is seen that SN~2010gx, SN~2016eay and SN~2019neq shows a factor of 2 higher $v_{\rm phot}$ near maximum than the rest of the sample, which decreases swiftly in the post-maximum phases. By $\sim$30 days after maximum their velocities become similar to those of the other 6 SLSNe. These rapidly evolving objects are plotted with different tones of red in Fig. \ref{fig:fazis_vs_vphot}, and they will be referred as Fast (F) Type I SLSNe from now. 
The fast evolution of these objects is consistent with previous studies \citep[e.g.][respectively]{Pastorello10,inserra18,ktr20-2}.

On the contrary, the velocity of PTF09cnd,PTF10nmn, SN~2015bn, SSS120810, SN~2016ard and LSQ12dlf evolves more slowly: it seems to be nearly constant throughout the observed epochs. 
It is seen also that these 6 objects, plotted with different tones of blue in Fig. \ref{fig:fazis_vs_vphot}, are significantly different from the Fast ones in terms of the photospheric velocity evolution, thus, they are to be called Slow (S) SLSNe. This classification is consistent with \citet{inserra18}, who pointed out that the fast-evolving Type I SLSNe tend to have larger velocity gradients and higher $v_{\rm phot}$ at maximum than Slow SLSNe-I.

It is also seen in Figure \ref{fig:fazis_vs_vphot} that the Fast SLSNe are not only swifter in their $v_{\rm phot}$ evolution, but their near-maximum velocity is significantly higher compared to the Slow ones, in good agreement with \citet{inserra18}. The photospheric velocity of Fast objects become similar to the $v_{\rm phot}$ of Slow SLSNe by $\sim$30 days after maximum. Therefore, one can distinguish between these two groups of SLSNe-I by comparing their photospheric velocity near maximum. In the followings, we designate the SLSNe-I having $v_{\rm phot} \geq 20000 $ km s$^{-1}$ as Fast, and the objects with $v_{\rm phot} \leq 16000 $ km s$^{-1}$ as Slow, as it can be seen in Table \ref{tab:mej}. Since no post-maximum data of SLSNe-I having $16000 \leq v_{\rm phot} \leq 20000 $ km s$^{-1}$ at maximum were available, 2 objects from our sample, SN~2016els and SN~2017faf, could not be classified into the F or S subgroup unambiguously. Thus, they are referred as "uncertain" (N) SLSNe-I in Table \ref{tab:mej}.

\begin{figure}
\centering
\includegraphics[width=8cm]{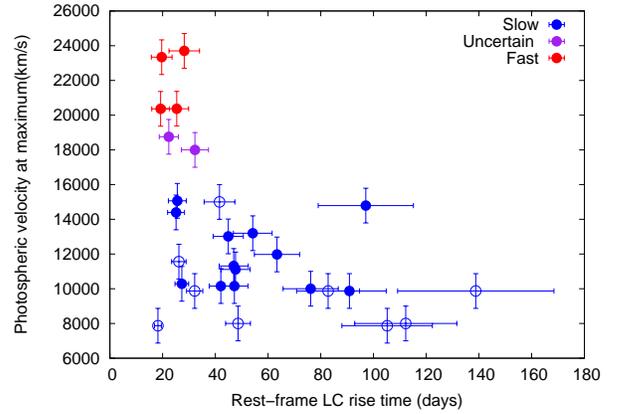}
\caption{The photospheric velocity estimates for the 28 SLSNe in our sample as a function of the light curve (LC) rise time. Filled dots denote ``Type W'' objects, while empty circles are the ``Type 15bn'' SLSNe-I. Red, purple and blue colors code the Fast (F), the "uncertain" (N), and the Slow (S) categories, respectively. }
\label{fig:trise_vs_vphot}
\end{figure}

In Figure \ref{fig:trise_vs_vphot}, the near-maximum $v_{\rm phot}$ estimates can be seen as a function of the rest-frame light curve (LC) rise time for all SLSNe in our sample. The latter was inferred from Eq.\ref{eq-trise} using the date of explosion ($t_0$) and the moment of the maximum ($t_{max}$) for each object shown in Table \ref{tab:mej}.  ``Type W'' SLSNe are plotted with filled symbols, while empty circles denote to the ``Type 15bn'' SLSNe. Red, purple and blue colors code the Fast, the "uncertain", and the Slow evolving objects, respectively. It is seen that the SLSNe classified as Fast by their photospheric velocity near maximum are showing short LC rise time scales as well. On the contrary, the objects having $v_{\rm phot} \leq 16000 $ km s$^{-1}$ at maximum exhibit quite diverse LC evolution time scales. It is also apparent that all ``Type 15bn'' events (at least those that are analyzed in this paper) belong to the slow-evolving SLSN-I group.

\section{Ejecta mass estimates}\label{sec:mej}

The photosheric velocity estimates presented in Section \ref{sec:vphot} open the door to derive the ejecta mass of the SLSNe in our sample by applying Eq. \ref{eq-mej1} and \ref{eq-mej2}. To infer the LC rise time ($t_{\rm rise}$), 
we used the date of the explosion and the moment of maximum light obtained from the Open Supernova Catalogue for all objects (see Table \ref{tab:mej}). 

Ejecta masses calculated from Eq.\ref{eq-mej1} and \ref{eq-mej2} can be found in Table \ref{tab:mej} amongst the estimated $t_{\rm rise}$ and $v_{\rm phot}$ values. We denote the masses inferred from Eq.\ref{eq-mej1} and \ref{eq-mej2} as $M_{\rm ej}$(\ref{eq-mej1}) and $M_{\rm ej}$(\ref{eq-mej2}), respectively. We consider their mean, named as $M_{\rm ej}$(mean) in Table \ref{tab:mej}, as our final mass estimate, and the difference between $M_{\rm ej}(\ref{eq-mej1})$ 
and $M_{\rm ej}(\ref{eq-mej2})$ as the systematic uncertainty of our ejecta mass estimate:
$\sigma_{\rm sys} \approx 0.5 \cdot (M_{\rm ej}(\ref{eq-mej1}) - M_{\rm ej}(\ref{eq-mej2}))$. 

The random errors of $M_{\rm ej}$, $\sigma_{\rm rnd}$, due to the uncertainty of the measured $v_{\rm phot}$ and $t_{\rm rise}$ (estimated as $\delta v_{\rm phot} \sim 1000$ km s$^{-1}$ and $\delta t_{\rm tise} \sim 3$ days) were also inferred using error propagation. Both $\sigma_{\rm sys}$ and $\sigma_{\rm rnd}$ are given in Table \ref{tab:mej} for each object.

It is seen that the ejecta masses for the whole sample are in the range from 2.9 ($\pm$0.8) to 208 ($\pm$60) $M_\odot$. The mean values are  $\langle M_{\rm ej}  \rangle_{\rm ALL}~=~42.96 \pm 12.50 ~ M_\odot$ for the 28 events,  $\langle M_{\rm ej}\rangle_{\rm S}~=~ 49.07\pm 14.80 ~ M_\odot$ for the Slow SLSNe, and $\langle M_{\rm ej}\rangle_{\rm F}~=~ 14.00\pm 6.20 ~ M_\odot$ for the Fast and uncertain ones.

\begin{figure}
\centering
\includegraphics[width=8cm]{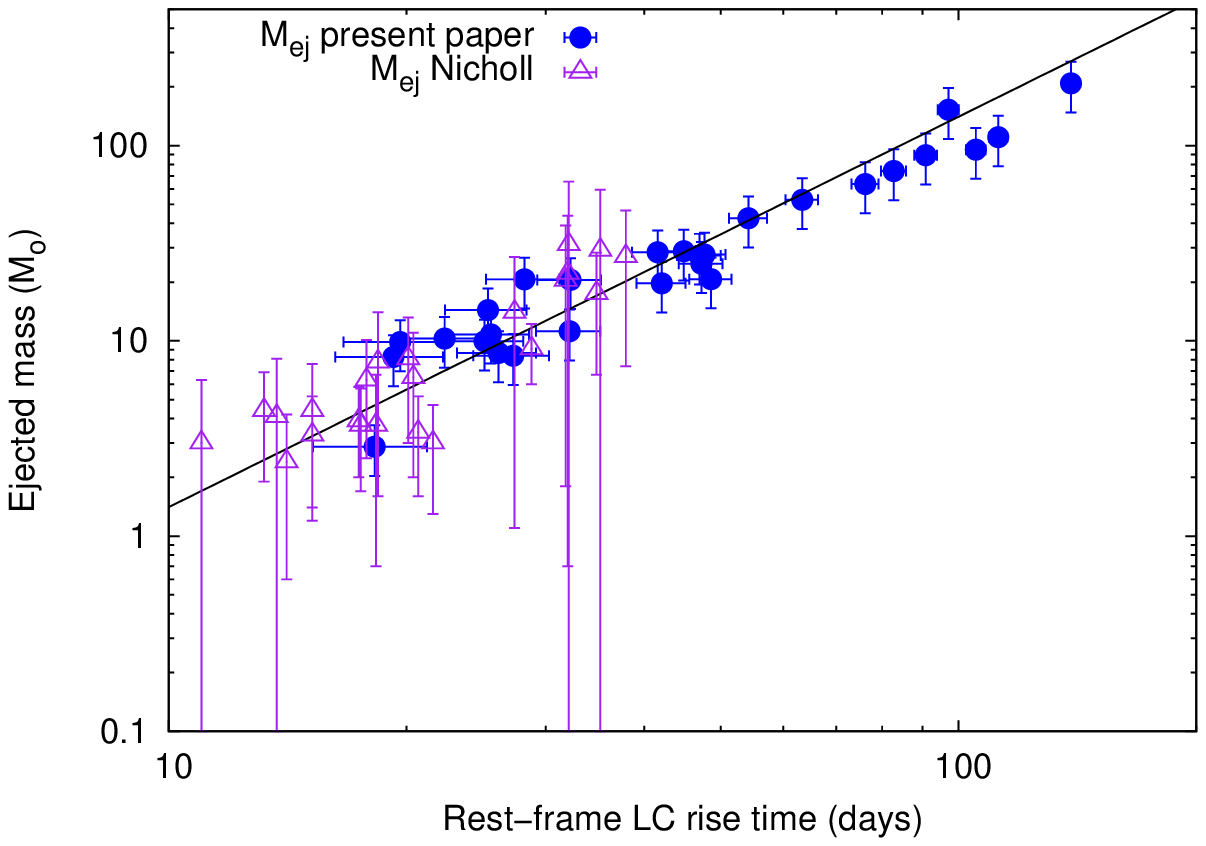}
\caption{Mean ejecta masses from Table~\ref{tab:mej}
(blue dots) as a function of the light curve rise time, illustrating that faster LC rise implies smaller ejecta mass. The black line represents the $M_{\rm ej}({\rm mean}) \sim t_{\rm rise}^2$ relation assuming a uniform expansion velocity of $10000$ km~s$^{-1}$ and $\kappa = 0.1$ cm$^2$~g$^{-1}$ for all objects. The ejecta mass estimates for Type I SLSNe from \citet{nicholl15} are plotted with purple triangles. It is seen that these are systematically lower compared to our mean ejecta mass estimates, but obey the same $M_{ej} \sim t_{rise}^2$ relation.}
\label{fig:trise_vs_mej2}
\end{figure}

Figure \ref{fig:trise_vs_mej2} displays the inferred ejecta masses (blue points) as a function of the light curve rise time scale. It is seen also that the logarithm of the ejecta mass is directly proportional to the logarithm of the LC rise time. It implies that SLSNe having longer rise time tend to have larger ejecta masses compared to the faster evolving objects.

\begin{table*}
\caption{Results for the light curve rise times and ejecta masses of the studied SLSNe-I. }
\label{tab:mej}
\begin{center}
\begin{tabular}{lcccccccccc}
\hline
\hline
SLSN & $t_{\rm obs}$[phase] &$t_{\rm rise}$  & $M_{\rm ej}$(\ref{eq-mej1}) & $M_{\rm ej}$(\ref{eq-mej2}) & $M_{\rm ej}$(mean) & $\sigma_{\rm sys}$ & $\sigma_{\rm rnd}$ & $v_{\rm phot}$ & W/15bn & F/S/N  \\ 
     & (days)   & (days)              & ($M_\odot$) & ($M_\odot$)  & ($M_\odot$) & ($M_\odot$)  & ($M_\odot$) & (km s$^{-1}$) & & \\
\hline
SN2005ap &  53436 [-3]& 19.64 & 12.74  & 6.99 &  9.86 &  2.87 &   3.93 &  23338 &  W & F  \\ 
SN2006oz &  54061 [-5]& 25.58 & 13.95  & 7.66 &  10.80 &  3.14 &   3.40 &  15064 &  W & S  \\ 
SN2010gx &55276 [-1]&  25.37 & 18.55  & 10.19 &  14.37 &  4.18 &   4.48 &  20371 &  W & F  \\ 
SN2010kd &55528 [-22] &  47.68 & 35.75  & 19.63 &  27.69 &  8.06 &   5.53 &  11112 &  W & S  \\ 
SN2011kg & 55926 [-10]& 26.17 & 11.20  & 6.15 &  8.68 &  2.53 &   2.75 &  11562 &  15bn & S  \\ 
SN2015bn & 57082 [-17] & 90.88 & 115.34  & 63.33 &  89.34 &  26.01 &   13.95 &  9870 &  W & S  \\ 
SN2016ard &  57449 [-4]& 25.11 & 12.85  & 7.06 &  9.95 &  2.90 &   3.20 &  14398 &  W & S  \\ 
SN2016eay &  57528 [-2]& 19.25 & 10.68  & 5.86 &  8.27 &  2.41 &   3.37 &  20362 &  W & F  \\ 
SN2016els & 57599 [-5]& 22.35 & 13.26  & 7.28 &  10.27 &  2.99 &   3.63 &  18754 &  W & N  \\ 
SN2017faf & 57934 [-7]& 32.26 & 26.51  & 14.56 &  20.54 &  5.98 &   5.15 &  18000 &  W & N  \\ 
SN2018bsz & 58259 [-16]& 76.17 & 82.09  & 45.08 &  63.58 &  18.51 &   10.45 &  10000 &  W & S  \\ 
SN2018hti & 58428 [-54] & 97.08 & 197.25  & 108.31 &  152.78 &  44.47 &   18.07 &  14790 &  W & S \\ 
SN2018ibb & 58453 [-11] & 112.24 & 142.61  & 78.30 &  110.45 &  32.15 &   19.39 &  8000 &  15bn & S \\ 
SN2019neq & 58722 [-4]& 28.21 & 26.69  & 14.66 &  20.68 &  6.02 &   5.79 &  23702 &  W & F  \\ 
DES14X3taz & 57059 [-29]& 44.90 & 37.13  & 20.39 &  28.76 &  8.37 &   5.72 &  13017 &  W & S  \\ 
iPTF13ajg & 56422 [-5]& 47.24 & 32.07  & 17.61 &  24.84 &  7.23 &   5.15 &  10155 &  W & S  \\ 
iPTF13ehe & 56658 [-14]& 82.78 & 95.69  & 52.54 &  74.12 &  21.58 &   11.92 &  9870 &  15bn & S \\ 
LSQ12dlf &  56149 [-1]& 41.59 & 36.72  & 20.16 &  28.44 &  8.28 &   5.84 &  15000 &  15bn & S  \\ 
LSQ14an & 56660 [0]& 18.23 & 3.70  & 2.03 &  2.87 &  0.83 &   1.31 &  7870 &  15bn & S \\ 
LSQ14bdq &  56784 [-11]& 46.99 & 35.35  & 19.41 &  27.38 &  7.97 &   5.49 &  11314 &  W & S  \\ 
LSQ14mo & 56694 [-1]& 27.29 & 10.84  & 5.95 &  8.40 &  2.44 &   2.61 &  10284 &  W & S  \\ 
PS1-14bj & 56744 [-42]& 138.81 & 269.11  & 147.76 &  208.44 &  60.67 &   29.64 &  9870 &  15bn & S  \\ 
PTF09atu &  55034 [-19]& 42.09 & 25.46  & 13.98 &  19.72 &  5.74 &   4.41 &  10155 &  W & S \\ 
PTF09cnd & 55068 [-14]& 54.20 & 54.86  & 30.12 &  42.49 &  12.37 &   7.36 &  13199 &  W & S  \\ 
PTF10nmn & 55384 [-1]& 105.19 & 123.22  & 67.66 &  95.44 &  27.78 &   17.16 &  7870 &  15bn & S \\ 
PTF12dam & 56072 [-17]& 63.39 & 68.11  & 37.40 &  52.75 &  15.36 &   8.60 &  11978 &  W & S  \\ 
PTF12gty &56135 [-4]& 48.61 & 26.74  & 14.68 &  20.71 &  6.03 &   4.70 &  8000 &  15bn & S \\ 
SSS120810 & 56135 [-4]& 32.18 & 14.46  & 7.94 &  11.20 &  3.26 &   3.07 &  9869 &  15bn & S  \\ 
\hline
\end{tabular}
\end{center}
\end{table*}

\subsection{Comparison with Nicholl et al. (2015)}

We compared our results to the calculations of \citet{nicholl15}, who inferred the ejecta mass of a sample of normal and superluminous supernovae via modeling their bolometric light curves. They utilized an alternative way of using the formula of \citet{arnett80} (Eq. \ref{eq-mej2}) by estimating the mean light curve time scale as $t_{\rm m}~=~ 0.5 \cdot (t_{\rm rise} + t_{\rm dec} )$, where $t_{\rm dec}$ is the LC decline time scale. They defined $t_{\rm rise}$, as the time ($t~<~0$) relative to maximum light ($L_{\rm max}$) at which $L_{\rm griz}~=~L_{\rm max}/e$, and $t_{\rm dec}$ as as the time ($t~>~0$) relative to maximum light ($L_{\rm max}$) at which $L_{\rm griz}~=~L_{\rm max}/e$. This is different from both the original definition of \citet{arnett80}, who used $t_{\rm m} = \sqrt{2 t_d t_h}$ (cf. Eq.\ref{eq-tdif} and \ref{eq-th}), and from our definition of $t_{\rm rise}$ (cf. Eq.\ref{eq-trise}) as well.

\citet{nicholl15} utilized a different method to estimate the photospheric velocity as well, based on the Fe II $\lambda$5169 lines in the spectra, obtaining significantly different values from the $v_{\rm phot}$ calculations presented in this study. However, as shown in e.g. \citet{ktr20-2}, the identification of the Fe II $\lambda$5169 line suffers from ambiguity. Thus, we believe that the modeling of the W-shaped O II feature or the whole spectra provide a more reliable method to estimate the photospheric velocities. 

In Figure \ref{fig:trise_vs_mej2}, the ejecta mass calculations of \citet{nicholl15} are plotted as a function of their LC rise time scales with purple triangles, in order to compare them to the $M_{\rm ej}$ calculations of this paper (shown with blue dots). It is seen that the ejecta masses published by \citet{nicholl15} are systematically smaller, than the $M_{\rm ej}$ values calculated in this study, due to the different method to calculate the light curve rise-time scales, and estimate the photospheric velocities.

\section{Discussion}\label{sec:discussion}

The main goal of this study was to derive the ejecta masses of all SLSNe having public pre-maximum photomeric and spectroscopic observational data in the Open Supernova Catalogue before 2020. To obtain $M_{\rm ej}$, we utilized the formulae of \citet{arnett80}, summarized in Section \ref{sec:theory}.  Pre- or near-maximum photospheric velocities were crucial to substitute into Eq. \ref{eq-mej1} and  Eq. \ref{eq-mej2}, thus we developed a method to determine the $v_{\rm phot}$ of each object in a fast and efficient way.

     We found that the W-shaped O II absorption blend, typically present between $\sim 3900$ and $\sim 4500$ \AA\ in the pre-maximum spectra of Type I SLSNe, is missing from the spectra of 9 SLSNe belonging to our sample. These events are found to be spectroscopically similar to SN~2015bn. Therefore, the studied 28 SLSNe were divided into two subtypes by the presence/absence of the W-shaped absorption: the ``Type W'' and the ``Type 15bn'' groups. 
     
     The expansion velocities around maximum light were then estimated for both groups by involving {\tt SYN++} synthetic models and cross-correlation, as described in Section~\ref{sec:vphot}. Furthermore, in order to distinguish between fast- and slow-evolving SLSNe, we repeated this procedure for those events that had public spectra taken around $\sim +30$ rest-frame days after maximum.    
     
     The fast or slow evolution of a SLSN can be decided from the photospheric velocity gradient between the $v_{\rm phot}$ measured at the maximum and +30 days phase. Fast SLSNe tend to have larger velocity gradients, while the objects belonging to the Slow group are characterized by much lower velocity gradients or nearly constant photospheric velocities through the observed epochs. This is consistent with the classification scheme of \citet{inserra18}.
     
    Fast evolving Type I SLSNe can also be distinguished from the Slow evolving objects by their $v_{\rm phot}$ at maximum light: Fast SLSNe-I tend to have  $v_{\rm phot} \geq 20000$ km s$^{-1}$, while, on the contrary, Slow SLSNe-I usually have $v_{\rm phot} \leq 16000$ km s$^{-1}$ instead (see Figure \ref{fig:fazis_vs_vphot}).
    
     In some cases, the Fast/Slow classification of a particular object presented in this paper differs from the results of other studies. Five objects out of 28 in our sample (PTF09cnd, PTF10nmn, LSQ14mo, SN~2016ard, and iPTF2016ajg) that were found to be Slow by their $v_{\rm phot}$ evolution are referred to as Fast SLSNe in e.g. \citet{inserra18,chen17,u}, and \citet{yu17}, respectively.
     
     The classification of LSQ12dlf is ambiguous as well: in this paper and according to \citet{yu17}, it seems to be  a slow-evolving SLSN, but \citet{inserra18} classified it as a Fast one. The cause of this inconsistency can be found in the different definition of the Fast or Slow evolution: \citet{inserra18} found that all objects having $v_{\rm phot} \geq 12000$ km s$^{-1}$ at maximum  belong to the Fast class according to their definition, while we found the threshold being at $v_{\rm phot} \geq 16000$ km s$^{-1}$, near $\sim 20000$ km s$^{-1}$ in this study. As displayed in Figure \ref{fig:fazis_vs_vphot}, SN~2016ard has $v_{\rm phot} \geq 12000$ km s$^{-1}$ with one of the flattest velocity gradients, while LSQ12dlf shows $v_{\rm phot} \leq 16000$ km s$^{-1}$ with a medium slope in velocity evolution.

     The SLSNe classified into the Fast evolving group by their photospheric velocity measured at maximum are belonging to the Fast Type I SLSN subgroup by their light curve rise times as well. On the contrary, the objects found to be slowly evolving by $v_{\rm phot}$ are quite diverse in $t_{\rm rise}$, ranging in between a few weeks and $\sim 150$ days (see Figure \ref{fig:trise_vs_vphot}).

    All SN~2015bn-like SLSNe are classified to the Slow group by their $v_{\rm phot}$, while amongst the ``Type W'' events both Fast and Slow objects are represented.

    The mean and the range of the estimated ejecta masses for the 28 SLSNe in our sample, $\langle M_{\rm ej}\rangle_{\rm ALL}~=~42.96 \pm 12.50~ M_\odot$ between 3 and 208 $M_\odot$, are significantly higher than the $M_{\rm ej}$ estimates presented by \citet{nicholl15} ($\langle M_{\rm ej}\rangle \sim 10$ $M_\odot$, between 3 and 30  $M_\odot$ for their sample). The difference is caused by the different method to calculate photospheric velocities and LC time scales.  

It is also interesting that the mean mass of the Fast events (including the uncertain ones) in our sample 
($\langle M_{\rm ej}\rangle_{\rm F}~=~ 14.00\pm 6.20 ~ M_\odot$) is significantly lower than that of the Slow events (
$\langle M_{\rm ej}\rangle_{\rm S}~=~ 49.07\pm 14.80 ~ M_\odot$). At first glance this might suggest that the physical cause of the Fast/Slow dichotomy could be related to the amount of the ejected envelope. However, as the sample is still very poor (only 28 objects), more data are strongly needed to be able to draw more reliable conclusion. 

\section{Summary}\label{sec:summary}

We have presented photospheric velocity estimates and ejecta mass calculations of a sample containing 28 Type I superluminous supernovae having publicly available photometric and spectroscopic data in the Open Supernova Catalogue \citep{guill17}. 

We utilized the formulae of the radiation-diffusion model of \citet{arnett80} to estimate the ejecta masses. The LC rise time and the photospheric velocity before or near the luminosity maximum was necessary to obtain $M_{\rm ej}$ values. 

The photospheric velocities of the sample SLSNe were estimated utilizing a method by combining the spectrum modeling with cross-correlation, similar to \citet{takats12}. It was found that the W shaped O II absorption blend, typically present in the pre-maximum spectra of Type I SLSNe is missing from the spectra of several objects that otherwise have very similar features to SN~2015bn. Thus, two groups of the sample SLSNe were created (called ``Type W'' and ``Type 15bn''), and their $v_{\rm phot}$ values were obtained using different {\tt SYN++} model spectra as templates in the cross-correlation. 

Post-maximum  $v_{\rm phot}$ values of 9 SLSNe with available spectra were also estimated in a similar way in order to to classify these events into the Fast or the Slow SLSN subtypes by calculating the velocity gradients between the maximum, and +30 rest-frame days. Fast SLSNe showed considerably higher velocity gradiens than Slow ones, in good agreement with \citet{inserra18}. 

These calculations also confirmed that Fast SLSNe generally show higher velocities close to maximum than Slow events. This allowed us to classify other SLSNe in our sample that did not have public spectra around +30 days. Thus, 
we considered the SLSNe having $v_{\rm phot} \geq 20000 $ km s$^{-1}$ near maximum as Fast (F), and the events with $v_{\rm phot} \leq 16000 $ km s$^{-1}$ as Slow (S) events. 

Amongst the studied SLSNe, the Fast evolving objects defined by the photospheric velocities were revealed to show a rapidly evolving light curve with a short LC rise time as well.
On the contrary, Slow evolving events having lower $v_{\rm phot}$ had more diverse LC rise time scales, ranging from a few weeks to $\sim$150 days. It was also found that all ``Type 15bn'' events belong to the Slow evolving SLSN-I subgroup defined by $v_{\rm phot}$, while ``Type W'' objects were represented in both the Fast and Slow groups.

Ejecta mass calculations of the SLSNe in our sample were carried our using Eq. \ref{eq-mej1} and \ref{eq-mej2}, resulting in  masses  within a range of 2.9 ($\pm 0.8$) - 208 ($\pm 61$) $M_\odot$, having a mean of  $\langle M_{\rm ej}\rangle = 42.96 \pm 12.50 ~M_\odot$. This is significantly larger than the $\langle M_{\rm ej} \rangle$ calculated by \citet{nicholl15}, who obtained $\langle M_{\rm ej} \rangle \sim 10 ~ M_\odot$ between 3 and 30  $M_\odot$ for a different sample of SLSNe I with different methods to estimate the photospheric velocities and LC evolution timescales.

The mean ejecta mass of Slow SLSNe in our sample ($\sim 49 \pm 15 ~M_{\odot}$) seems to be 
higher than that of the Fast ones ($\sim 14 \pm 6 ~M_{\odot}$), suggesting a physical link between the Fast/Slow dichotomy and the ejecta mass. However, since it is based on only 28 (24 Slow and 4 Fast) objects, more data are inevitable for a more reliable conclusion. 

Our ejecta mass estimates further strengthen the long-standing concept that SLSNe probably originate from a range of moderately massive to very massive progenitors, and their (still uncertain) explosion mechanism is able to eject large amount of their envelope mass.

\acknowledgments
Our study is supported by the project ``Transient Astrophysical Objects" GINOP 2.3.2-15-2016-00033 of the National Research, Development and Innovation Office (NKFIH), Hungary, funded by the European Union.

{}

\section{Appendix}

\setcounter{table}{0}
\renewcommand{\thetable}{A\arabic{table}}

\setcounter{figure}{0}
\renewcommand{\thefigure}{A\arabic{figure}}

Table \ref{tab:removed} summarizes the selection process of out sample. Here, we describe in details the cause of removing 13 SLSNe-I having pre-maximum spectra.  
\begin{itemize}
    \item {The pre-maximum spectra of {\bf SN~2019szu, SCP-06F6} and {\bf OGLE15qz} were so noisy that spectral features could not be identified at all.}
    \item {{\bf DES15E2mlf, SNLS-06D4eu} and  {\bf SNLS-07D2bv} were observed only in the UV bands up to 3000 \AA. }
    \item {The spectra taken of {\bf SN~2010md, PTF10vqv, SN~2016aj,} and {\bf SN~2010uhf} did not contain the typical SLSN-I spectral features, or the W-like absorption between 3900 and 4500 \AA, which is usually present in the pre-maximum spectra of SLSNe-I.}
    \item {The selected spectrum of {\bf SN~2007bi} was actually taken after the maximum.}
    \item {{\bf SN~2015L}, the most luminous "SLSN" ever seen is presumably a tidal disruption event \citep[e.g.][]{lelo16,margutti17,coughlin18}. If we assume it to be a SLSN, it interacts so robustly that the photosphere is not even visible, thus it is impossible to estimate the photospheric velocity in the maximum.}
    \item {{\bf SN~2017gir} had a spectrum more similar to a Type II SLSN. }
\end{itemize}

\begin{figure*}
\centering
\includegraphics[width=8.8cm]{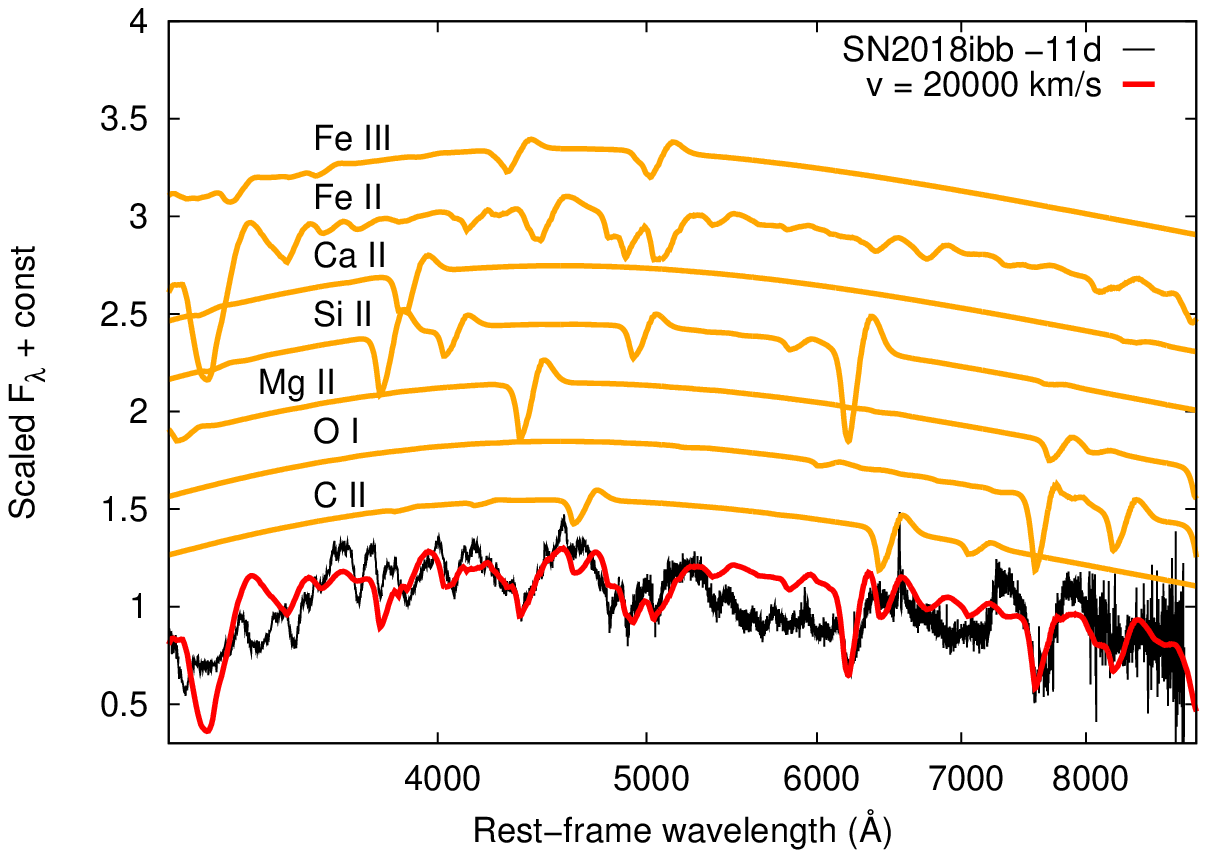}
\includegraphics[width=8.8cm]{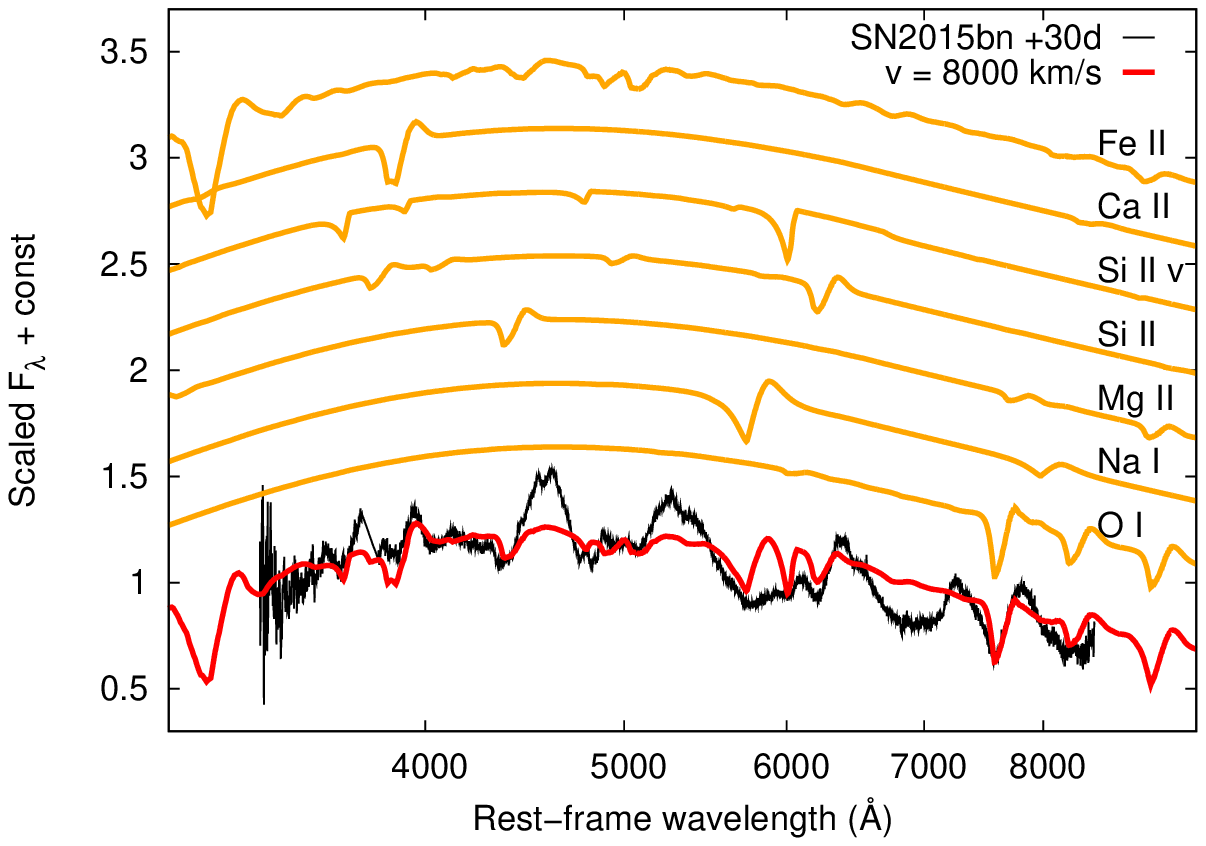}
\caption{Left panel: The -11 days rest-frame phase spectrum of SN~2018ibb (black), belonging to the Type 15bn group of SLSNe-I together with its best-fit model obtained in SYN++ (red). Single ion contributions to the overall model spectrum are plotted with orange, shifted vertically to guide the eye. Right panel: The +30 days phase spectrum of SN~2015bn plotted the same way as the figure on the left. }
\label{fig:18ibb15bn_models}
\end{figure*}

In Fig. \ref{fig:18ibb15bn_models}, the SYN++  modeling of the -11 days phase spectrum of SN~2018ibb (Type 15bn; left panel) and  the +30 days phase spectrum of SN~2015bn (right panel) is plotted.

Table \ref{tab:syn_alltypes} summarizes the global and local SYN++ parameters obtained for Type W, Type 15bn, and post-maximum SLSN spectra.

\begin{table*}
\caption{SLSNe removed from our sample. }
\label{tab:removed}
\begin{center}
\scriptsize
\begin{tabular}{ll}
\hline
\hline
Reason for exclusion [number] & SLSN \\ 
\hline

SLSNe-II [18] & SN2006gy, SN1000+0216, SN2008am, SN2008es, CSS121015:004244+132827, PTF12mkp, SN2013hx,  PS15br, \\ 
& LSQ15abl, SN2016aps, 
SN2016ezh, SN2016jhm, SN2016jhn, SN2017bcc, SN2017egm, SN2018jkq, SN2019cmv, \\ &
SN2019meh \\

\hline 

Without pre-maximum spectra [39] & SN2213-1745, SDSS-II SN 2538, SDSS-II SN17789, SN2009cb, SN2009jh, PTF10bfz, PTF10bjp, PS1-10pm, PS1-10ky, \\ &

PS1-11tt, PS1-10ahf, SN2010hy, PS1-10awh, PTF10aagc, PS1-10bzj, 
PS1-11ap, SN2011ke, PS1-11afv, PTF11hrq, \\ & SN2011kl, SN2011kf,
SN2012il, PTF12mxx, SN2013dg, SN2013hy, CSS130912:025702-001844, PS15cjz, \\ & OGLE15sd, PS16yj, iPTF16bad, DES16C2nm, AT2016jho, SN2017jan, DES17C3gyp, SN2018bgv, SN2018gkz, \\ &  SN2018lfd, SN2019meh, SN2019szu \\

\hline

Problem with spectra [13] & SN~2019szu, SCP-06F6, OGLE15qz, DES15E2mlf, SNLS-06D4eu,  SNLS-07D2bv, SN~2010md, PTF10vqv, SN~2016aj, \\ & SN~2010uhf,  SN~2007bi,  SN~2015L, SN~2017gir \\
\hline
\end{tabular}

\end{center}
\end{table*}

\begin{table*}
\caption{Global and local SYN++ parameters for the studied Type W, Type 15bn, and post-maximum phase SLSNe.}
\label{tab:syn_alltypes}
\centering
\begin{tabular}{cccccc}
\hline
\hline
\multicolumn{6}{c}{{\bf Type W SLSNe}} \\
\hline
\multicolumn{6}{c}{Global parameters} \\

 $a_0$ & $v_{\rm phot}$ & $T_{\rm phot}$ & \multicolumn{3}{c}{} \\
  & (km~s$^{-1}$) & ($10^3$ K) & \multicolumn{3}{c}{} \\

  1.0 &  10000-30000 & 15000 & \multicolumn{3}{c}{}                               \\
\hline
\multicolumn{6}{c}{Local parameters} \\

Element & $\log\tau$ & $v_{\rm min}$ & $v_{\rm max}$ & $aux$ & $T_{\rm exc}$ \\
     &           & ($10^3$ km~s$^{-1}$) & ($10^3$ km~s$^{-1}$) & ($10^3$ km~s$^{-1}$) & ($10^3$ K) \\
\hline

O II & -2.0 & $v_{\rm phot}$ & 50.0 & 2.0 & 15.0 \\

\hline
\hline
\multicolumn{6}{c}{{\bf Type 15bn SLSNe}} \\
\hline
\multicolumn{6}{c}{Global parameters} \\

 $a_0$ & $v_{\rm phot}$ & $T_{\rm phot}$ &\multicolumn{3}{c}{} \\
  & (km~s$^{-1}$) & ($10^3$ K) &\multicolumn{3}{c}{} \\

  0.7 &  8000-30000 & 11000 &\multicolumn{3}{c}{}                              \\
\hline
\multicolumn{6}{c}{Local parameters} \\

Element & $\log\tau$ & $v_{\rm min}$ & $v_{\rm max}$ & $aux$ & $T_{\rm exc}$ \\
     &           & ($10^3$ km~s$^{-1}$) & ($10^3$ km~s$^{-1}$) & ($10^3$ km~s$^{-1}$) & ($10^3$ K) \\
\hline

C II & -1.4 & $v_{\rm phot}$ & 50.0 & 1.0 & 10.0 \\
O I & 0.3 & $v_{\rm phot}$ &  50.0 & 1.0 & 10.0 \\
Mg II & 0.0  & $v_{\rm phot}$ & 50.0 & 1.0 & 10.0 \\
Si II & 0.5 & $v_{\rm phot}$ & 50.0 & 1.0 & 10.0 \\
Ca II & 0.0  & $v_{\rm phot}$ & 50.0 & 1.0 & 10.0 \\
Fe II & 0.0 & $v_{\rm phot}$ & 50.0 & 1.0 & 12.0 \\
Fe III & -0.5 & $v_{\rm phot}$ & 50.0 & 1.0 & 10.0 \\
\hline
\hline
\multicolumn{6}{c}{{\bf SLSNe after maximum}} \\
\hline
\multicolumn{6}{c}{Global parameters} \\

 $a_0$ & $v_{\rm phot}$ & $T_{\rm phot}$ &\multicolumn{3}{c}{} \\
  & (km~s$^{-1}$) & ($10^3$ K) &\multicolumn{3}{c}{} \\

  0.7 &  5000-15000 & 9000 &\multicolumn{3}{c}{}                              \\
\hline
\multicolumn{6}{c}{Local parameters} \\

Element & $\log\tau$ & $v_{\rm min}$ & $v_{\rm max}$ & $aux$ & $T_{\rm exc}$ \\
     &           & ($10^3$ km~s$^{-1}$) & ($10^3$ km~s$^{-1}$) & ($10^3$ km~s$^{-1}$) & ($10^3$ K) \\
\hline
O I & 0.1 & $v_{\rm phot}$ & 50.0 & 1.0 & 10.0 \\
Na I & -0.5 & $v_{\rm phot}$ & 50.0 & 4.0 & 10.0 \\
Mg II & -0.5 & $v_{\rm phot}$ & 50.0 & 1.0 & 10.0 \\
Si II & -0.3 & $v_{\rm phot}$ & 50.0 & 1.0 & 10.0 \\
Si II v & 0.7 & 17.0 & 50.0 & 3.0 & 10.0 \\
Ca II & 0.0 & $v_{\rm phot}$ & 50.0 & 1.0 & 10.0 \\
Fe II & -0.5 & $v_{\rm phot}$ & 50.0 & 1.0 & 12.0 \\
\hline

\end{tabular}
\end{table*}

\end{document}